\documentclass[reprint,superscriptaddress,amsmath,amssymb,aps,pra]{revtex4-2}
\usepackage{graphicx}
\usepackage{mathtools}
\usepackage{xcolor}
\usepackage{physics}
\usepackage{amsmath}
\usepackage{float}
\usepackage{hyperref}
\usepackage[caption=false]{subfig}
\usepackage{lipsum}
\usepackage{bbold}
\usepackage{feynmp}
\usepackage{natbib}
\usepackage{accents}
\usepackage{leftindex}
\usepackage{dcolumn}
\usepackage{bm}

\newcommand{\blue}[1]{#1}

\begin{document}

\title{\blue{Far from equilibrium field theory for  strongly coupled light and matter:\\  dynamics of frustrated multi-mode cavity QED}}

\author{Hossein Hosseinabadi}
\email[hhossein@uni-mainz.de]{}
\affiliation{Institut f{\"u}r Physik, Johannes Gutenberg-Universit{\"a}t Mainz, 55099 Mainz, Germany}
\author{Darrick E. Chang}
\affiliation{ICFO—Institut de Ci{\`e}ncies Fot{\`o}niques, The Barcelona Institute of Science and Technology, 08860 Castelldefels, Spain}
\affiliation{ICREA—Instituci{\'o} Catalana de Recerca i Estudis Avan{\c c}ats, 08015 Barcelona, Spain}
\author{Jamir Marino}
\affiliation{Institut f{\"u}r Physik, Johannes Gutenberg-Universit{\"a}t Mainz, 55099 Mainz, Germany}

\begin{abstract}
Light-matter interfaces have now entered a new stage marked by the ability to engineer quantum correlated states under driven-dissipative conditions. To propel this new generation of experiments, we are confronted with the need to model non-unitary many-body dynamics in   strongly coupled regimes, by transcending traditional approaches in quantum optics.
In this work, we contribute to this program by adapting a functional integral technique, conventionally employed in high-energy physics, in order to obtain non-equilibrium dynamics for interacting light-matter systems. Our approach is grounded in constructing 'two-particle irreducible' (2PI) effective actions, which provide a non-perturbative and conserving framework for describing quantum evolution at a polynomial cost in time. We apply our method to complement the analysis of spin glass formation in the context of frustrated multi-mode cavity quantum electrodynamics, initiated in our accompanying work [H. Hosseinabadi, D. Chang, J. Marino, arXiv:2311.05682]. Finally, we   outline the capability of the technique  to describe other   near-term platforms in    many-body quantum optics, and its potential to make predictions for this new class of experiments. 

\end{abstract}

\maketitle

\section{Introduction}

 The field of quantum simulation has recently undergone a   transformation, evolving into a new realm of research where the worlds of condensed matter physics and quantum optics merge. Today, an increasing array of platforms hosting many-body systems can accommodate both unitary and dissipative dynamics in a controlled fashion~\cite{browaeys2020many,chang2014quantum,byrnes2014exciton,monroe2021programmable,blais2021circuit,PRXQuantum.2.017003}. This circumstance paves the way for the exploration of phases of matter and strongly correlated behavior that have no counterparts in either thermodynamic equilibrium or isolated non-equilibrium conditions.

In contrast to conventional solid-state systems, driven-dissipative condensed matter (a.k.a many-body quantum optics) systems exhibit a host of innovative characteristics. In the former, dissipation poses the primary challenge to quantum coherence, while in the latter, dissipation is at times intentionally harnessed or even engineered to drive the system into entangled states~\cite{verstraete2009quantum,harrington2022engineered,diehl2008quantum}.

Traditional solid-state physics focuses on understanding equilibrium phases of matter that result from the interplay of interactions and thermal fluctuations~\cite{sachdev2023quantum,altland2010condensed,mahan2000many,chaikin1995principles}. In the realm of driven-dissipative condensed matter, instead, emergent behavior can occur in dynamics or in the non-equilibrium steady state where the system settles. These novel responses usually arise from the intricate interplay of classical and quantum noise within the strongly coupled limit of a many-particle system.
Finally, the nature of interactions fostering strong correlations takes on a fundamentally different character in the domain of quantum many-body optics. Here, short-range interactions originating from atoms or molecules coexist with long-range or even all-to-all interactions mediated by   light, leading to fundamentally distinct cooperative mechanisms.\\

These distinctive features find natural occurrence in cavity quantum electrodynamics (cQED) experiments, which have occupied a central position in the field of driven-open quantum simulators for over a decade. In cavity QED, atoms and photons experience couplings as a consequence of their confinement within high-finesse optical cavities~\cite{walther2006cavity}. This confinement enables repeated light-matter scattering in nearly isolated conditions, achieving effectively strong interactions.
Leveraging the flexibility, tunability, and engineering capabilities inherent in quantum simulators, these platforms have become ideal environments for realizing non-equilibrium phases of matter under driven-open conditions. Beyond fundamental research, applications extend to quantum information, where cavity QED currently holds the world record for spin squeezing~\cite{hosten2016measurement,hosten2016quantum},    an entangled state which surpasses classical limits for metrology and sensing.

In the majority of these experiments, the dynamical behavior of the system can be reduced to a few degrees of freedom: typically, these are  photon amplitudes and numbers in conjunction with the dynamics of a collective spin that encapsulates the motion of the entire assembly of atoms~\cite{kelly2022resonant,kirton2019introduction,keeling2010collective,valencia2023crafting,chelpanova2023intertwining,kirton2017suppressing,marino2022dynamical,muniz2020exploring,lewis2019unifying,norcia2018cavity,dogra2019dissipation,ferri2021emerging,dreon2022self,baumann2010dicke,kongkhambut2022observation,kessler2021observation,klinder2015dynamical}. 
This usually results from the all-to-all nature of photon-mediated interactions in the cavity, which allows for an effective mean-field description. Quantum fluctuations are sub-leading in the number of atoms when the dynamics of such macroscopic degrees of freedom are considered and the system's behavior can be effectively characterized through semi-classical descriptions~\cite{kirton2017suppressing}.\\

This situation presents significant advantages when it comes to modeling cavity QED experiments. Simultaneously, it naturally prompts the exploration of conditions where the strong correlations inherent in these systems become dominant. In this context, we are witnessing the emergence of a novel category of experiments, where the true many-body nature of the platform emerges and it defies a description using only a few macroscopic degrees of freedom. These experiments encompass a variety of setups, such as Rydberg tweezer arrays integrated into optical cavities~\cite{kong2021melting}, atomic ensembles with adjustable loading capacities~\cite{periwal2021programmable,kroeze2022high}, and the fermionic variant of traditional cavity QED experiments~\cite{sauerwein2023engineering}.

All these experiments necessitate the integration of methods traditionally employed for addressing strongly correlated problems with the extra complication that detailed balance is broken and dynamics are non-unitary. Accessing the long-term evolution of open many-body systems subject to external (coherent or incoherent) drives as well as strong interactions, is one of the most challenging computational frontiers. Nevertheless, it has now become increasingly important in order to steer this new generation of cavity QED platforms.

Indeed, as numerous established platforms push the boundaries of the NISQ (Noisy Intermediate-Scale Quantum) era~\cite{bharti2022noisy}, a cavity QED platform endowed with strong correlations presents an intriguing opportunity. It would hold the potential for innovative strategies in quantum processing, leveraging the combined advantages   offered by cooperative behavior~\cite{sorensen2002entangling}, the interplay of long and short-range interactions~\cite{kong2021melting,rudelis2023cavity,chen2022high},   the manipulation of controllable quantum fluctuations~\cite{Marsh_PRX21}, including the potential to manipulate decoherence channels~\cite{seetharam2022correlation}.\\

As of today, state of art methods to model these experiments would encompass a number of options.  Refined versions of semi-classical techniques, such as discrete phase space representations of the Hilbert space, have been tailored for addressing the driven-dissipative dynamics of spin and bosonic systems~\cite{schachenmayer2015many,mink2022hybrid,huber2022realistic,huber2021phase,PRXQuantum.4.030304,kelly2021effect}. Tensor network ans\"{a}tze, which retain the most informaive correlations to describe dynamics, have been used to study phase transitions in driven open systems~\cite{mc2021stable,cui2015variational,gangat2017steady,kilda2021stability,kshetrimayum2017simple,mascarenhas2015matrix,weimer2021simulation,kelly2020exploring}. At the same time,  cluster mean field methods~\cite{PhysRevX.6.031011} or self-consistent Gaussian approximations~\cite{zhu2019dicke} remain a straightforward, and, in some circumstances, competitive way, to qualitatively capture non-unitary dynamics.\\

\begin{figure*}
    \centering
    \includegraphics[width=.98\linewidth]{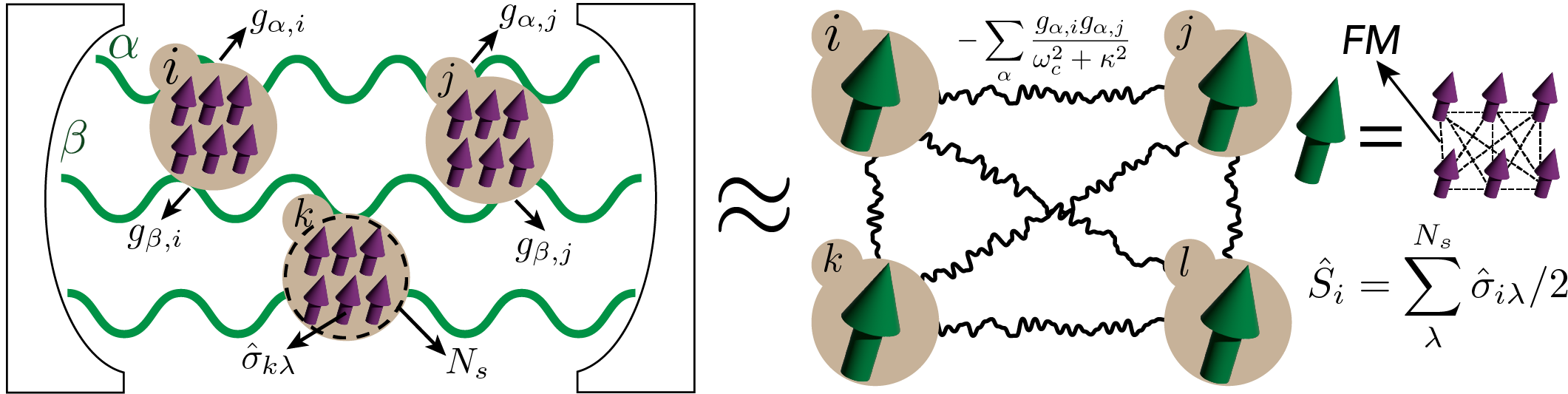}
    \caption{Schematics of the model considered in this work: $N$ clusters (tan circles) of $N_s$ two-level atoms (spins) interact with $M$ cavity modes (green lines) via static disordered couplings $g_{\alpha,i}$. Cavity modes mediate effective all-to-all interactions modulated among clusters. The effective interaction is modulated by static disorder and generates frustration and glassy dynamics. Each cluster can be taken as a large spin $S$ with amplitude $N_s/2$. Cavity modes induce ferromagnetic (FM) interactions between spins within the same cluster.}
    \label{fig:setup}
\end{figure*}

\blue{In this paper, we introduce a novel method to tackle the dynamics of strongly interacting light-matter systems in the many-body limit \cite{Berges_02,Berges2004introduction}. It is an adaptation of the 2-particle irreducible (2PI) effective action formalism which has been used  in high energy physics and cosmology \cite{Cornwall_74,Aarts_brokensymm2002,Cooper_phi4symmbreak2003,Arrizabalaga_preheating2004,Berges_PRL04,Berges_Preheating2011,BERGES_defectFormation2011,GASENZER_GagueTurbulance2014,Nowak_2014,Orioli_nonThermalFP2015,Berges_relBEC2017,Berges_nonThFP2017,Walz_largeN2018}, as well as in condensed matter  and atomic, molecular and optical (AMO) physics\cite{Rey_DynBEC2004,Balzer_2009,Kronenwett_FermiGas2011,Babadi_PRX15,Bock_kondo2016,Erne_1dBoseGas2018,schuckert2018nonequilibrium,Bruchard_hydro2022,lang2023field}. The method involves deriving a quantum effective action for the system as a function of its 2-point correlation functions. Such action gives exact equations of motion of two-point functions, and with some ingenuity also the dynamics of higher point correlation functions~\cite{babadi2013non,carrington2014four}. It is via a set of controlled non-perturbative approximations (large $N$ limit, dilute expansion, loop expansion) for the effective action \cite{Berges_02,Berges2004introduction,Babadi_PRX15,schuckert2018nonequilibrium} that the dynamics is numerically solved. }

Equations of motion derived from 2PI effective actions, known as Dyson equations (DE)~\cite{Berges2004introduction}, offer numerous advantages. They give rise to self-consistent dynamics of two-point functions free from secular effects \cite{Berges2004introduction,Rammer_2007}. Since approximations are directly performed at the level of the action, the resulting DE are \textit{conserving}, in the sense they cannot spoil the  conserved quantities of the original model. 
 They also can seamlessly incorporate both coherent and dissipative dynamics, as functional integral methods do not markedly differentiate between the two~\cite{Sieberer_2016}. They do not suffer from limitations when   degrees of freedom with unbounded Hilbert space, like photons or phonons, are included in dynamics.  The dynamics governed by   DE have polynomial time costs in system size and  they can be run on a personal computer. The price to pay is formulating educated guesses on the physics of the problem, which are crucial for selecting the proper approximation scheme. Furthermore, these DE are known to semi-quantitatively reproduce dynamics, and they are therefore excellent for elucidating the mechanisms at work in a given problem of interest, but less suited to fit with accuracy  experimental curves, for instance. With these caveats, the method is highly flexible and  applicable virtually to any driven-dissipative many particle system, made of fermions, bosons or spins, as we also expand in the conclusions of this manuscript. \\

 In this work, we initiate our research program by applying this method to examine the dynamics of strongly correlated light-matter systems within multi-mode cavity QED. Our model is inspired by the experimental setup presented in Refs.~\cite{kroeze2022high,vaidya2018tunable,kollar2015adjustable,PhysRevLett.122.193601} which involves several photonic modes connecting nodes (Fig. \ref{fig:setup}). Each of these nodes houses atomic ensembles with adjustable loading capacities. By manipulating the number of atoms in each node (ranging from a few to thousands in the experiment) by using optical tweezers, one can introduce tunable quantum fluctuations in the platform. These fluctuations enable the exploration of system dynamics, from strongly correlated to semi-classical regimes. Importantly, this flexibility is not unique to this platform~\cite{periwal2021programmable,bentsen2019integrable} and represents a promising starting point for delving into many-body cavity QED beyond the domain of collective dynamic responses~\cite{kirton2019introduction,keeling2010collective}.  During the final stages of the current work, the experiment detailed in Ref. \cite{kroeze2023replica} verified the existence of a spin glass (SG) phase within the quantum gas microscope platform of Refs.~\cite{kroeze2022high,vaidya2018tunable,kollar2015adjustable,PhysRevLett.122.193601} for the first time through direct measurement of the configuration of spins in the system. \blue{In this paper, we complement the findings presented in our accompanying work on dynamical spin glass formation~\cite{hosseinshort}   by providing a comprehensive derivation of non-equilibrium field theory for frustrated cavity QED and investigating the full spectrum of non-equilibrium phases and crossovers inherent to these systems. Our objective is to bridge the domains of AMO and the many-body community working at the interface of condensed matter and field theory. }

\section{Outline of the paper}
The goal of this work is to understand the far from equilibrium dynamics of SG phases in frustrated cavity QED with strong disorder, where fluctuations cannot be omitted and mean field treatments are not applicable. We remark that our approach is distinct from those of Refs. \cite{Strack_PRL11,Buchhold_PRA13}  which by construction, are suitable only for the universal SG behavior at steady state, in two key aspects. First, the formalism developed here is applicable in far from equilibrium situations such as quench dynamics,  while keeping track of the quantum nature of spins. Second, the platform of Refs. \cite{kroeze2022high,vaidya2018tunable,kollar2015adjustable,PhysRevLett.122.193601} naturally includes extra FM interactions which are unimportant in the steady state of the system \cite{Marsh_PRX21}, but as shown in this work, can qualitatively modify conventional SG dynamics away from equilibrium.
We now briefly outline our key results which expand upon the results of our accompanying work~\cite{hosseinshort}.

\textit{Introduction to the model} -- We commence the paper by introducing the model in Section~\ref{sec:model}. We will briefly review previous works on the behavior of the model in different regimes of parameters at the steady state, and will motivate using our approach to treat its far from equilibrium dynamics.

\textit{Introduction to the method} -- We provide a brief introduction to effective action methods and non-equilibrium field theory in Section \ref{sec:NEQFT}. This serves as a foundation for the detailed derivation of our formalism in the context of frustrated light-matter interactions in cavity QED, covered in Section \ref{sec:2pi_for_cavityQED}, where we develop a versatile approach to address real-time dynamics in a model for disordered cavity QED given in Fig. \ref{fig:setup}. To enhance accessibility, this section and its accompanying appendices are crafted to be reproducible from scratch by the interested reader.

\begin{figure*}[!t]
    \centering
    \includegraphics[width=0.8\textwidth]{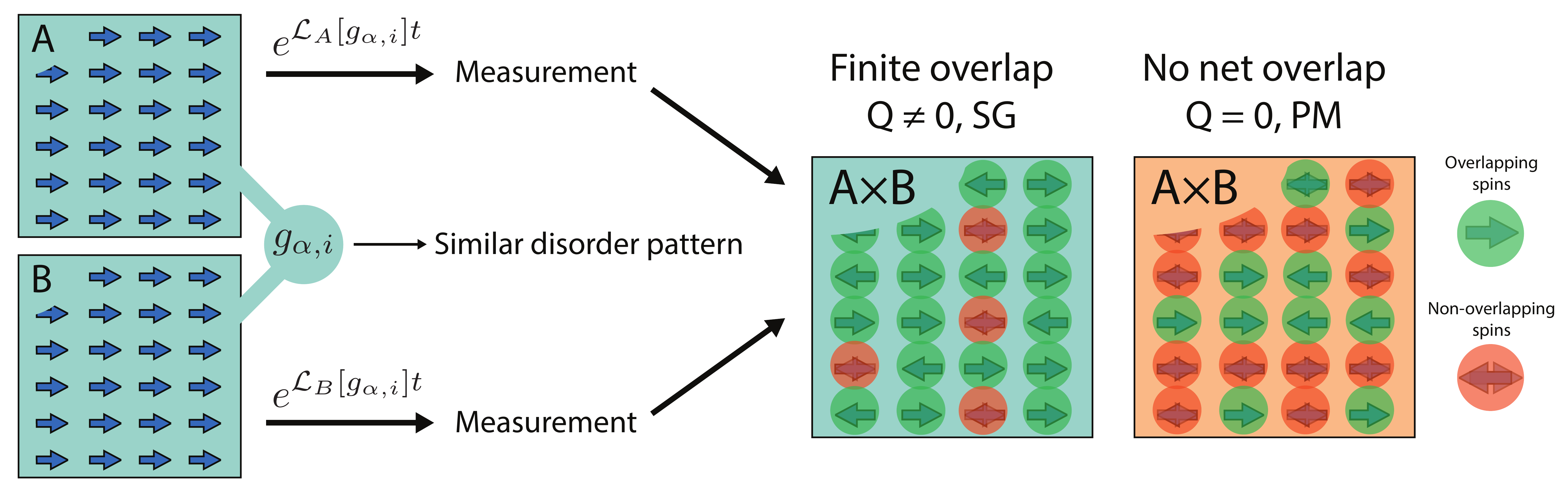}
    \caption{\blue{Measuring SG order through the overlap of spin configurations ($Q$) between two similar, but independent, systems $A$ and $B$ which share the same pattern of disordered couplings $g_{\alpha,i}$. Starting from the same initial state for both systems and evolving both of them under similar Lindblad dynamics, we can still obtain different spin configuration for each of them by doing measurements and applying projections to the their states. If the systems are SG, the overlap remains finite at long times. For a PM, the relative orientations of spins in the two copies are random and only half of the spins share the same orientation as their counterparts.}}
    \label{fig:overlap_cartoon}
\end{figure*}

\textit{Magnetization dynamics} -- In Section \ref{sec:LO_results}, we show that the mean-field (MF) approximation, given by the leading order contribution in our approach, predicts a paramagnetic (PM) to ferromagnetic (FM) phase transition of effective spin degrees of freedom in the universality class of infinite range Ising model, but it completely omits the effect of frustrated interactions generated by static disorder in the system. FM interactions are mediated by virtual photon exchange processes among spins within the same cluster, and in contrast to inter-cluster couplings, are not frustrated. In Section \ref{sec:NLO_results} we demonstrate that, upon the non-perturbative incorporation of disorder and fluctuations, our approach provides a dramatic improvement of MF results. As the focus of this work is the far from equilibrium dynamics of this system after interaction quenches, we first look at the dynamics of simple spin observables, such as global magnetization. We find that if the system is initiated in a symmetry broken state with a finite total magnetization, the relaxation of magnetic order substantially depends on the coupling strength and the size of atomic ensembles $N_s$ or equivalently, the amplitude of large spins $\hat{S_i}$ per each cluster. For weak couplings (Section~\ref{sec:dyn_PM}), global magnetization displays paramagnetic oscillations which are weakly damped due to the dephasing generated by static disorder. Upon increasing the coupling (Section~\ref{sec:dyn_FM}), spin relaxation changes from underdamped dynamics to overdamped dynamics without oscillations. In Section~\ref{sec:spin_size} we address the effect of ensemble size $N_s$ on magnetization dynamics. We show that in the overdamped regime and for small $N_s$, global magnetization $\langle \hat{S}^x \rangle$ relaxes quickly to zero while for large $N_s$, after an initial collapse to a finite value, it stays in a transient prethermal state with a slow spiral decay of the magnetization vector along the axis of temporary FM order. \blue{We benchmark 2PI with a semi-classical phase space approximation~\cite{Polkovnikov_reviewTWA,schachenmayer2015many} and demonstrate excellent agreement in the limit of large spins between the two, where the latter becomes exact. This indicates the viability of 2PI for approximating quantum dynamics by starting from the limit of large spins, and systematically incorporating quantum fluctuations as the spin size is lowered down to $S=1/2$.}

\textit{Dynamics of SG} -- To probe into the nature of the transition in the system as magnetization dynamics change from underdamped to overdamped, we consider more complex and richer spin observables in Section~\ref{sec:dynamics_of_SG}. Particularly, and with the expectation of exploring SG order in the system, we consider the dynamics of two different order parameters for SG in Sections \ref{sec:q_def} and \ref{sec:edwards-anderson}.  \blue{First, we consider the time evolution of the overlap of spin configurations between two identical systems (replicas) with a similar disorder profile but otherwise decoupled from each other. This overlap $Q$, which measures the statistical correlations between the two replicas only due to their shared disorder pattern, is a finite quantity in the SG phase and vanishes for a PM (Fig.~\ref{fig:overlap_cartoon}), as has been demonstrated experimentally~\cite{kroeze2023replica}.} To obtain dynamics of the overlap, we extend the formalism of Keldysh field theory, which is done for the first time in this work, with technical similarity between our approach and the one introduced by Ref.~\cite{Buchhold_MIPT2021} to study measurement-induced phase transitions. In PM phase, the overlap parameter relaxes to zero after experiencing small temporary fluctuations, in contrast to the SG phase where the order parameter relaxes to a finite value. We show that weak cavity losses stabilize SG order by effectively cooling down the system. Stronger photon losses, on the other hand, suppress the SG phase. As a second measure of SG order, we consider the temporal correlations of spins over long times, conventionally known as the Edwards-Anderson (EA) \cite{Edwards_75,SG_RMP} order parameter, corresponding to the overlap of two snapshots of the same system taken at long time intervals. We confirm that EA order parameter changes from zero in the PM phase, when the system loses its memory quickly, to a finite value in the SG phase, proving that the system is glassified. We proceed to show that spin fluctuations violate the fluctuation-dissipation theorem \cite{kamenev,Rammer_2007}, a phenomenon conjectured to be closely related to replica symmetry breaking in SG systems \cite{Marinari_FDT_RSB98}.

\textit{Effect of resonant photons} -- In Section \ref{sec:resonance}, we consider the effect of   photon frequency on the glass phase by looking at SG order parameter in a wide range of photon frequencies from fast photons to the resonance limit, where atomic and cavity detunings are close, and below. We see that the SG order is peaked close to the resonance, while it saturates in the adiabatic limit. At frequencies below the resonance, SG order is dramatically suppressed, resembling the suppression of various types of order by low frequency lattice distortions (phonons) in solid state physics. We also address briefly the spectrum of low-lying excitations in the SG phase, showing a continuum of sub-Ohmic modes at small energies. 

The capability to include in the same set of dynamical equations variable ranges of coupling, tunable values of $N_s$ and active photons, is one of the key merits of our approach. It allows us to solve for the dynamics of the full platform without the need to invoke large energy scale separations, effective descriptions suited only to      atomic or photonic degrees of freedom, or to treat distinctly the quantum and semi-classical regimes. In this regard, the method has a degree of flexibility that appears   promising to treat other strongly correlated driven-dissipative systems, as we discuss in the concluding Section  \ref{sec:conclusion}. 

\section{The Model}\label{sec:model}
The experiment in~\cite{vaidya2018tunable,kroeze2022high,kollar2015adjustable,PhysRevLett.122.193601} can be modeled by a system of $N$  clusters, each one containing $N_s$ two-level atoms encoded by the spin-1/2 operators $\sigma_{i\lambda}$, with cluster $1\leq i \leq N$ and atom indices $1\leq \lambda \leq N_s$ , as shown in Fig. \ref{fig:setup}. 
The couplings between the atoms and the $M$ photonic modes of the cavity are spatial-dependent and uncorrelated  from each other, which justifies their modeling via  random spin-boson couplings~\cite{Strack_PRL11,Buchhold_PRA13,wierzchucka2023integrability}. Starting from the same initial state for all spins, each cluster is equivalent to a single spin $S_i = \sum_{\lambda}\sigma_{i\lambda}/2$ with amplitude $S=N_s/2$.
The parameter $S$ can be tuned by loading   few or several atoms in each cluster, and it   dictates the strength of quantum fluctuations. For instance, at large $S$ each cluster would be effectively described by a classical angular momentum, since its quantum noise would scale down as    $1/S$~\cite{kirton2019introduction, Fiorelli_PRL20, Marsh2023}.   A minimal model for the   system is given by the 
random Dicke model whose evolution is governed by $\partial_t \rho = -i \comm{H}{\rho} + \sum_{\alpha=1}^M \mathcal{D}\qty[a_\alpha]\rho$, where
\begin{multline}\label{H}
    H=\frac{\Delta}{2} \sum_{i,\lambda} \sigma^z_{i\lambda}+ \sum_{\alpha}\omega_\alpha a^\dagger_\alpha a_\alpha  \\ + \frac{1}{\sqrt{(M+N)N_s}}\sum_{\substack{i,\lambda,\alpha}} g_{\alpha i} \qty(a_\alpha+a_\alpha^\dagger) \sigma^x_{i\lambda},
\end{multline}
and 
\begin{equation}
    \mathcal{D}\qty[a_\alpha]\rho=\kappa_\alpha \qty(2 a_\alpha \rho a_\alpha^\dagger - \acomm{a_\alpha^\dagger a_\alpha}{\rho}).
\end{equation}
We assume that cavity modes are nearly degenerate such that $\omega_\alpha=\omega_c$ and $\kappa_\alpha=\kappa$. The couplings $g_{\alpha i}$ are assumed to be random and chosen from a Gaussian distribution:
\begin{equation}
    \overline{g_{\alpha i}}=0, \quad \overline{g_{\alpha i}g_{\beta j}}=\delta_{\alpha\beta}\delta_{ij}g^2.
\end{equation}
Couplings for spins in the same cluster are similar as we assume that the spatial size of each cluster is smaller than the wavelength of cavity modes. The scaling of the interaction term guarantees that the total energy is extensive in system size~\cite{Kac}. For $\omega_c\gg g,\Delta$ photons are the fastest degree of freedom in the problem; they quickly relax to stationary value and approximately, they mediate instantaneous interactions among spins.  Assuming $\kappa=0$, photons can be adiabatically eliminated~\cite{agarwal1997atomic, asboth2005self,PhysRevA.82.043612, breuer2002theory, Marsh_PRX21} and the model in (\ref{H}) is mapped to
\begin{align}
    H_\mathrm{eff}&= \sum_{i} H_i + H_\mathrm{int}, \label{H_eff}\\
    H_i &\equiv \Delta S^z_i - \frac{4}{(N+M)N_s\omega_c} \qty(\sum_\alpha g_{\alpha i}^2)\qty(S^x_i)^2, \label{H_LMG}\\
    H_\mathrm{int} &\equiv -\frac{4}{(N+M)N_s\omega_c} \sum_{i\neq j} \sum_\alpha g_{\alpha i}g_{\alpha j}S^x_i S^x_j, \label{H_int}
\end{align}
where $S^x_i=\sum_\lambda \sigma^x_{i\lambda}/2$ is the total spin operator for each cluster. For $N_s>1$,   each cluster in Eq. (\ref{H_LMG}) is an infinite range quantum Ising model also known as the Lipkin-Meshkov-Glick \cite{LMG_65} (LMG) model with Hamiltonian
\begin{equation}\label{H_LMG_S}
    H_\mathrm{LMG}=\Delta S^z - \frac{J}{N_s} (S^x)^2.
\end{equation}
$H_\mathrm{LMG}$ admits an exact solution using mean-field theory in the limit $N_s\to \infty$, and features a paramagnet (PM) to FM phase transition \cite{Sciolla_2011,Zunkovic_PRL18,lerose2019impact,marino2022dynamical,defenu2023long,berdanier2019universal} at $\Delta=J$. The effective interactions between atoms within the same cluster in Eq. (\ref{H_LMG}) are purely ferromagnetic and, to leading order, we can identify $J=4g^2\eta/(1+\eta)\omega_c$ after disorder averaging, with $\eta=M/N$. Each cluster is further coupled to other clusters via  Eq. (\ref{H_int}), which is expected to generate frustration in the system. For $N_s=1$, the ferromagnetic interaction vanishes since $(S^x)^2=(\sigma^x)^2=1$ and this model becomes the quantum Hopfield model (QHM) \cite{Hopfield}. The QHM has a PM ground-state for sufficiently large $\Delta$ while for small $\Delta$, the ground-state crucially depends on the ratio $\eta$ \cite{Amit_PRL85}. For small values $\eta < \eta_c\sim O(10^{-1})$, the system is in the memory retrieval phase \cite{Gopal_PRL11,Rotondo_hopfield2015,Fiorelli_PRL20,Marsh_PRX21}, which is a Dicke model in disguise with multiple superradiant/FM ground-states. When the number of photon modes ($M$) surpasses a critical limit $\eta > \eta_c$ \cite{Amit_PRL85}, frustrations dominate and turn the system into a quantum glass \cite{Strack_PRL11,Buchhold_PRA13,Rotondo_Glass2015,Marsh2023}, arguably in the same universality class of the quantum Sherrington-Kirkpatrick \cite{SK_PRL75,Ray_PRB89} (SK) model. In this paper, we are interested in SG dynamics and will only consider the limit $\eta=1$.

The many-body nature of the model (\ref{H}) when photons participate in the dynamics prevents us to use exact diagonalization. Moreover, because of the frustrated couplings, mean-field (MF) methods or dynamics of cumulants expansions (CE)~\cite{kirton2017suppressing} are inapplicable. Instead, we attack this problem using methods of non-equilibrium quantum field theory (NEQFT). In the next section, we will introduce the method using a simple example first, and then will proceed to apply it to the model in Eq. (\ref{H}). A treatment of the Dicke model without disordered couplings is also provided in Appendix \ref{app:Dicke}, for comparison with the random Dicke model and pedagogical purposes.

\section{Non-Equilibrium Field Theory}\label{sec:NEQFT}

\subsection{‌Basics of 2PI Formalism}\label{sec:basics}

\blue{Similar to classical mechanics, the dynamics of a quantum system can also be obtained from an action principle~\cite{Heisenberg_1936,Cornwall_74,Jackiw_79}. In particular, one can define a quantum effective action (EA) $\Gamma[\varphi,G]$ in terms of one-point $\varphi=\expval{\phi}$ and two-point $iG(t,t')=\expval{\phi(t)\phi(t')}_c$ correlation functions of the system, known as 2PI-EA, whose stationary solution with respect to correlation functions yields the equations of motion for those correlation functions~\cite{Berges2004introduction}. The equations of motion include all quantum effects, and in principle, can be solved to obtain the exact correlation functions of the system. However, this is only true if the full expression of 2PI-EA is known. For a system of real-valued bosonic fields, the general expression of  $\Gamma$ is given by
\begin{equation}
    \Gamma[\varphi,G]=S_\mathrm{cl}[\varphi] -\frac{i}{2}\Tr \ln G + \frac{i}{2} \Tr(G_0^{-1}G) + \Gamma_2[\varphi,G].
\end{equation}
The first term is the classical action, and the rest of the terms capture fluctuations (both quantum and statistical~\cite{Berges_PRL04}) and $G_0$ is the Green's function of the non-interacting system. The last term $\Gamma_2$ usually admits an expansion in terms of connected Feynman diagrams which cannot be disconnected by cutting at most two of their lines (hence the name 2-particle irreducible). Often, it is only possible to do an approximation for $\Gamma_2$, by keeping only a finite number of diagrams or, similar to this work, an infinite subset of diagrams. This in turn yields an approximate solution for correlation functions and field expectation values. The main advantage of 2PI is that, despite the inevitable use of approximations, it is a conserving method~\cite{Berges2004introduction,babadi2013non}. This means that, the approximated dynamics respect all of the conservation laws of the original problem and therefore, is immune to the instabilities that many other methods of approximation for dynamics have. In addition to being a conserving method, in certain problems, including the one considered here, it is possible to obtain non-perturbative approximations for $\Gamma_2$, which produce qualitatively valid results for dynamics at long times~\cite{Berges_02,berges_ParametricResonance2003,Berges_PRL04,Berges2004introduction,Berges_BEC2012,Babadi_PRX15}.}

\subsection{Comparison with Other Approaches}
There are various methods to explore many-body quantum dynamics theoretically and each one has its own advantages and shortcomings. Exact diagonalization (ED) gives accurate results but is often limited to very small system sizes, especially for open quantum systems where the size of the vector space grows even faster due to the necessity of working with mixed states. Methods based on matrix product states (MPS) \cite{Verstraete_MPS2004} are mostly limited to one spatial dimension and systems with local interactions and weak entanglement. Among the most frequently used methods in the AMO community are cumulants expansion (CE) and truncated Wigner approximation (TWA) \cite{Polkovnikov_reviewTWA} together with its extension, discrete truncated Wigner approximation (DTWA) \cite{Schachenmayer_DTWA}. The main advantages of CE are simplicity and cheap computational cost. On the other hand, it is an uncontrolled approximation \cite{sanchez2020cumulant,pavskauskas2012equilibration}. There is no a priori knowledge of its domain of applicability before solving the equations and checking the physical consistency of the results. Furthermore, to calculate correlation functions at different times in CE, one needs to resort to the quantum regression theorem \cite{breuer2002theory,Gardiner_noise}, which complicates the calculations. TWA approximates quantum dynamics with classical statistical mechanics by sampling the initial probability distribution function from the system's initial wave function and subsequently, evolving the system according to the classical equations of motion. The advantage of TWA is that it is a controlled approximation, since it can be expressed as the leading order contribution in the expansion of dynamics in powers of $\hbar$, which turn out to be equivalent to classical statistical mechanics \cite{Polkovnikov_reviewTWA,Berges2004introduction}. The drawback of TWA is its limitation to systems with weak quantum fluctuations and when quantum effects are not built up over time. For instance, TWA is unable to capture tunneling phenomena \cite{Polkovnikov_reviewTWA} even at the level of qualitative accuracy. While the calculation of non-local symmetric correlation functions is straightforward in TWA, to evaluate quantities such as response functions one needs to go to higher order terms in $\hbar$ \cite{Polkovnikov_reviewTWA}, considerably increasing the required effort to use the method. A detailed comparison of DTWA and 2PI for our system is provided in Section ~\ref{sec:spin_size} and Appendix~\ref{app:2pi_dtwa}.

In comparison to CE and TWA, 2PI can be used to perform controlled approximations, provided that a control parameter exists in the system. In this case, the 2PI action admits an expansion in powers of the control parameter \cite{Berges2004introduction}. For instance, this parameter can be $\hbar$, similar to the example given before, or inverse of the components of a vector field or in our case, the inverse of the spin amplitude per cluster and the number of photon modes to the number of clusters. Moreover, symmetric and anti-symmetric correlation functions are the quantities in terms of which the formalism is built and are the direct outcomes of calculations. 2PI also excels in capturing quantum effects, mainly because it involves a resummation of the perturbative expansion to infinite order. For instance, it has been used to study dynamics in strongly  correlated systems of electrons \cite{Eberlein_PRB17,Haldar_PRR20,Kuhlenkamp_DrivenSYK} and phonons \cite{HH_PRB23,Grunwald2023} with non-Fermi liquid (NFL) behavior and critical fluctuations. 2PI works well also when small quantum effects are accumulated over time, leading to drastic changes in the system at long times such as in tunneling phenomena \cite{Cooper_phi4symmbreak2003,Batini_tunnel}. Despite numerous advantages, 2PI has some limitations. First, the approximations that are usually made to retain only a subset of the diagrams in the effective action, such as $1/N$ expansions \cite{Aarts_brokensymm2002,Berges_02,Berges2004introduction,Babadi_PRX15}, give  qualitatively valuable results about the universal trend of the dynamics, but are not tailored to have quantitative accuracy, i.e., they are not suitable for a point-wise comparison with experimental data. Sometimes, one needs to make educated guesses about which diagrams have to be kept to capture a certain aspect of the physics which is of interest. The exception to these is working in the weak coupling or the dilute limit, where collisions can be incorporated perturbatively \cite{kamenev,Rammer_2007}. Second, working with non-Gaussian initial states is difficult in 2PI as these require the inclusion of extra interaction vertices \cite{Berges2004introduction} that complicate the approximation. We emphasize that this restriction only holds for initial states. 2PI is not limited to Gaussian dynamics (such as second order cumulants) and in fact, captures non-Gaussian correlations generated over time after initializing the systems in a Gaussian state.

\section{2PI for Disordered Cavity-QED}\label{sec:2pi_for_cavityQED}

\begin{figure}[!t]
\includegraphics[width=0.18\textwidth]{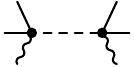}
\caption{The spin-photon interaction vertex after averaging over the disorder. The dashed line indicates the average $\overline{g_{\alpha,i}g_{\beta,j}}$.}
\label{fig:S_eff_bad}
\end{figure}

\begin{figure*}[!t]
     \centering
    $\Gamma_2 \hspace{5pt} \approx \hspace{5pt}\underset{\mbox{\scalebox{1.0}{$\Gamma_2^\mathrm{LO} \sim N_s$}}}{\vcenter{\hbox{\includegraphics[height=0.05\textwidth]{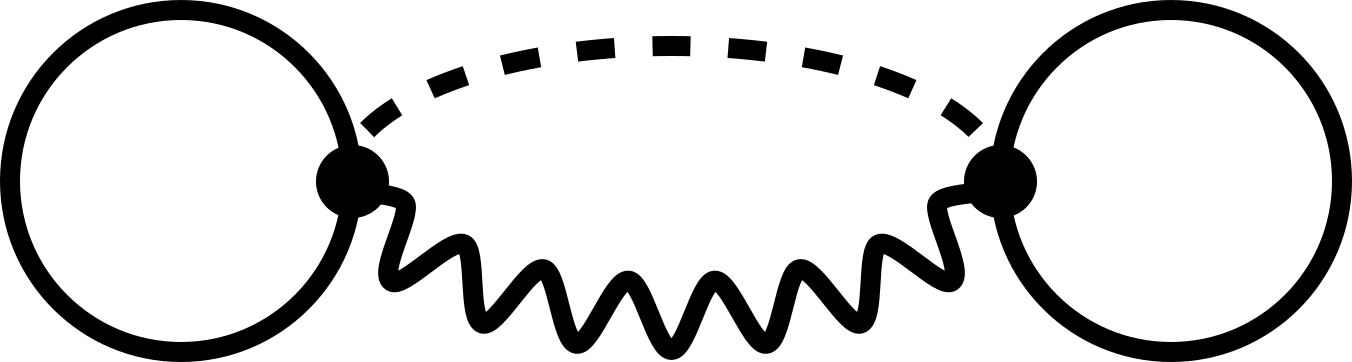}}}} \hspace{5pt} +$  $\underbrace{\Bigg(\vcenter{\hbox{\includegraphics[height=0.08\textwidth]{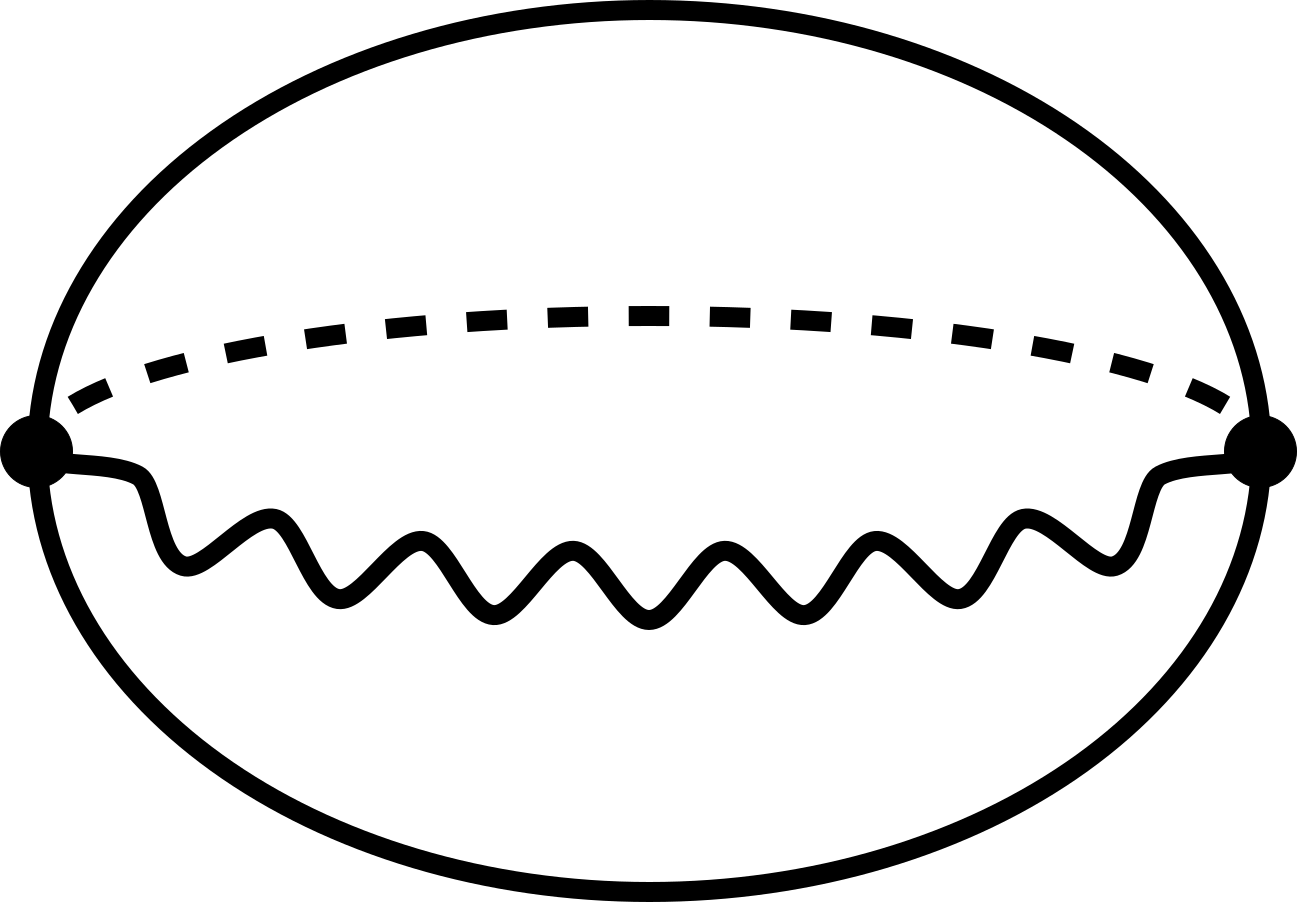}}} \hspace{2pt} + \hspace{2pt} \vcenter{\hbox{\includegraphics[height=0.08\textwidth]{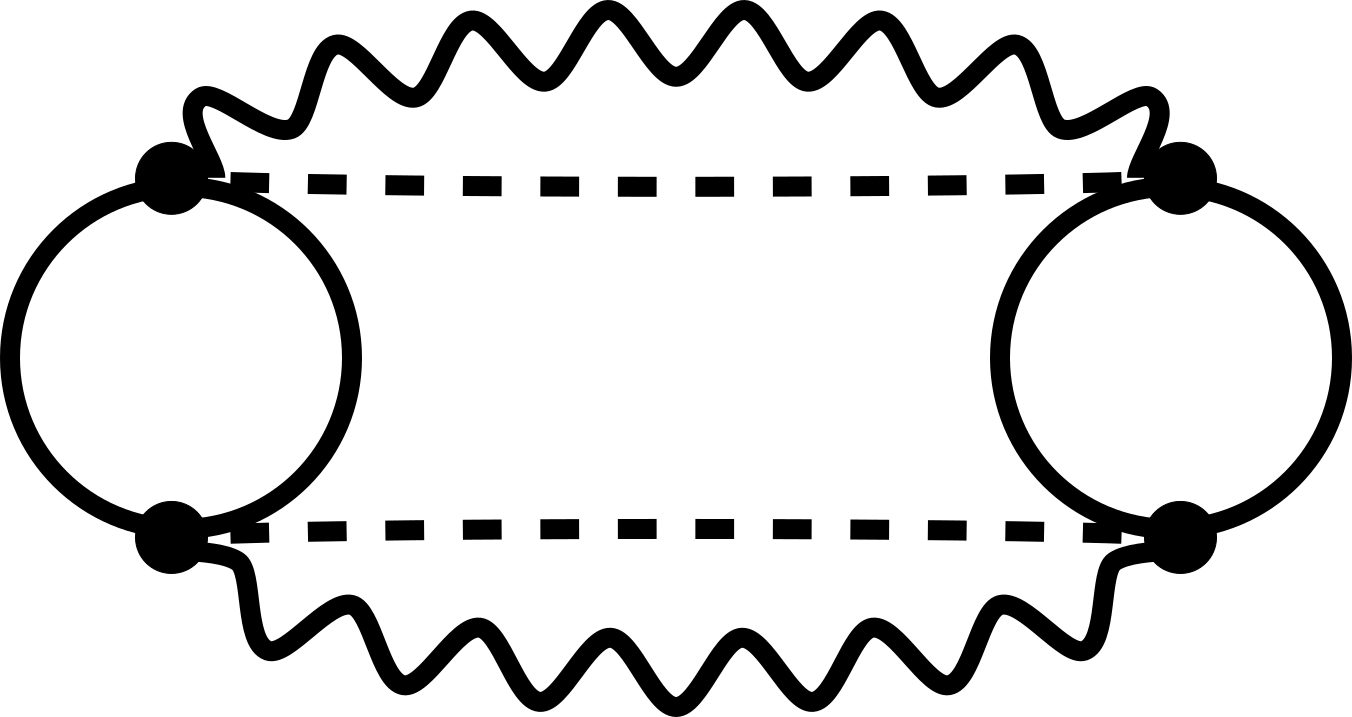}}} \hspace{2pt} + \hspace{2pt} \vcenter{\hbox{\includegraphics[height=0.08\textwidth]{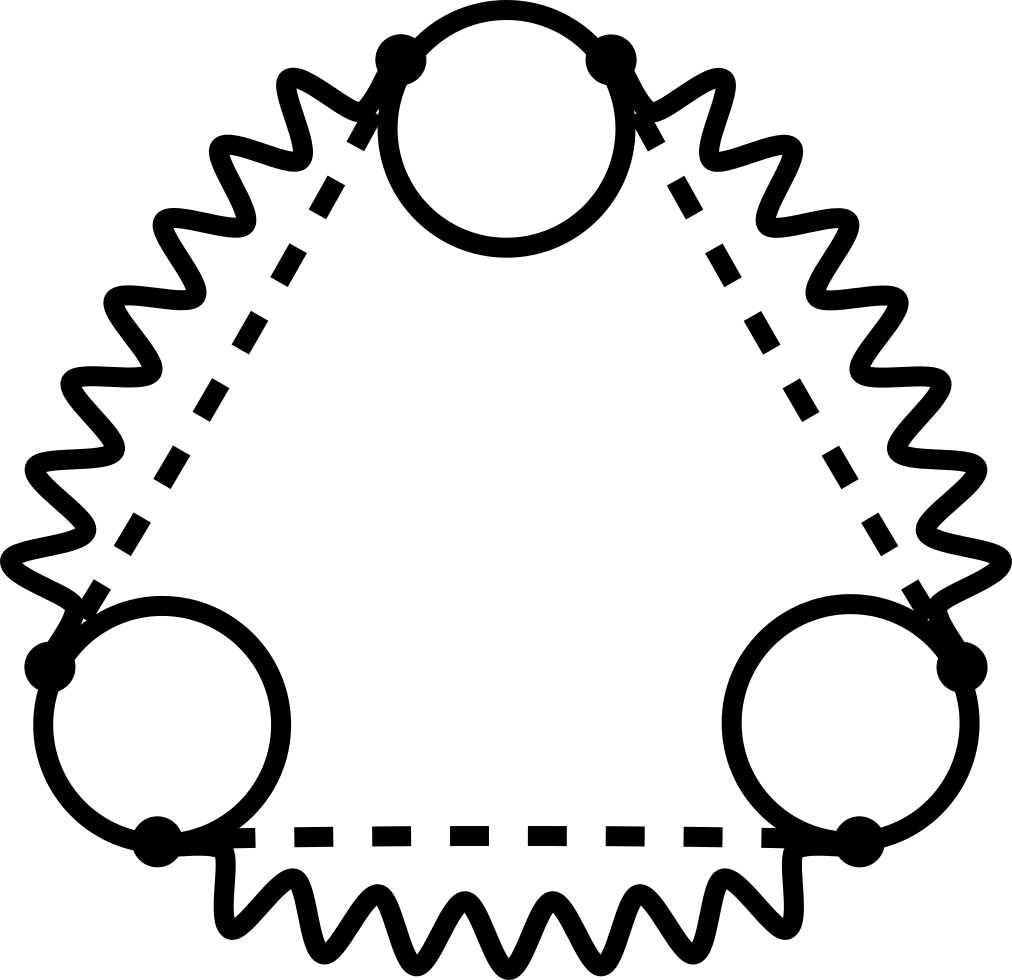}}} \hspace{2pt} + \hspace{2pt} \vcenter{\hbox{\includegraphics[height=0.08\textwidth]{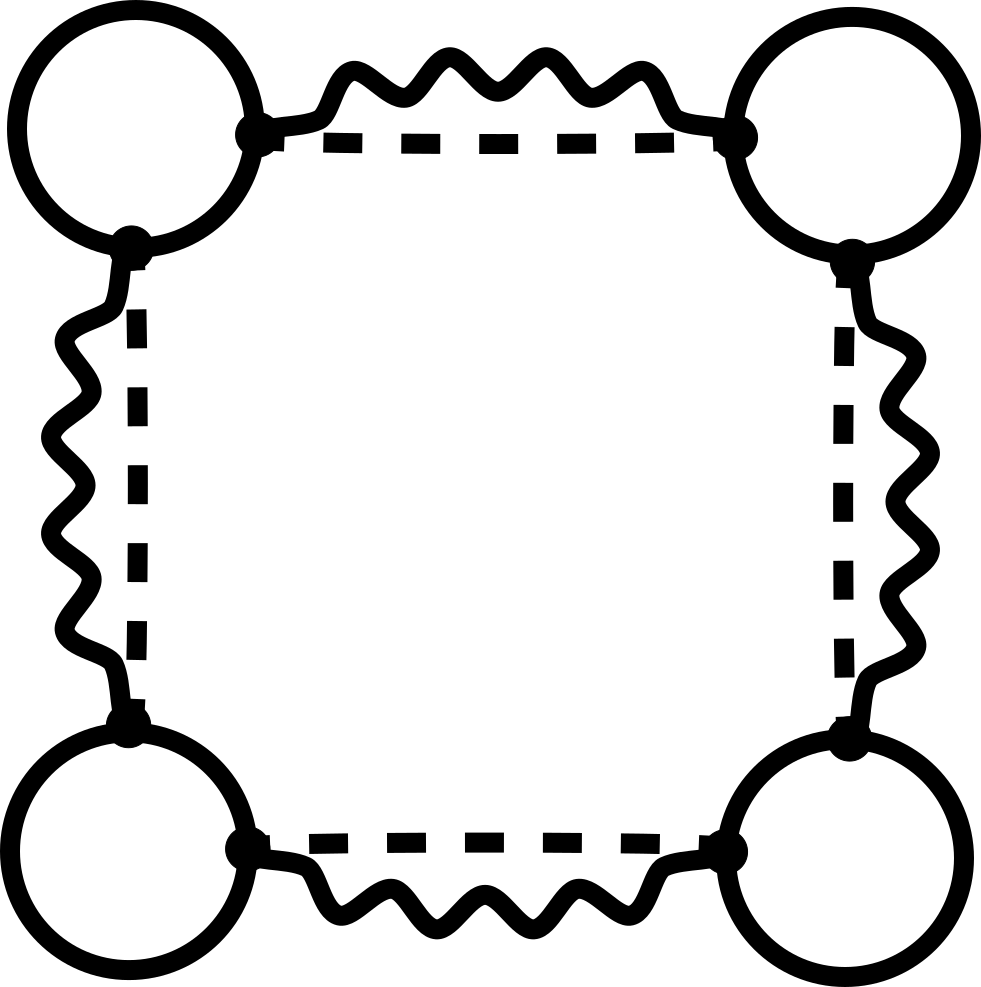}}} \hspace{2pt} + \, \,\dots\Bigg)}_{\mbox{\scalebox{1.0}{$\Gamma_2^\mathrm{NLO}\sim N_s^0$}}}$
     \caption{The leading-order and next-to-leading-order diagrams in the $1/N_s$ expansion of 2PI action.}
     \label{fig:Gamma_orig}
 \end{figure*}

\subsection{Keldysh Action for Spins}
In order to treat Eq. (\ref{H}) using field theory, we need a path integral representation for spin operators. Different spin representations include spin coherent state path integral, the Holstein-Primakoff transformation and particularly, spinon representations in terms of Abrikosov fermions or Schwinger bosons~\cite{fradkin_2013,auerbach1998interacting}. In this work, we represent each spin-half operator in terms of three Majorana fermions $(\psi^x,\psi^y,\psi^z)$ as  \cite{Mao_PRL03,Schnirman_PRL03, Martin_59, Spencer_67, Spencer_68, Berezin_77, Shastry_Majorana97, Biswas_Majorana2011, Schad_Majorana2014, Schad_Majorana2015, Schad_Majorana2016, DallaTorre_DickeMajorana2016,shchadilova2020fermionic} given by
\begin{equation}\label{Majorana_rep}
    \sigma^\alpha= - i \epsilon_{\alpha \beta \gamma} \psi^\beta \psi^\gamma, \qquad \acomm{\psi^\alpha}{\psi^\beta}=\delta_{\alpha \beta},
\end{equation}
where we have assumed summation over repeated indices. It is easy to check that (\ref{Majorana_rep}) satisfies spin commutation relations $\comm{\sigma^\alpha}{\sigma^\beta}=2i\epsilon_{\alpha\beta\gamma}\sigma^\gamma$. This representation was used by Refs.~\cite{DallaTorre_DickeMajorana2016,shchadilova2020fermionic} to study the onset of superradiance in the steady state of the Dicke model with different types of external baths. \blue{We note that, although fermionic and bosonic spinons are formally equivalent (after projection into the physical sector of the Hilbert space), they can yield different results upon using further approximations. For instance, in 2PI we mostly start from ``simple'' initial states for which the values of correlation functions $G(t,t')$ are known only at a single initial time $t=t'=0$. This corresponds to a Gaussian state for bosons and fermions (more precisely, the Gibbs state of a quadratic fermionic Hamiltonian~\cite{ferm_gaussianstates}). In Appendix.~\ref{app:schwinger_bosons}, we show that a Gaussian state for Schwinger bosons is always a mixed state at least for one of the boson species, and generates a relative error of $\mathcal{O}((2S)^0)$ for the values of extensive quantities such as energy or the effective action. However, Gaussian states for fermionic spinons (complex or Majorana) can be pure states, and do not introduce any errors in representing spin coherent states. As we will show below, diagrammatic corrections beyond mean-field dynamics start at $\mathcal{O}(1/2S)$, and are sub-leading to the error of using Gaussian states for Schwinger bosons. Hence, using fermionic spinons is in fact essential for the consistency of the approximation for all spin sizes and not only for $S\approx 1$, as long as we use Gaussian initial states.}

The Keldysh action for ``free" spins , corresponding to the first term in Eq. (\ref{H}) and written in terms of Majorana fermions, has two parts
\begin{align}
    S_\sigma &= S_\mathrm{B} + S_\Delta, \label{S_f} \\
    S_\mathrm{B}&= \sum_{i}^N \sum_\lambda^{N_s} \sum_{\alpha}^{x,y,z} \oint \frac{i}{2} \psi^\alpha_{i\lambda}\partial_{t_c}\psi^\alpha_{i\lambda}\,dt_c , \\
    S_\Delta &=\sum_{i}^N \sum_\lambda^{N_s} \oint i\Delta \psi^x_{i\lambda}\psi^y_{i\lambda}\,dt_c ,
\end{align}
where $S_B$ is the contribution of the Berry phase of spins to the action \cite{sachdev_book}. The action in Eq. (\ref{S_f}) can be compactly written as
\begin{equation}
    S_\sigma = \frac12 \sum_i^N \sum_\lambda^{N_s} \oint \Psi^T_{i\lambda} \hat{G}_0^{-1} \Psi_{i \lambda}\, dt_c,
\end{equation}
with $\Psi^T_{i\lambda}\equiv (\psi^x_{i\lambda},\psi^y_{i\lambda},\psi^z_{i\lambda})$. The inverse bare Green's function for fermions $\hat{G}_0^{-1}$ is defined as
\begin{equation}\label{G0}
    \hat{G}_0^{-1}\equiv \begin{bmatrix}
        i\partial_{t_c} & i \Delta & 0 \\ -i\Delta & i\partial_{t_c} & 0 \\ 0 & 0 & i\partial_{t_c}
    \end{bmatrix}.
\end{equation}
Finally, we define the fermion Green's function and its diagrammatic representation as
\begin{equation}
    iG_{i \lambda,j\lambda'}^{\alpha,\beta}(t,t')\equiv \expval{\psi^\alpha_{i \lambda}(t)\psi^\beta_{j \lambda'}(t')}: \includegraphics{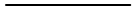}.
\end{equation}
Note that $G$ is the dressed fermion Green's function.

\subsection{Keldysh Action for Photon Sector}
The dissipative Keldysh action for photons can be obtained directly from the Liouvillian, by following the prescription given in Ref. \cite{Sieberer_2016}
\begin{multline}\label{S_ph_a}
    S_\mathrm{ph}= \int \big[\bar{a}_+ \qty(i\partial_t - \omega_c + i \kappa )a_+ - \bar{a}_- \qty(i\partial_t - \omega_c - i\kappa)a_- \\  - 2i\kappa \bar{a}_- a_+ \big]\, dt.
\end{multline}
{Since the spin-photon coupling in Eq.~(\ref{H}) depends on the combination $(a+a^\dagger)$ of photon operators, dealing with interactions is simpler when we make the following transformation to real-valued photon fields $(\phi,\pi)$ given by:}
\begin{equation}
    a_\pm = \sqrt{\frac{\omega_c}{2}}\qty(\phi_\pm + i \frac{\pi_\pm}{\omega_c}).
\end{equation}
Substitution in Eq. (\ref{S_ph_a}) gives
\begin{equation}\label{S_phi}
    S_\mathrm{ph}= \frac12 \sum_\alpha^M\int \Phi_\alpha^T \hat{D}_0^{-1} \Phi_\alpha \, dt,
\end{equation}
where $\Phi_\alpha^T \equiv \qty(\phi_{\alpha+}, \pi_{\alpha+}, \phi_{\alpha-}, \pi_{\alpha-})$ and $\hat{D}_0^{-1}$ is a $4\times 4$ matrix defined as
\begin{align}
    \hat{D}_0^{-1} &= \begin{bmatrix}
        \qty(\hat{D}_0^{-1})^{++} & \qty(\hat{D}_0^{-1})^{+-} \\ \qty(\hat{D}_0^{-1})^{-+} & \qty(\hat{D}_0^{-1})^{--}
    \end{bmatrix}, \label{D0}\\
    \qty(\hat{D}_0^{-1})^{++} &= \begin{bmatrix}
        -\omega_c^2 + i \kappa \omega_c & - \partial_t \\
        \partial_t & -1 + i \frac{\kappa}{\omega_c}
    \end{bmatrix}, \\
    \qty(\hat{D}_0^{-1})^{+-} &= \begin{bmatrix}
        - i \kappa \omega_c & - \kappa \\
        \kappa & - i \frac{\kappa}{\omega_c}
    \end{bmatrix}, \\
    \qty(\hat{D}_0^{-1})^{-+} &= \begin{bmatrix}
        - i \kappa \omega_c &  \kappa \\
        -\kappa & - i \frac{\kappa}{\omega_c}
    \end{bmatrix}, \\
    \qty(\hat{D}_0^{-1})^{--} &= \begin{bmatrix}
        \omega_c^2 + i \kappa \omega_c & \partial_t \\
        -\partial_t & 1 + i \frac{\kappa}{\omega_c}
    \end{bmatrix}.
\end{align}
Photon Green's functions are defined according to
\begin{equation}
    iD^{\rho, \rho'}_{\alpha , \beta}(t,t')\equiv \expval{\Phi^\rho_{\alpha}(t)\Phi^{\rho'}_{\beta}(t')}_c,
\end{equation}
where $\rho,\rho'=(\phi,\pi)$. We will see that only the $D^{\phi \phi}$ component   appears explicitly in the diagrams for the effective action and self-energies. Hence, only $D^{\phi \phi}$ requires a diagrammatic representation which is given by
\begin{equation}
    iD^{\phi \phi}_{\alpha , \beta}(t,t')= \expval{\phi_{\alpha}(t)\phi_{\beta}(t')}_c: \includegraphics{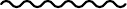}.
\end{equation}

\subsection{Spin-Photon Interaction}

Using the conventions introduced above, the Keldysh action for the interaction term reads
\begin{equation}\label{S_int}
    S_\mathrm{int}=2i\sqrt{\frac{2\omega_c}{(N+M)N_s}}\sum_{\alpha,i,\lambda}\oint dt_c\, g_{\alpha i }\phi_{\alpha}\psi^y_{i\lambda}\psi^z_{i\lambda}.
\end{equation}
In principle, we can proceed by {taking the average of the Keldysh action} over the random couplings $g_{\alpha i}$. {This yields an effective interaction defined by $e^{iS_\mathrm{eff}}=\overline{e^{iS_\mathrm{int}}}$ where
\begin{multline}\label{S_eff_bad}
S_\mathrm{eff}\equiv - \frac{4ig^2\omega_c}{(N+M)N_s}\sum_{\alpha,i}\sum_{\lambda,\lambda'} \oint \oint dt_c \, dt'_c \, \phi_\alpha(t_c)\phi_\alpha(t'_c) \\ \times \psi^y_{i\lambda}(t_c)\psi^z_{i\lambda}(t_c) \psi^y_{i\lambda'}(t'_c)\psi^z_{i\lambda'}(t'_c).
\end{multline}
We have assumed that the initial state is not correlated with disorder profile (see comments in Sections~\ref{sec:dis_avg} and \ref{sec:symmetries} for more details). The diagrammatic form of $S_\mathrm{eff}$ is given in Fig. \ref{fig:S_eff_bad}.} To have a systematic and controlled approximation in $1/N_s$ that captures the frustrated nature of the problem, we have to keep an infinite subset of 2PI diagrams shown in Fig. \ref{fig:Gamma_orig}.  {A closed form for the corresponding summation can be found, as shown for example for the quantum $O(N)$ model in Refs.~\cite{Aarts_brokensymm2002,Berges_02,Berges2004introduction,lang2023field}. An easier approach is the auxiliary field method \cite{Aarts_brokensymm2002,Berges_02,Berges2004introduction}, based on the Hubbard-Stratonovich (HS) transformation \cite{kamenev,altland2010condensed,Babadi_PRX15,schuckert2018nonequilibrium}. HS transformation finds various applications in the study of collective effects in many body systems such as plasmons~\cite{altland2010condensed}, superconductivity~\cite{altland2010condensed, Feigelman_DisorderSC2000}, superfluidity~\cite{sachdev_book} and quantum spin liquids~\cite{fradkin_2013}. The basic idea is to introduce a new field which we label as $\chi$, that mediates the original interaction in $S_\mathrm{int}$. In our case, $\chi$ decouples the interaction between spins and cavity modes as diagrammatically illustrated in Fig.~\ref{fig:vertex}. The action of $\chi$ and its coupling to other degrees of freedom are given by (see Appendix~\ref{app:HS} for a mathematical derivation)}
\begin{equation}\label{S_int_decomp}
    S_\mathrm{int} \to S_\chi + S_{\chi \psi} + S_{g \chi \phi},
\end{equation}
where $S_\chi$ is the action of the HS defined as
\begin{multline}\label{S_chi}
    S_\chi \equiv \frac12 \sum_{\alpha,\beta}^M\sum_{i,j}^N \sum_{\sigma,\sigma'}^{1,2}\oint \oint dt_c \,dt'_c \, \\ \times \chi^\sigma_{\alpha i}(t_c) \cdot\qty(\hat{W}^{-1}_0)^{\sigma,\sigma'}_{\alpha i,\beta j}(t_c,t'_c)\cdot \chi^{\sigma'}_{\beta j}(t'_c).
\end{multline}
We name $\chi$ as the Ising field, as it mediates the Ising-type interaction amongst spins as we will see below.  $\chi$ is a two-component real valued scalar field defined as
\begin{equation}
    \vec{\chi}_{\alpha i}\equiv \qty(\chi^1_{\alpha i},\chi^2_{\alpha i}),
\end{equation}
together with its inverse bare Green's function
\begin{equation}\label{W0}
    \qty(\hat{W}^{-1}_0)^{\sigma,\sigma'}_{\alpha i,\beta j}(t_c,t'_c)\equiv \sqrt{N+M}\delta_{\alpha \beta}\delta_{i j}\qty(\sigma^x)_{\sigma \sigma'}\delta(t_c,t'_c),
\end{equation}
and its full Green's function
\begin{equation}
    iW_{\alpha i , \beta j }^{\sigma,\sigma'}(t,t')\equiv \expval{\chi^{\sigma}_{\alpha i}(t)\chi^{\sigma'}_{\beta j}(t')}_c: \includegraphics{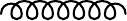}.
\end{equation}
{Note that the ``free" part of the action for the Ising field $S_\chi$ is local in time and does not contain any time derivatives of $\chi$. This makes the equations of motion for $\chi$ algebraic rather than differential. The latter is the generic case where the equations of motion for correlation functions form a system of coupled differential equations, and adiabatic elimination is equivalent to approximately ignoring the time derivatives of some of the dynamical variables, which are assumed to have a quick response compared to other timescales in the system. For the HS field, adiabatically eliminating $\chi$ is exact, which is equivalent to taking the Gaussian integral over $\chi$ in Eq.~(\ref{S_chi}), and the result is given by $S_\mathrm{int}$ in Eq.~(\ref{S_int}).}
The next term in Eq.~(\ref{S_int_decomp}) is $S_{\chi \psi}$ which describes the coupling of fermions to the first component of $\chi$:
\begin{equation}\label{S_chipsi}
    S_{\chi \psi} \equiv -\frac{2i}{\sqrt{N_s}} \sum_{\alpha,i,\lambda}\oint dt_c\, \chi^1_{\alpha i}\psi^y_{i \lambda}\psi^z_{i \lambda},
\end{equation}
and $S_{g \chi \phi}$ describes the disordered interaction of photons with the second component of $\chi$
\begin{equation}\label{S_chiphi}
    S_{g \chi \phi}\equiv \sqrt{2\omega_c} \sum_{\alpha,i} \oint dt_c\, \chi^2_{\alpha i } g_{\alpha i}\phi_\alpha.
\end{equation}
The diagrammatic representations of the original vertex in Eq. (\ref{S_int}) and the transformed ones (Eqs. (\ref{S_chipsi}) and (\ref{S_chiphi}) are given in Fig. \ref{fig:vertex}.

We will show later that the two components of the Ising field correspond to different physical quantities. As will be shown in Section~\ref{sec:eom_at_LO}, $\chi^1$ is related to the effective magnetic field each cluster experiences and $\chi^2$ is connected to magnetization. Similarly, the Green's functions of Ising fields are not just mathematical objects and have physical meanings. $W$ can be expressed in terms of the original Green's functions as shown in Fig. \ref{fig:W_to_GD}. $W^{22}$ is related to the spin-spin correlation function or equivalently, the 4-point function of Majorana fermions
\begin{equation}
    \expval{S^x_i(t)S^x_i(t')}=-4\sum_{\lambda,\lambda'}^{N_s}\expval{\psi^y_{i\lambda}(t)\psi^z_{i\lambda}(t)\psi^y_{i\lambda'}(t')\psi^z_{i\lambda'}(t')}
\end{equation}
whose leading order expansion is given by the same set of diagrams as $W^{22}$ in Fig.~\ref{fig:W_to_GD}, up to multiplication by an overall constant, as given by Eq.~(\ref{ss_to_V}). Therefore, spin-spin correlation functions are natural byproducts of our formalism. Therefore, there is no need to solve the Bethe-Salpeter equations to obtain 4-point functions of fermions, usually a cumbersome task particularly for out of equilibrium systems \cite{Babadi_PRX15,arrizabalaga2004quantum,carrington2014four}. 

\begin{figure*}[!t]
    \centering
    \subfloat[\label{fig:vertex}]{\includegraphics[height=0.1\textwidth]{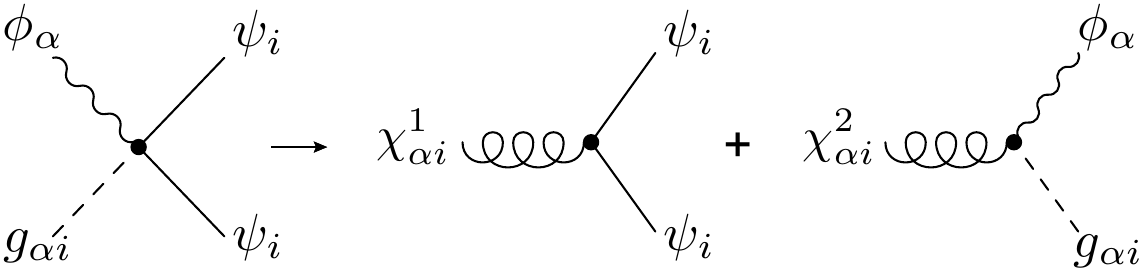}}
    \hspace{30pt} \subfloat[\label{fig:vert_avg}]{\includegraphics[height=0.1\textwidth]{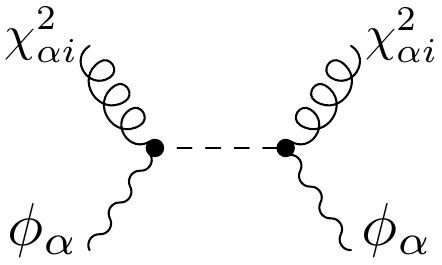}}\\
    \subfloat[\label{fig:W_to_GD}]{\begin{minipage}{0.95\textwidth}
    \includegraphics[width=\textwidth]{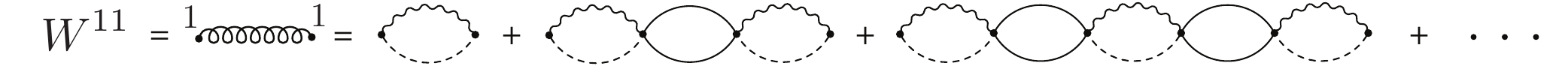}\vspace{5pt}\\ \includegraphics[width=\textwidth]{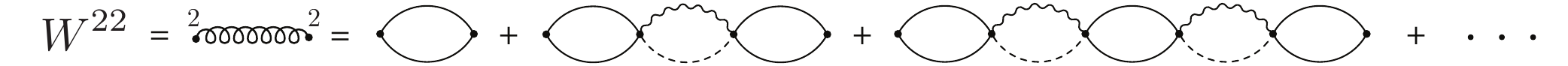}
    \end{minipage}}
    \caption{(a) The original interaction vertex and the decoupled interactions after HS transformation. (b) The effective interaction between photons and the Ising field after disorder averaging. (c) Diagrammatic representation of the Ising (HS) field propagators. The appearing fermion and photon lines are assumed to be renormalized by interactions. $W^{22}$ coincides with spin-spin correlation function, up to an overall multiplicative factor.}
\end{figure*}

\subsection{Disorder Averaging}\label{sec:dis_avg}
We can now take the average over the disordered couplings. In the Keldysh formalism, the average can be taken without resorting to the replica trick \cite{kamenev}. The only term in the action depending on $g$ is $S_{g \chi \phi}$. The effective interaction after disorder averaging is given by $e^{iS_{\chi \phi}}=\overline{e^{i S_{g \chi \phi}}}$, where
\begin{equation}\label{S_chiphi2}
    S_{\chi \phi}=ig^2 \omega_c \sum_{\alpha i} \oint \oint dt_c\,dt'_c\, \chi^2_{\alpha i }(t_c) \phi_\alpha(t_c) \chi^2_{\alpha i }(t'_c) \phi_\alpha(t'_c),
\end{equation}
is shown diagrammatically in Fig. \ref{fig:vert_avg}. We have to make an important remark about the process of disorder averaging. In obtaining Eq. (\ref{S_chiphi2}) we have assumed that the initial state of the system is not correlated with the disorder. This is valid for the initial states we consider in this paper. However, to study phenomena such as associative memory in multi-mode cavity QED~\cite{Hopfield,Amit_PRL85,Rotondo_hopfield2015,Fiorelli_PRL20,Carollo_MemoryRet2021,Marsh_PRX21}, where a significant overlap of the initial spin configuration is required for memory retrieval, one has to assume that the initial state depends on $g_{\alpha i}$. In that case, disorder averaging will generate more terms than Eq. (\ref{S_chiphi2}), which couple the initial state to the interaction vertex in Eq. (\ref{S_chiphi}).

\subsection{Symmetry Considerations}\label{sec:symmetries}

The symmetry structure of the model helps us to simplify the study of its dynamical response. Originally, the Hamiltonian in Eq. (\ref{H}) is invariant only under a global $Z_2$ transformation that maps all spins and cavity modes simultaneously according to
\begin{equation}
    \sigma^x_{i\lambda} \to - \sigma^x_{i\lambda}, \qquad a_\alpha \to - a_\alpha.
\end{equation}
According to the language of Ref.~\cite{PhysRevA.89.022118,Sieberer_2016,buvca2012note,lieu2020symmetry}, in the absence of photon loss this is a quantum symmetry of the system with a conserved $Z_2$ charge. A quantum symmetry is a symmetry of the fields on each individual Keldysh contour, while a classical symmetry is the invariance of the Keldysh action under a simultaneous transformation of the fields on forward and backward contours~\cite{Sieberer_2016}. With photon loss, the quantum symmetry is demoted to a classical symmetry without a conserved charge. However, starting from a symmetric initial state, a classical symmetry still guarantees that the symmetry will remain unbroken in the absence of symmetry breaking perturbations.

The symmetry structure of the model is enriched after disorder averaging and using the fermion representation in Eq. (\ref{Majorana_rep}). It can be easily verified that the disorder-averaged Keldysh action has the following sets of symmetries
\begin{enumerate}
    \item A local $Z_2$ gauge symmetry under the transformation
    \begin{equation}
        \vec{\psi}_{i\lambda} \to - \vec{\psi}_{i\lambda},
    \end{equation}
    which holds for each spin separately. This symmetry is an artifact of representing spins in terms of quadratic fermion operators, and is not physical. The initial state or external forces cannot break this symmetry. The important consequence of this symmetry is that
    \begin{equation}
        G^{\alpha,\beta}_{i\lambda,j\lambda'}\propto \delta_{ij}\delta_{\lambda \lambda'}.
    \end{equation}
    \item A $Z_2$ symmetry for each separate cluster $i$ and Ising fields coupled to it:
    \begin{align}
        \sigma^x_{i\lambda} &\to -\sigma^x_{i\lambda}, \quad (\lambda=1,...,N_s), \\
        \chi^\sigma_{\alpha i} &\to - \chi^\sigma_{\alpha i}, \quad (\sigma=1,2), \quad (\alpha=1,...,M).\label{z2_chi}
    \end{align}
    (\ref{z2_chi}) holds because the effective interaction in Eq. (\ref{S_chiphi2}) is quadratic in $\chi$. This symmetry is a classical symmetry with no conserved quantities.
    \item A $Z_2$ symmetry of each photon mode given by
    \begin{equation}
        \phi_\alpha \to - \phi_\alpha, \quad \pi_\alpha \to -\pi_\alpha.
    \end{equation}
    This symmetry is also a result of $S_{\chi\phi}$ being quadratic in photon fields and is a weak symmetry. This symmetry implies that
    \begin{equation}
        D_{\alpha,\beta}^{\rho,\rho'} \propto \delta_{\alpha \beta}.
    \end{equation}
\end{enumerate}
We see that the $Z_2$ symmetries of spin and photon sectors are decoupled. This means that, even if the initial state of spins breaks the symmetry, no photon coherence will be generated ($\overline{\expval{a_\alpha(t)}}=0$). On the other hand, a finite value for $\sigma^x_{i \lambda}$ results in a finite value for $\chi^{1,2}_{\alpha i}$.

We again remark that the above arguments hold true only if the initial state of the system is not correlated with the disorder pattern, such that Eq. (\ref{S_chiphi2}) is the only outcome of disorder averaging. Otherwise, a $Z_2$ broken initial state can in principle break the $Z_2$ symmetry of some of the photon modes. This happens for example, if the initial spin configuration has a strong overlap with a single disorder pattern corresponding to the photon mode $\alpha$, such that
\begin{equation}\label{overlap_sigma_g}
    \lim_{N\to \infty} \frac{1}{N N_s}\abs{\sum_{i}^N\sum_{\lambda}^{N_s} \expval{\sigma^x_{i \lambda}}_0g_{\alpha i}} > 0,
\end{equation}
{or if a symmetry breaking perturbation that favors a single pattern such as
\begin{equation}
\delta H = \epsilon \sum_{i,\lambda} g_{\alpha,i}\sigma^x_{i\lambda},
\end{equation}
is applied to the system.} In this case, one expects that for sufficiently small $M/N$, the pattern $\alpha$ to be activated and retrieved \cite{Amit_PRL85,Fiorelli_PRL20}.

For the fully polarized initial states of spins considered in this problem and in the thermodynamic limit, we can safely take $\overline{\expval{\phi_\alpha(t)}}=0$ throughout the evolution. Even starting from a state with $\expval{\phi}\neq0$, its value will decay to zero as it cannot align itself with any of the disorder patterns.

\subsection{2PI Action}

\begin{figure}[!t]
    \centering
    \subfloat[\label{fig:Gamma_LO}]{\includegraphics[height=0.13\textwidth]{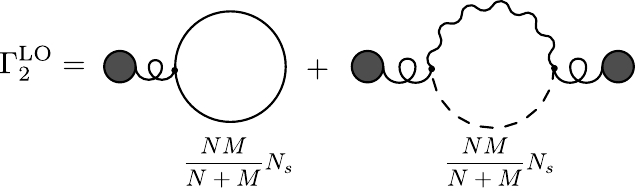}}\\
    \subfloat[\label{fig:Gamma_neg}]{\includegraphics[height=0.075\textwidth]{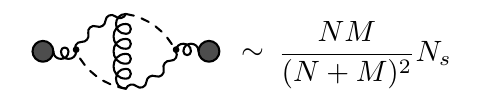}}
    \caption{(a) The leading-order contribution of interactions to 2PI action. The dashed line in the right diagram is not a Green's function and cannot be cut. (b) A diagram which linearly scales with $N_s$ but is not extensive and neglected. Black circles connected to springs represent the expectation value of Ising field $\chi$.}
\end{figure}

The 2PI action for the model given above is a functional of fermion, photon and Ising field correlation functions together with the expectation values of Ising fields and has the general form given by~\cite{Berges2004introduction}
\begin{multline}\label{Gamma_DisDick}
    \Gamma\qty[\tilde{\chi},G,D,W] = S_\chi\qty[\tilde{\chi}] + \frac{i}{2} \Tr \ln G - \frac{i}{2} \Tr \qty(G_0^{-1}G) \\ - \frac{i}{2} \Tr \ln D + \frac{i}{2} \Tr \qty(D^{-1}_0 D) - \frac{i}{2} \Tr \ln W  \\ + \frac{i}{2} \Tr \qty(W^{-1}_0 W) + \Gamma_2\qty[\tilde{\chi},G,D,W].
\end{multline}
The expressions for $S_\chi$, $G_0$, $D_0$ and $W_0$ were respectively given in Eqs. (\ref{S_chi}), (\ref{G0}), (\ref{D0}) and (\ref{W0}). $\tilde{\chi}$ is the expectation value of the Ising field
\begin{equation}
    \tilde{\chi}^\sigma_{\alpha i }(t) \equiv \overline{\expval{\chi^\sigma_{\alpha i}(t)}},
\end{equation}
shown by a black circle connected to a spring in (Fig.~\ref{fig:Gamma_LO}). The last term in Eq. (\ref{Gamma_DisDick}) captures interactions and as we mentioned in Sec.~\ref{sec:basics}, is given by the sum of 2PI diagrams. In order to systematically expand $\Gamma_2$, we need to specify how the expectation values and Green's functions of the Ising field scale with parameters of the system. According to Eq. (\ref{W0}) to the leading order in $(N+M)^{-1}$ we have
\begin{equation}
    W^{12}_{i\alpha, j \beta}\sim \delta_{ij}\delta_{\alpha \beta} \,\mathcal{O}\qty(\frac{1}{\sqrt{N+M}}),
\end{equation}
{The diagonal elements $W^{11}$ and $W^{22}$ are zero at the bare level in Eq.~(\ref{S_chi}). However, they become non-zero when the couplings of $\chi$ to $\psi$ and $\phi$ are taken into account (Fig.~\ref{fig:W_to_GD}). For $W^{22}$ we have}
\begin{equation}
    W^{22}_{i\alpha,j\beta}\sim \delta_{ij}\,\mathcal{O}\qty(\frac{1}{N+M}).
\end{equation}
As will be shown later, $W^{11}$ has a sub-leading term due to interactions which contributes at leading order when it is summed over photon modes
\begin{multline}\label{w11_scaling}
    W^{11}_{\alpha \beta}\sim \delta_{ij}\delta_{\alpha\beta}\, \mathcal{O}\qty(\frac{1}{N+M})\\ + \delta_{ij}(1-\delta_{\alpha \beta}) \,\mathcal{O}\qty(\frac{1}{(N+M)^2}).
\end{multline}
Furthermore, $\tilde{\chi}$ will have the following scalings {(Eqs.~(\ref{chi1_EOM}) and (\ref{chi2_EOM}))}
\begin{equation}\label{chi_scaling}
    \tilde{\chi}^1 \sim \mathcal{O}\qty(\frac{\sqrt{N_s}}{N+M}), \quad \tilde{\chi}^2 \sim \mathcal{O}\qty(\sqrt{\frac{N_s}{N+M}}).
\end{equation}
At last, the fermion-Ising vertex in Eq.~(\ref{S_chipsi}) has
\begin{equation}
    S_{\chi\psi} \sim \frac{1}{\sqrt{N_s}}.
\end{equation}
We now have all of the necessary ingredients to perform a systematic expansion of $\Gamma_2$.

\subsection{Diagrammatic Evaluation of 2PI Action}

We are interested in the thermodynamic limit of the system in Eq. (\ref{H}) where $N,M \to \infty$ while the ratio $\eta=M/N$ is kept fixed. Moreover, the number of spins per cluster $N_s$ is assumed to be larger than one and will play the role of the control parameter for the expansion. Also, we assume that $N_s \ll N$, which is a valid assumption in the thermodynamic limit of the problem. The interaction part of the 2PI action, given by $\Gamma_2$ in Eq.~(\ref{Gamma_DisDick}), admits a diagrammatic expansion in terms of the connected vacuum bubbles of the theory which cannot be split into half by cutting one or two of their Green's function lines, also known as two-particle irreducible (2PI) graphs~\cite{Berges2004introduction}. Below, we will classify these diagrams for our system as leading-order (LO) terms
\begin{equation}
    \Gamma_2^\mathrm{LO}\sim N_s,
\end{equation}
and next-to-leading-order (NLO) terms 
\begin{equation}
    \Gamma_2^\mathrm{NLO}\sim N_s^0,
\end{equation}
and higher order terms which are ignored in this work. We will also ignore terms which are sub-extensive. 

The following discussion will also elucidate how the parameter controlling the strength of quantum fluctuations, $N_s$, enters naturally in the field theory description and in the DE derived from it. This is one of the key merits of the approach, at variance with more numerical oriented methods which have to deal with growing computational complexity as $N_s$ is decreased.

\subsubsection{Leading-order contributions}

The LO terms have linear scaling with $N_s$ and at the same time, scale extensively with system size. Two of these diagrams exist and both involve the expectation values of Ising fields, as shown in Fig. \ref{fig:Gamma_LO}. Their mathematical expressions are given by
\begin{multline}\label{Gamma_LO}
    \Gamma_2^\mathrm{LO}=\frac{2}{\sqrt{N_s}}\sum_\alpha^M\sum_i^N \sum_\lambda^{N_s} \oint dt_c\, \tilde{\chi}^1_{\alpha i}(t_c) G^{y,z}_{i \lambda, i \lambda}(t_c,t_c) \\   -g^2 \omega_c \sum_\alpha^M \sum_i^N \oint\oint dt_c\,dt'_c\, \tilde{\chi}^2_{\alpha i}(t_c) D_{\alpha,\alpha}^{\phi \phi}(t_c,t'_c) \tilde{\chi}^2_{\alpha i}(t'_c).
\end{multline}
Note that the disorder (dashed) line in Fig. \ref{fig:Gamma_LO} cannot be cut, as it is a part of the disorder averaged interaction vertex in Fig. \ref{fig:vert_avg}. Accordingly, the  right diagram in Fig.~\ref{fig:Gamma_LO} is not 2-particle reducible. It can be shown, by solving the resulting equations of motion derived from Eq. (\ref{Gamma_LO}), that the LO terms describe a dynamical mean field interaction of spin expectation values $\overline{\expval{\sigma^x}}$ mediated by photons through their response function (see Section~\ref{sec:LO_results}). Therefore, the LO contribution describes the LMG coupling in Eq. (\ref{H_LMG}) with the inclusion of retardation effects due to photon dynamics.

We note that there are other terms that scale linearly with $N_s$, such as the one given in Fig. \ref{fig:Gamma_neg}, but all of them scale non-extensively with system size and can be neglected in the thermodynamic limit.

\subsubsection{Next-to-leading-order contributions}

\begin{figure}[!t]
    \centering
    \includegraphics[height=0.15\textwidth]{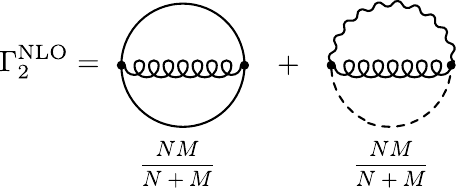}
    \caption{Next-to-leading-order contributions to 2PI action.}
     \label{fig:Gamma_NLO}
\end{figure}

\begin{figure}[!t]
    \centering
    \includegraphics[height=0.14\textwidth]{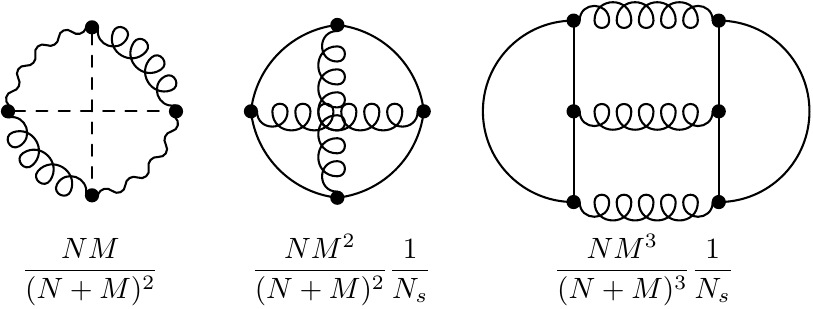}
    \caption{Some of the neglected diagrams in this work which appear after NLO terms. The left diagram is sub-extensive in system size and the rest are NNLO in $1/N_s$.}
    \label{fig:Gamma_NNLO}
\end{figure}

The NLO diagrams do not scale with $N_s$ but still  scale linearly with the number of clusters $N$. There are two NLO diagrams shown in Fig. \ref{fig:Gamma_NLO} and their formulae are given by
\begin{multline}\label{Gamma_NLO}
    \Gamma_2^\mathrm{NLO}= -\frac{2}{N_s}\sum_{\alpha,\beta}^M \sum_i^N \sum_\lambda^{N_s} \oint \oint dt_c\,dt'_c\, W^{11}_{\alpha i,\beta i}(t_c,t'_c) \\ \times \Big( G^{y,z}_{i\lambda}(t_c,t'_c)G^{z,y}_{i\lambda}(t_c,t'_c) - G^{y,y}_{i\lambda}(t_c,t'_c)G^{z,z}_{i\lambda}(t_c,t'_c)\Big) \\
    -ig^2 \omega_c \sum_\alpha^M \sum_i^N \oint \oint dt_c\,dt'_c\, W^{22}_{\alpha i ,\alpha i}(t_c,t'_c) D^{\phi \phi}_{\alpha \alpha}(t_c,t'_c).
\end{multline}
The rest of the terms in $\Gamma_2$ are either next-to-next-to-leading-order (NNLO) in $1/N_s$ or scale sub-extensively with system size (Fig. \ref{fig:Gamma_NNLO}). In this work we neglect these terms and take
\begin{equation}
    \Gamma_2 \approx \Gamma_2^\mathrm{LO}+ \Gamma_2^\mathrm{NLO}.
\end{equation}

It is worth showing which diagrams we are keeping in terms of the original action prior to HS transformation (Fig. \ref{fig:Gamma_orig}). As mentioned before, the LO part describes a retarded self-interaction of the spin expectation value expressed in terms of fermion Green's function (according to Eq.~\ref{Majorana_rep})
\begin{equation}\label{sx_to_Gzy}
\overline{\expval{\sigma^x}}=2G_{-+}^{z,y}(t,t),
\end{equation}
given by closed loops in the first term of Fig. \ref{fig:Gamma_orig}, within the same cluster and mediated by photons. In the regime of fast photons and at steady state, this reduces to the ferromagnetic interaction in Eq. \ref{H_int}. For NLO terms, the two diagrams in Fig. \ref{fig:Gamma_NLO} are equivalent to the sum of an infinite number of diagrams in the original representation of the theory, shown in the brackets of Fig. \ref{fig:Gamma_orig}. {This can be verified by plugging the diagrammatic expression of $W$ given in Fig.~\ref{fig:W_to_GD} into the diagrams of Fig.~\ref{fig:Gamma_NLO}, and subsequently substitute the lines for bare $W$ with simple dots, as the latter is just a constant.} This infinite series is a byproduct of disordered couplings. For the Dicke model without disorder (see Appendix \ref{app:Dicke}), only the first NLO diagram would contribute since in the absence of disorder, dashed lines would disappear. This makes the rest of NLO diagrams 2-particle-\textit{reducible} and hence, forbidden in the expansion of $\Gamma_2$.

\subsection{Equations of Motion}
Green's functions and field expectation values are obtained from the stationary solution of QEA which is found from
\begin{align}
    \frac{\delta \Gamma}{\delta \tilde{\chi}^\sigma_{\alpha i}(t)}=0, 
    \qquad \frac{\delta \Gamma}{\delta W^{\sigma \sigma'}_{\alpha i ,\beta j}(t_c,t'_c)}&=0,  \label{chi_eom}\\
    \frac{\delta \Gamma}{\delta D^{\rho \rho'}_{\alpha ,\beta}(t_c,t'_c)}=0,
    \qquad\frac{\delta \Gamma}{\delta G^{\alpha \beta}_{i\lambda ,j\lambda'}(t_c,t'_c)}&=0.
\end{align}
After taking functional derivatives of Eq. (\ref{Gamma_DisDick}), we see that the equations of motion for Green's functions can always be cast compactly as
\begin{align}
    \hat{G}^{-1}&= \hat{G}_0^{-1} - \hat{\Sigma}, \label{Dy_G}\\
    \hat{D}^{-1}&= \hat{D}_0^{-1} - \hat{\Pi},\label{Dy_D}\\
    \hat{W}^{-1}&= \hat{W}_0^{-1} - \hat{\Omega}, \label{Dy_W}
\end{align}
known as Dyson equations \cite{Berges2004introduction, Rammer_2007,kamenev}. The matrices $\hat{\Sigma}$, $\hat{\Pi}$ and $\hat{\Omega}$ are fermion, photon and Ising field self-energies, respectively. They are given in terms of the functional derivatives of $\Gamma_2$ as
\begin{align}
    \Sigma^{\alpha ,\beta}_{i\lambda,j\lambda'}(t_c,t'_c) &\equiv -2i \frac{\delta \Gamma_2}{\delta G^{\beta,\alpha}_{j\lambda',i\lambda}(t'_c,t_c)},\\
    \Pi^{\rho ,\rho'}_{\alpha,\beta}(t_c,t'_c) &\equiv +2i \frac{\delta \Gamma_2}{\delta D^{\rho',\rho}_{\beta,\alpha}(t'_c,t_c)},\label{Pi_def}\\
    \Omega^{\sigma ,\sigma'}_{\alpha i,\beta j}(t_c,t'_c) &\equiv +2i \frac{\delta \Gamma_2}{\delta W^{\sigma',\sigma}_{\beta j,\alpha i}(t'_c,t_c)}.
\end{align}
Due to their large size, the expanded forms of Eqs. (\ref{Dy_G}-\ref{Dy_W}), required for numerically solving them, are given in Appendix \ref{app:DE}. We will only mention the important details here. The fermion and photon Green's functions will be diagonal in the spin-site and photon-mode bases, respectively. Due to permutation symmetry~\cite{kirton2019introduction,keeling2010collective,shammah2018open} we also have
\begin{align}
    G^{\alpha ,\beta}_{i\lambda,j\lambda'}(t_c,t'_c)&= \delta_{ij}\delta_{\lambda \lambda'} G^{\alpha,\beta}(t_c,t'_c), \\
    D^{\rho ,\rho'}_{\alpha,\beta}(t_c,t'_c) &= \delta_{\alpha \beta} D^{\rho, \rho'}(t_c,t'_c).
\end{align}
The same is true for their self-energies. The Ising field's Green's function and its self-energy will acquire off-diagonal elements only in the photon indices. Due to the emergent permutation symmetry after disorder averaging, all diagonal elements of $W$ and $\Omega$ are the same. This holds also for the off-diagonal elements of $W$ and $\Omega$:
\begin{align}
    W^{\sigma,\sigma'}_{\alpha i,\alpha j}(t_c,t'_c)&= \delta_{ij} V^{\sigma,\sigma'}(t_c,t'_c), \\
    W^{\sigma,\sigma'}_{\alpha i,\beta j}(t_c,t'_c)&= \delta_{ij} U^{\sigma,\sigma'}(t_c,t'_c), \quad \alpha \neq \beta.
\end{align}
A similar argument applies to $\tilde{\chi}$, and it has the same value for all sites and photon modes:
\begin{equation}
    \tilde{\chi}^\sigma_{\alpha i}(t)=\tilde{\chi}^\sigma(t)
\end{equation}
We proceed similarly to the previous section. We find the equations of motion at LO first and then consider the NLO corrections.

\subsubsection{Equations of motion at leading-order (LO)}\label{sec:eom_at_LO}
For $\tilde{\chi}$, one always finds that it has the same value on forward and backward Keldysh contours, as it should be since these are classical variables. From Eqs.~(\ref{Gamma_LO}) and (\ref{chi_eom}) we have
\begin{equation}\label{chi2_EOM}
    \frac{\delta \Gamma^\mathrm{LO}}{\delta \tilde{\chi}^1}=0 \to \tilde{\chi}^2(t)=-2\sqrt{\frac{N_s}{N+M}}G_{-+}^{y,z}(t,t),
\end{equation}
\begin{equation}\label{chi1_EOM}
     \frac{\delta \Gamma^\mathrm{LO}}{\delta \tilde{\chi}^2}=0 \to \tilde{\chi}^1(t)=\frac{2g^2 \omega_c}{\sqrt{N+M}} \int_0^t dt'\, D_R^{\phi \phi}(t,t')\tilde{\chi}^2(t'),
\end{equation}
where $D_R^{\phi\phi}$ is the photon response function defined as (for details see Appendix \ref{app:DE})
\begin{equation}\label{DR_def}
    D_R^{\phi\phi}(t,t')\equiv -i \Theta(t-t') \overline{\expval{\comm{\hat{\phi}(t)}{\hat{\phi}(t')}}},
\end{equation}
which describes the response of the photons to an external linear perturbation. Note that the time variables used in Eq. (\ref{DR_def}) are normal variables and are not defined on the time contour. Substituting (\ref{chi2_EOM}) into (\ref{chi1_EOM}) gives
\begin{equation}\label{chi1_final}
    \tilde{\chi}^1(t)=-\frac{4g^2\omega_c \sqrt{N_s}}{N+M}\int_0^t dt'\, D_R^{\phi \phi}(t,t') G_{-+}^{y,z}(t',t'),
\end{equation}
in agreement with the scaling relations for $\tilde{\chi}$ given in Eq. (\ref{chi_scaling}).

For $\Sigma$ we have at the leading order
\begin{equation}\label{Sigma_LO}
    \Sigma_\mathrm{LO}^{\alpha,\beta}(t_c,t'_c)=\parbox{1.2cm}{\includegraphics[height=\linewidth]{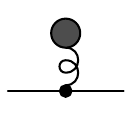}}=-\frac{2iM}{\sqrt{N_s}} \epsilon_{x\alpha \beta}  \delta(t_c,t'_c) \tilde{\chi}^1(t_c).
\end{equation}
The temporally local elements of fermion self-energy describe the effective magnetic field spins experience due to interactions. Accordingly, Eq. (\ref{Sigma_LO}) describes a time-dependent magnetic field $B_x$ in the $\hat{x}$-direction experienced by each spin. This sheds light on the physical meaning of the first component of the HS field:  it is the magnetic field experienced by each spin and its Green's functions give the fluctuations of this field. This could be inferred also from Eq.~(\ref{S_chipsi}) which describes the coupling of $\sigma^x=-2i\psi^z\psi^y$ to $\chi_1$. Similarly, Eqs.~(\ref{chi2_EOM}) and (\ref{sx_to_Gzy}) relate $\chi^2$ to $\expval{\sigma^x}$ as
\begin{equation}
    \overline{\expval{S^x_i(t)}}=-\sqrt{(N+M)N_s} \,\tilde{\chi}^2(t)/2.
\end{equation}
Furthermore, it can be shown that $W^{22}$ is related to the 2-point correlation function of the large spins $S_i^x=\sum_\lambda \sigma^x_{i\lambda}/2$ through
\begin{equation} \label{ss_to_V}
    \overline{\expval{S^x_i(t_c)S^x_i(t'_c)}_c}=i(N+M)N_s \, V^{22}(t_c,t'_c)/4.
\end{equation}

Eq.~(\ref{ss_to_V}) can be rigorously proven by introducing source terms in the Keldysh action which are coupled to $\sigma^x$, and then taking functional derivatives with respect to the sources twice (cf. Appendix~\ref{app:HS}). By substituting $\tilde{\chi}^1$ from Eq. (\ref{chi1_final}) into Eq. (\ref{Sigma_LO}), the effective magnetic field $B_x$ is found to be
\begin{equation}\label{Bx_def}
    B_x(t)=-\frac{4g^2\omega_c M}{N+M} \int_0^t dt'\, D_R^{\phi \phi}(t,t') G_K^{y,z}(t',t').
\end{equation}
 Interpreting Eq. (\ref{Bx_def}) is straightforward now; spins perturb photons whose displacement is given by $\phi$ which acts as an effective field applied to spins through the Dicke coupling in Eq. (\ref{H}), creating a self-interaction for spins.

Last, we investigate the photon sector at LO by calculating the photon self-energy $\Pi$. Since all terms in $\Gamma_2$ only depend on the $\phi \phi$ component of $D$, the only non-zero element of photon self-energy is $\Pi^{\phi \phi}$ given by
\begin{multline}
    \Pi_\mathrm{LO}^{\phi \phi}(t_c,t'_c)=\parbox{1.2cm}{\includegraphics[height=\linewidth]{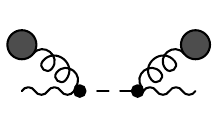}}\\= -2i N g^2\omega_c \tilde{\chi}^2(t_c) \tilde{\chi}^2(t'_c).
\end{multline}
Written in terms of normal time variables, it is easy to show that $\Pi^\mathrm{LO}$ does not alter the spectrum or equivalently \cite{kamenev,altland2010condensed}, the response function of photons. Hence, its only effect is to increase the photon population. At this order, spins pump photons, but without generating any finite values for $\expval{\phi}$. As couplings to different clusters have different signs, their MF contributions cancel out each other in the thermodynamic limit and photon pumping is realized only at the level of fluctuations. Furthermore, no changes of $D^{\phi \phi}_R$ at this order of approximation means that the kernel of the effective interaction between spins is given by its non-interacting form, i.e. it is the response function of a damped (for $\kappa\neq 0$) harmonic oscillator. We note that the self-energy of the Ising field vanishes at this order.

We summarize the physics of the problem in LO approximation. There is an effective Ising-like interaction of spins within the same cluster with a retarded kernel given by the response of cavity photons in the non-interacting limit. Photons are coherently pumped by spins, but  their energy levels and loss rates remain unaffected. At this order of approximation, the model behaves very similar to the MF solution of the Dicke model. The effective retarded spin-spin interaction is generated also for the Dicke model, if we formally solve the equation of motion for the photon mode in terms of $\sigma^x$ and then, substitute them back in the equations of motion for spins.

\subsubsection{Equations of motion at next-to-leading-order}
As is evident from Eq. (\ref{Gamma_NLO}), $\Gamma^\mathrm{NLO}_2$ does not depend on $\tilde{\chi}$. Therefore, Eqs. (\ref{chi2_EOM}) and (\ref{chi1_EOM}) describe $\tilde{\chi}$ to NLO. Accordingly, the general picture of an effective MF interaction between spins remain unaltered. Although the interaction kernel given by $D^{\phi\phi}_R$ will be renormalized by fluctuations at NLO.
\\
\\
The self-energies at NLO are found from the functional derivatives of $\Gamma_2^\mathrm{NLO}$ and are given by
\begin{multline}\label{Sigma_NLO}
    \Sigma_\mathrm{NLO}^{\alpha,\beta}(t_c,t'_c)=\parbox{1.2cm}{\includegraphics[height=\linewidth]{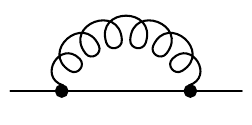}} \\=\frac{4iM}{N_s}\qty(V^{11}(t_c,t'_c)+(M-1)U^{11}(t_c,t'_c))\\ \times \sum_{\gamma,\delta} \epsilon_{x\alpha \gamma}\epsilon_{x\beta \delta}G^{\gamma,\delta}(t_c,t'_c), 
\end{multline}
\begin{multline}\label{Pi_NLO}
    \Pi_\mathrm{NLO}^{\phi \phi}(t_c,t'_c) =\parbox{1.2cm}{\includegraphics[height=\linewidth]{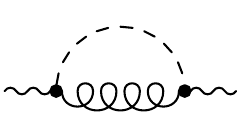}}\\ =2 N g^2\omega_c \,V^{22}(t_c,t'_c).
\end{multline}  
Ising self-energies are non-vanishing at this order:
\begin{multline}\label{Omega_11_NLO}
    (\Omega_\mathrm{NLO})_{\alpha \beta}^{11}(t_c,t'_c) =\parbox{1.2cm}{\includegraphics[height=\linewidth]{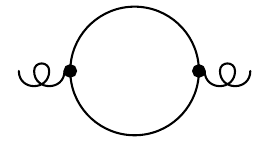}} \hspace{25pt} =4i g^2\omega_c \\ \times \Big(G^{yy}(t_c,t'_c)G^{zz}(t_c,t'_c) -G^{yz}(t_c,t'_c)G^{zy}(t_c,t'_c)\Big),
\end{multline}
\begin{multline}\label{Omega_22_NLO}
    (\Omega_\mathrm{NLO})_{\alpha \beta}^{22}(t_c,t'_c) =\parbox{1.2cm}{\includegraphics[height=\linewidth]{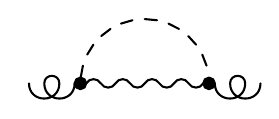}} \\ =2 g^2\omega_c \delta_{\alpha \beta} D^{\phi\phi}(t_c,t'_c).
\end{multline}
We see that the off-diagonal element of $W$ given by $U$ is multiplied by an extra factor of $\mathcal{O}(M)$ in Eq. (\ref{Sigma_NLO}), and it has to be kept though it is sub-leading compared to the diagonal element $V$. The resulting equations of motion are a system of 36 coupled integro-differential equations for different components of Green's functions and $\tilde{\chi}$. The complete expressions for these equations are given in Appendix \ref{app:DE} in terms of the symmetric (Keldysh) and anti-symmetric (retarded and advanced) correlation functions.

\subsubsection{Evaluation of glass order parameter}\label{sec:q_eval}

The formalism developed so far is sufficient to calculate some of the correlation functions of our system which are usually the quantities of interest for quantum dynamics. However, for systems with static disorder, we can define new types of expectation values depending on the order of calculating operator expectation values and disorder averaging \cite{SG_RMP}. \blue{As will be explained in Sec. \ref{sec:q_def}, the spin overlap quantity
\begin{align} \label{q_def}
    Q(t)&\equiv \frac{1}{N_s^2} \overline{\expval{S^x_{i,A}(t)S^x_{i,B}(t)}_c},
\end{align}
 between two similar copies (A and B) of the system, which measures the statistical correlations between A and B generated by the disorder, is of particular importance in our analysis (Cf. Fig.~\ref{fig:overlap_cartoon}). $Q(t)$ cannot be evaluated directly in terms of the ``normal" correlation functions, simply because of the non-commutativity of taking expectation values and disorder averaging in Eq. (\ref{q_def}). However, it can still be calculated thanks to the versatility  of the Keldysh approach in dealing with quenched disorder~\cite{altland2010condensed,kamenev}. The correlation between the value of an observable $\hat{O}$ in two systems before disorder averaging can be written as
\begin{multline}\label{OA_OB}
    \expval{\hat{\mathcal{O}}_A(t)\hat{\mathcal{O}}_B(t)} = \int \mathcal{D}\qty[\phi_A] \mathcal{D}\qty[\phi_B] \\ \times \mathcal{O}_A(t) \mathcal{O}_B(t) e^{iS\qty[\phi_A,g_{\alpha i }]+iS\qty[\phi_B,g_{\alpha i }]}.
\end{multline}
Note that up to this point fields belonging to different copies, $\phi_A$ and $\phi_B$, do not interact with each other. We can now straightforwardly find
\begin{equation}\label{expval_replica}
    Q(t)= \int \mathcal{D}\qty[\phi_A] \mathcal{D}\qty[\phi_B]  \mathcal{O}_A(t) \mathcal{O}_B(t)  \overline{e^{iS\qty[\phi_A,g_{\alpha i }]+iS\qty[\phi_B,g_{\alpha i }]}}.
\end{equation}}
Since the actions of both copies depend on the same realization of $g_{\alpha i}$, averaging over disorder couples fields of different replicas, and generates an effective interaction between them. However, the effective interaction only affects inter-replica quantities of the form given in Eq.~\eqref{OA_OB}. Before applying Eq. (\ref{expval_replica}) to our problem, we make some remarks about our finding. Clearly, there is a strong resemblance between our result and the replica trick \cite{Bray_1980,SG_RMP,mezard1987spin,altland2010condensed} as they both involve more than one copy of the system. However, there are also crucial differences between the two. In the replica trick the number of replicas is taken to zero via analytical continuation while here we are strictly working with 2 copies of the system. Depending on the particular system and other parameters such as temperature, RSB may or may not occur while the system is nevertheless a glass \cite{SG_RMP,Ray_PRB89}.  However, as we will explain later (cf. Sec.~\ref{sec:q_def}), if 1-point functions are vanishing while $Q$ is finite, the system has glassy behavior. Despite these differences, we call Eq. (\ref{expval_replica}) the ``replicated model" for convenience. {We note that a similar replica approach has been used by Ref.~\cite{Buchhold_MIPT2021} to study measurement-induced phase transitions due to continuous-time measurements.}

We apply Eq. (\ref{expval_replica}) to the HS transformed interaction part of the action in Eq. (\ref{S_int_decomp}). $S_{g\chi \phi}$ is the only term depending on $g_{\alpha i}$ and has to be averaged in Eq. (\ref{expval_replica}). The result of disorder averaging are three terms, two of them couple fields from the same copies and correspond to the effective vertex $S_{\chi \phi}$ in Eq. (\ref{S_chiphi2}). The third term gives an interaction between fields of different copies and reads as
\begin{multline}
    S^{AB}_{\chi \phi} = 2ig^2 \omega_c \sum_{\alpha i} \iint dt_c\,dt'_c\, \chi^{2}_{\alpha i ,A}(t_c) \phi_{\alpha,A}(t_c) \\ \times \chi^{2}_{\alpha i, B }(t'_c) \phi_{\alpha,B}(t'_c).
\end{multline}
Using this ``replicated Keldysh field theory", we can obtain $Q(t)$ in terms of the diagonal ($t=t'$) elements of inter-replica correlation functions. To distinguish inter-replica correlators from the normal ones, we represent the former with a tilde mark below them such as $\underaccent{\sim}{D}^{\rho ,\rho'}_{\alpha,\beta}(t_c,t'_c)$. For example, for $Q$ we have
\begin{equation}
    Q(t)=i\frac{N+M}{8 N_s}\underaccent{\sim}{V}^{22}(t,t).
\end{equation}
For the replicated theory, there will be 4 more independent equations of motion that have to be solved in addition to those of the previous section. Crucially, these extra equations do not alter the dynamics of replica-diagonal quantities, as expected, since replicas are just abstractions and they are not ``aware" of each other. Non-replica diagonal quantities, on the other hand, depend both on replica diagonal and non-replica diagonal correlation functions. Furthermore, all of the non-replica diagonal response functions turn out to vanish, as perturbing one replica cannot leave any effects on the other one. We leave the details of the calculations and the extra equations of motion to Appendix \ref{app:q}.

\begin{figure}[!t]
    \centering
    \includegraphics[width=0.48\textwidth,trim={0 1.2cm 0 0},clip]{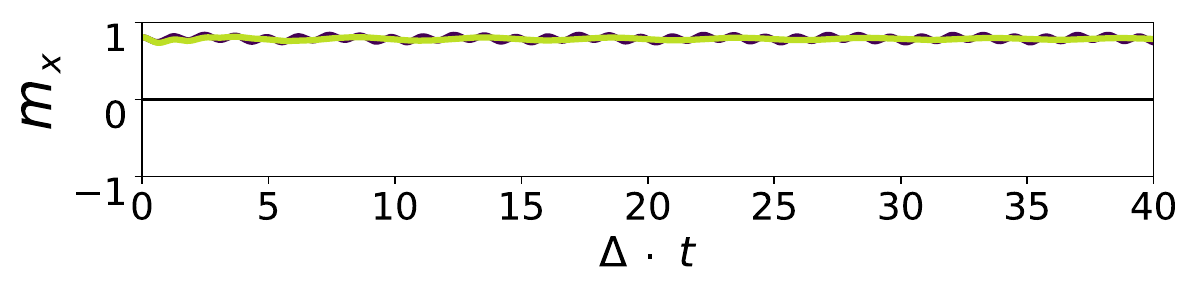}\\
    \includegraphics[width=0.48\textwidth]{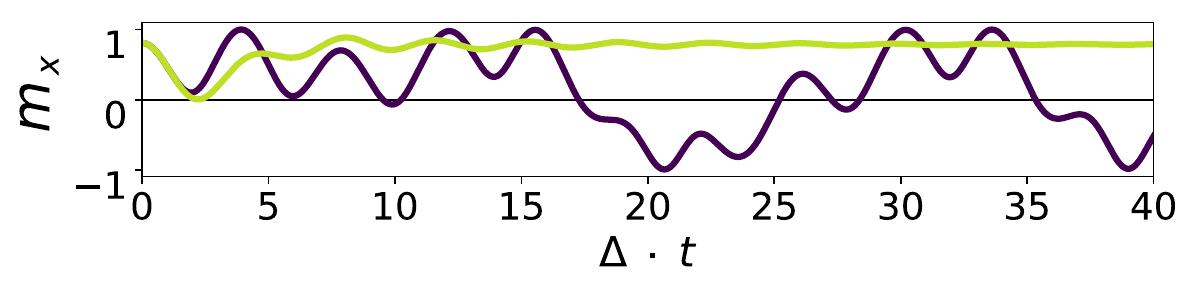}
    \caption{Dynamics at the leading-order of approximation describing the motion of classical spins coupled via a retarded interaction, for $g/g_c=1.27$ and $(\theta_0,\varphi_0)=(0.7\pi,0)$. Top panel: dynamics of spins in the adiabatic limit $\omega_c/\Delta=5.0$ and for $\kappa=0$ (dark purple curve) and $\kappa/\Delta=0.5$ (light green curve). Bottom panel: the same as top panel only for slow photons $\omega_c=\Delta$, showing tunneling events when photon loss is weak. }
    \label{fig:lo_sx}
\end{figure}

\section{Results}\label{sec:results}

In this section, we will report our findings in the following order. First, we discuss in Sec. \ref{sec:LO_results} the results of approximating the effective action only to LO. In Sec. \ref{sec:NLO_results}, we demonstrate that the NLO corrections significantly change the dynamics for all values of $N_s$, motivating the necessity of keeping NLO effects. In Sec. \ref{sec:dynamics_of_SG}, we discuss the formation of SG phase in the system by studying various physical characteristics of SG in our system. We will also address the effect of spin size $N_s$ and photon frequency on the glassy behavior of the system.

We will study the quench dynamics starting from a polarized spin state specified by the angles $(\theta_0,\varphi_0)$ such that
\begin{equation}\label{spin_init_state}
     \expval{\vec{\sigma}}_0=(\sin{\theta_0}\cos{\varphi_0},\sin{\theta_0}\sin{\varphi_0},\cos{\theta_0}),
\end{equation}
and the vacuum state for photons $\ket{0}$ satisfying $\hat{a}\ket{0}=0$. We turn on the spin-photon coupling $g$ at $t=0$ and let the system evolve.

\subsection{Results at LO}\label{sec:LO_results}

The equations of motion for the diagonal elements of fermion Green's functions become decoupled from the non-diagonal elements at LO. This allows us to write the dynamics of magnetization $\vec{m}=\expval{\vec{\sigma}}$ in a transparent form as (Appendix~\ref{app:lo_eom})
\begin{align}
    \dv{}{t}m_x(t) &= - \Delta m_y(t), \label{LO_Sx_EOM} \\
    \dv{}{t} m_y(t) &=  \Delta m_x(t) \nonumber \\ &- J\omega_c^2 \int_0^t m_z(t) D^{\phi \phi}_R(t,t') m_x(t') \,dt', \label{LO_Sy_EOM} \\
    \dv{}{t} m_z(t)&=  J\omega_c^2 \int_0^t m_y(t)D^{\phi \phi}_R(t,t')m_x(t')  \,dt', \label{LO_Sz_EOM}
\end{align}
where $J$ is the coupling of LMG model defined in Eq. (\ref{H_eff}). The photon response function $D^{\phi \phi}_R$ is given by its bare value at LO which is (the minus of) the response function of a damped harmonic oscillator 
\begin{equation}
    D^{\phi \phi}_R(t-t')=-\frac{\Theta(t-t')}{\omega_c^2} e^{-\kappa(t-t')}\sin{\omega_c(t-t')}.
\end{equation}\begin{figure}[!t]
    \centering
    \includegraphics[width=0.48\textwidth,trim={0 1.2cm 0 0},clip]{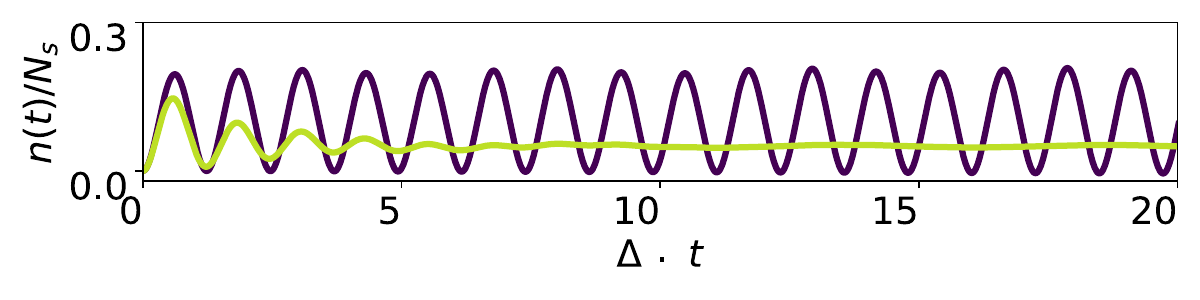}\\
    \includegraphics[width=0.48\textwidth]{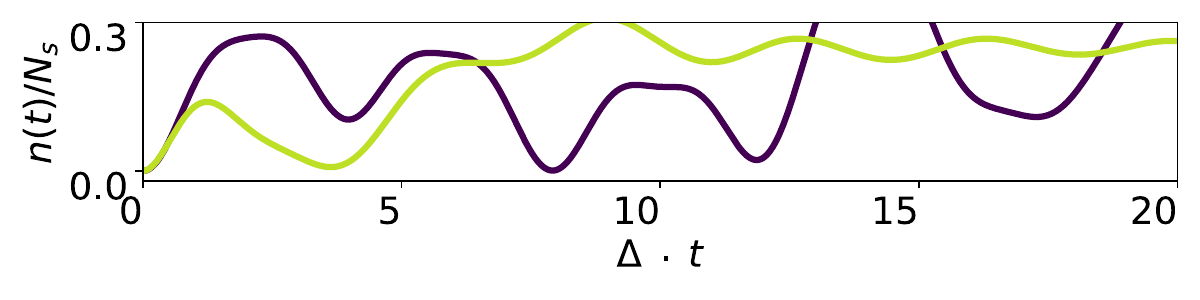}
    \caption{Evolution of photon population per each mode normalized by cluster size $N_s$, for $g/g_c=1.27$ and $(\theta_0,\varphi_0)=(0.7\pi,0)$. Top panel: the adiabatic limit $\omega_c/\Delta=5.0$ and for $\kappa=0$ (dark purple curve) and $\kappa/\Delta=0.5$ (light green curve). Bottom panel: the same as top panel only for slow photons $\omega_c=\Delta$.}
    \label{fig:lo_N}
\end{figure}
Eqs (\ref{LO_Sx_EOM})-(\ref{LO_Sz_EOM}) describe the motion of classical angular momentum variables with a conserved vector length $\abs{\vec{m}}=1$.  Adiabatic elimination of photons \cite{jager2022lindblad} amounts to approximate the integral in Eq. (\ref{LO_Sz_EOM}) as
\begin{multline}\label{adiab_elim}
     \approx m_y(t) m_z(t)  \int_0^t  D^{\phi \phi}_R(t-t')\,dt'\approx m_y(t) m_z(t)  \\ \times \int_0^\infty  D^{\phi \phi}_R(t')\,dt' \approx - \frac{1}{\omega_c(\omega_c^2+\kappa^2)}m_y(t) m_z(t) .
\end{multline}
Substituting (\ref{adiab_elim}) into Eqs. (\ref{LO_Sy_EOM}) and (\ref{LO_Sz_EOM}) results in the MF equations of motion for LMG model with a coupling modified by photon loss according to $ J \to J \omega_c^2/(\omega_c^2+\kappa^2)$. From a physical point of view, the above derivation shows that at LO, our approximation maps each cluster to a classical LMG system  \textit{without coupling to other clusters}. Therefore, the LO approximation describes the FM to PM transition of an infinite range Ising model with retarded interactions. The critical coupling of this system is determined according to the condition $\Delta = J$ \cite{Sciolla_2011} to be
\begin{equation}\label{gc}
    g_c = \frac12 \sqrt{\frac{\Delta(\omega_c^2+\kappa^2)}{\omega_c}\frac{N+M}{M}}.
\end{equation}
Although only valid in the adiabatic regime of the MF solution, we will use $g_c$ throughout this paper and scale the coupling $g$ according to it when  comparing our results for different values of $\Delta$, $\omega_c$ or $\kappa$. Clearly, dynamics at LO do not have any features unexplored in the past, and we only report the results of our simulations as a consistency check of our approach. In Fig. \ref{fig:lo_sx} we have shown the dynamics of spins initiated close to an equilibrium state of Eqs. (\ref{LO_Sx_EOM})-(\ref{LO_Sz_EOM}) inside the FM phase. We see that for fast photons (top panel of Fig. \ref{fig:lo_sx}), the system remains close to the minimum with oscillations which are smoothed by photon loss. For slow photons when adiabatic elimination does not work (bottom panel of Fig. \ref{fig:lo_sx}), the fluctuations induced by photons' dynamics relieve FM correlations and can create rare tunneling events. With photon loss, the destructive effect of slow photons on FM order is reduced due to the effective damping that slows down the spins towards the bottom of the nearby energy minimum.

\begin{figure}[!t]
    \centering
    \includegraphics[width=.48\textwidth]{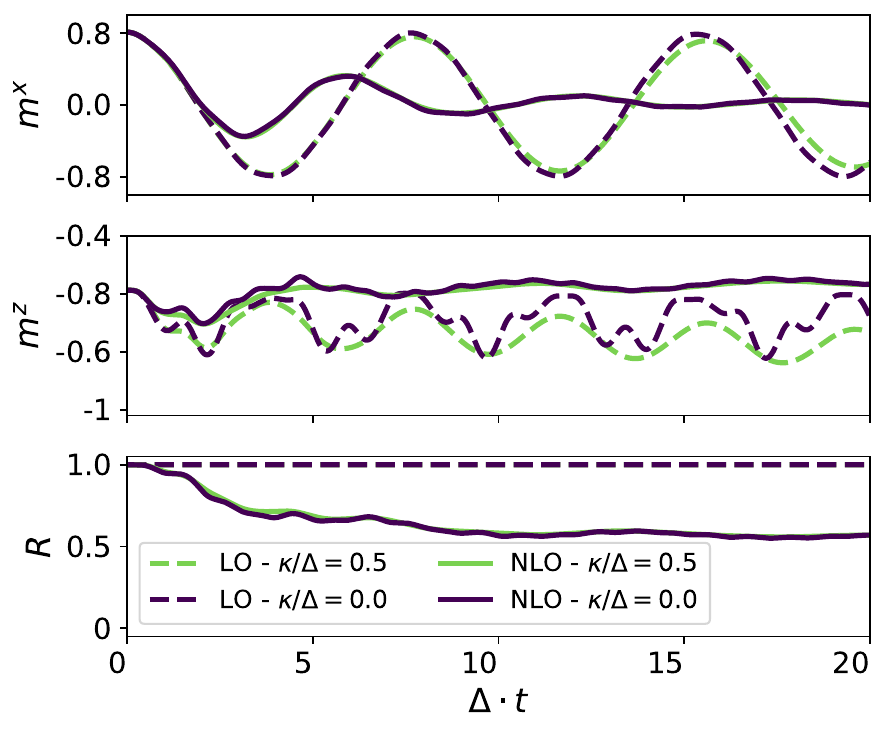}
    \caption{ Comparison of spin dynamics at LO and NLO  approximations for a large spin size ($N_s=10^5$), following a PM quench ($g/g_c=0.71$) and in the limit of fast photons $\omega_c/\Delta=5$ with ($\kappa/\Delta=0.5$) and without photon loss. Top, middle and bottom rows show the evolution of $m_x$, $m_z$ and the expectation value of spin size, respectively. Spins dephase to a single point due to random static couplings in the system. }
    \label{fig:largeS_PM_adia_RS}
\end{figure}

Having explored spin dynamics at LO, we now discuss the evolution of the photon sector. As we said before, the response of photons given by the retarded function $D^{\phi \phi}_R$ is unaltered by interactions at LO. However, the symmetric correlation function of photons and photon population are affected by the coupling to spins. Spins pump cavity modes due to the presence of the transverse field $\Delta$, and as shown in Fig. \ref{fig:lo_N}, create large oscillations (if $\kappa=0$) or a superradiant burst (if $\kappa\neq 0$) in the population of photons $n(t)$. We do not discuss dynamics of the PM phase in LO approximation here due to their trivial nature, and will show them in comparison with NLO results later.

The insufficiency of the LO approximation is clear by above observations. At this order of approximation, the size of each cluster $N_s$ only appears as an overall scaling factor in photon number $n(t) \propto N_s$, and spin dynamics do not depend on $N_s$. The latter is not physically valid, as we expect fluctuations to affect spin dynamics noticeably when the spin per each cluster $\hat S_i = \sum_\lambda \hat{\sigma}_{i\lambda}/2$ becomes smaller as $N_s$ is decreased. The other important shortcoming of LO approximation is visible in the dynamics of photon population in the adiabatic limit without loss, where the amplitude of the oscillations in $n(t)$ does not change and remains constant (Fig.~\ref{fig:lo_N}). In reality, photons experience dissipation even in a perfect cavity, as they can be reabsorbed by spins over longer timescales and the system is expected to equilibrate eventually. It is clear from the discussion above that LO approximation fails to capture this effect. As we will see next, NLO corrections take into account fluctuations and the reabsorption of photons, in addition to non-trivial regimes of dynamics such as glassy behavior.

\subsection{Results at NLO for Large Spins}\label{sec:NLO_results}

Below we will report on the dynamics at NLO for quenches to PM and FM phases of the system, in the limit of large spin per cluster ($N_s=10^5$). We note that we label these phases according to the behavior of the system at the MF level. The dynamics may change significantly when going beyond MF theory, making these labels inaccurate a posteriori. 

\subsubsection{Dynamics in the paramagnetic phase}\label{sec:dyn_PM}

\begin{figure}[!t]
    \centering
    \includegraphics[width=.48\textwidth]{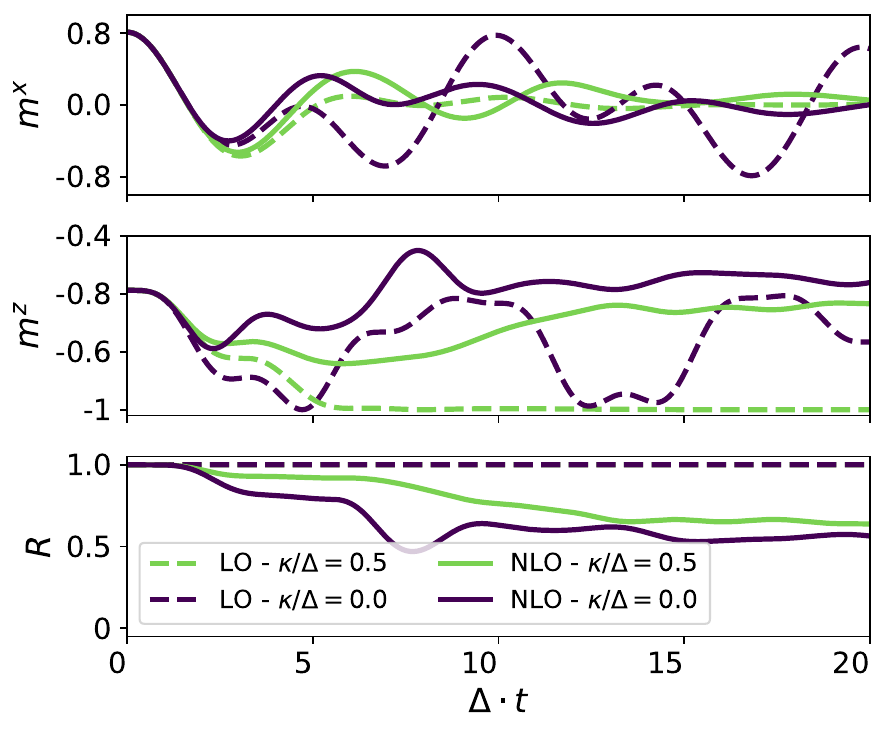}
    \caption{ Comparison of spin dynamics at LO and NLO approximations  for a large spin size ($N_s=10^5$), following a PM quench ($g/g_c=0.71$) and in the limit of resonant photons $\omega_c=\Delta$ with ($\kappa/\Delta=0.5$) and without photon loss. Top, middle and bottom rows show the evolution of $m_x$, $m_z$ and the expectation value of spin size, respectively. Spins dephase to a single point due to random static couplings in the system. }
    \label{fig:largeS_PM_dia_RS}
\end{figure}

We initialize the system in the ground-state of photons and the spin state with $(\theta_0,\varphi_0)=(0.7\pi,0)$ in Eq. (\ref{spin_init_state}). We choose a coupling below $g_c$ given in Eq. (\ref{gc}), and consider both cases of perfect and lossy cavities. 

The results for fast photons ($\omega_c/\Delta=5$) are shown in Fig. (\ref{fig:largeS_PM_adia_RS}), in comparison with LO results. We see that the evolution of $m_x$ is considerably altered by taking fluctuations into account. Spin dynamics is now damped even in the absence of photon loss. This is expected, as dissipation is a natural byproduct of interactions in a many-body system. We also see that photon loss has a minor impact on spin dynamics as it is weaker than the fluctuation-induced dissipation. As shown in the middle panel of Fig. \ref{fig:largeS_PM_adia_RS}, $\sigma^z$ shows a surprising behavior at NLO by not relaxing to its minimum value, in contrast to the LO result which at least, when $\kappa \neq 0$, approaches $-1$. However, there is a clear explanation for this phenomenon. The value of $g^2$ gives the variance of the disordered coupling in the system. This means that, while most of the individual couplings for each realization of the disorder are smaller than $g$, some of them are still large enough to weaken the PM configuration of the ground-state without causing a phase transition in the system. The fact that this behavior is purely due to random interactions is supported by our observation that the NLO approximation for non-disordered Dicke model results in the decay of spins into a state with $\sigma^z=-1$ in the PM phase. We also have shown the evolution of the spin vector size defined by
\begin{equation}
    R(t)\equiv \sqrt{(m_x)^2+(m_y)^2+(m_z)^2},
\end{equation}
in Fig. \ref{fig:largeS_PM_adia_RS}. While $R$ is a constant of motion of Eqs. (\ref{LO_Sx_EOM})-(\ref{LO_Sz_EOM}), it changes when fluctuations are considered. The final state of the system is a PM with a smaller spin size. Note that $R(t\to \infty)$ is independent of the initial state and is always smaller than one even at the lowest temperatures, only due to the frustrated nature of the system. 
In the following discussion,   $R(t)$ will be  a useful proxy for assessing the impact of correlations in dynamics. It is   the radius of the Bloch sphere, given at $t=0$ by the spin coherent state in Eq.~\eqref{spin_init_state}, and a constant of motion for
 all-to-all interacting spin systems with homogeneous couplings and collective dissipation~\cite{defenu2023long,PhysRevB.104.014307,kirton2019introduction}. It stays constant over time because in this class of systems, the MF approximation is exact in the thermodynamic limit and therefore no higher order cumulants are formed, which would make $R(t)$ shrink,  signaling the onset of a strongly correlated regime. 

\begin{figure}[!t]
    \centering
    \includegraphics[width=.48\textwidth]{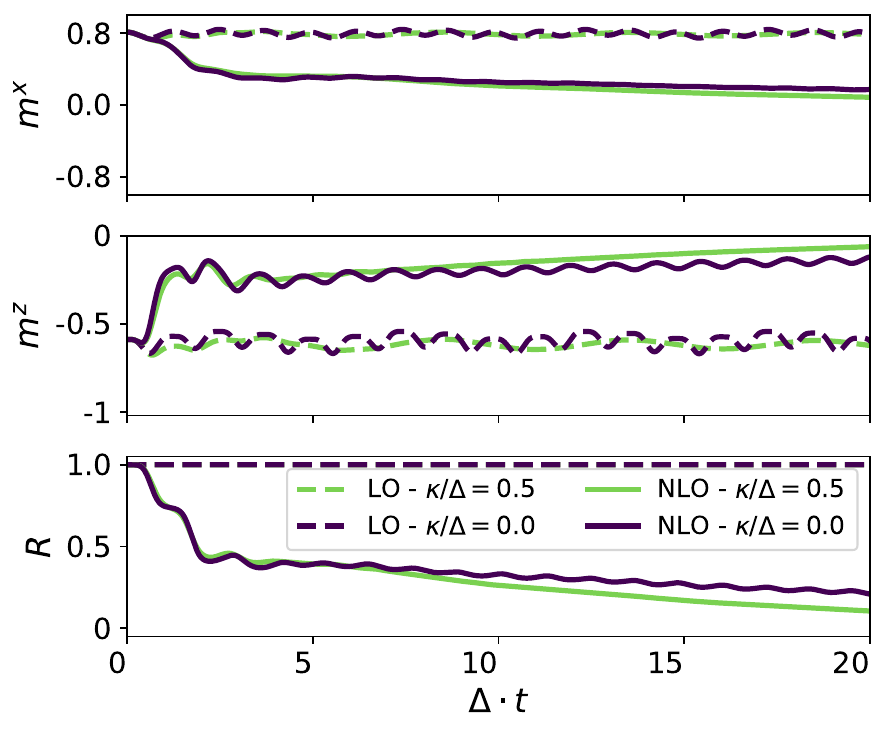}
    \caption{  Comparison of spin dynamics at LO  and NLO  approximations  for a large spin size ($N_s=10^5$), following a FM quench ($g/g_c=1.27$) and in the limit of fast photons $\omega_c/\Delta=5$ with ($\kappa/\Delta=0.5$) and without  photon loss. Top panel: following a quick collapse, the decay of FM order parameter strongly slows down and a prethermal plateau is formed. The middle and bottom panels show the evolution of $m_z$ and the radius of the Bloch sphere, respectively.}
    \label{fig:nlo_FM_largeS_adia}
\end{figure}

Spin dynamics at NLO approximation for the case of resonant photons $\omega_c=\Delta$ are illustrated in Fig. \ref{fig:largeS_PM_dia_RS}. As expected, dynamics become more irregular both at LO and NLO and photon loss has a more dramatic effect on the dynamics.

\subsubsection{Dynamics in the ferromagnetic phase}\label{sec:dyn_FM}

\begin{figure}[!t]
    \centering
    \hspace{-20pt}\includegraphics[width=.33\textwidth]{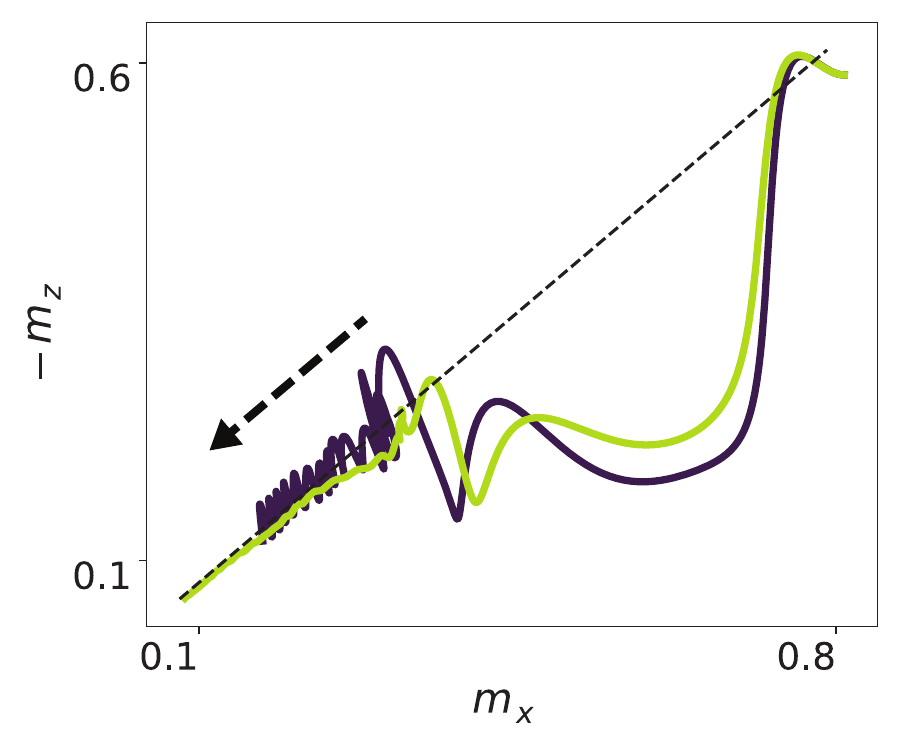}
    \caption{Evolution of spin vectors projected to the $xz$ plane for large cluster size $N_s=10^5$ and  for $\kappa/\Delta=0$ (dark purple curve) and $\kappa/\Delta=0.5$ (light green curve). Spins feature a spiral relaxation around the axis of FM ordering (dashed line) and towards the origin (parallel to the arrow).}
    \label{fig:nlo_FM_xz}
\end{figure}

We quench the coupling to $g=1.27 g_c$ where a FM state is realized at LO as shown previously,. The initial spin vector is again taken to be $(\theta_0,\varphi_0)=(0.7\pi,0)$, such that spins are close to the ground-state of the classical model in Eqs. (\ref{LO_Sx_EOM})-(\ref{LO_Sz_EOM}) at the chosen coupling strength. Similar to the PM case, we take a large spin size ($N_s=10^5$) where the effect of NLO corrections is supposed to be small. It will become clear that this is an incorrect assumption and NLO contributions are significant.

\begin{figure}[!t]
    \centering
    \subfloat[\label{fig:nlo_FM_N_adia}]{\includegraphics[width=.48\textwidth]{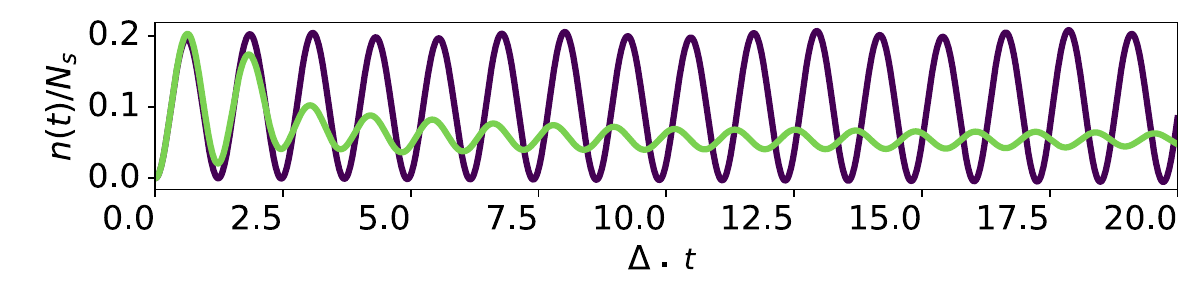}}\\
    \subfloat[\label{fig:nlo_n_vs_S}]{\includegraphics[width=.48\textwidth]{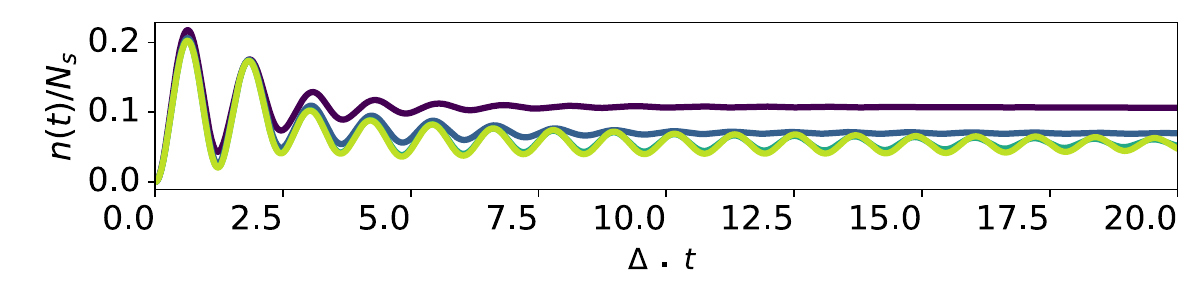}}
    \caption{(a) Photon population for a quench into the ordered phase with $g/g_c=1.27$, $\omega_c/\Delta=5$ and $\kappa=0$ at LO (dark purple curve) and NLO (light green curve). (b) Photon number per mode for $g/g_c=1.27$ and for $N=1000,\,200,\,20,\,5$ (light to dark curves). Relaxation of late-time oscillations is faster for smaller $N_s$.}
\end{figure}

The results of the numerics for the adiabatic limit $\omega_c/\Delta = 5$ are depicted in Fig. \ref{fig:nlo_FM_largeS_adia} at LO and NLO approximations. At LO, $m_x$ and $m_z$ show small oscillations around the equilibrium of the MF dynamics and the spin vector is confined to the surface of the Bloch sphere ($R=1$). Fluctuations captured at NLO drastically alter the dynamics. All components of spin decay and the spin vector shrinks toward the center of the Bloch sphere. The spin decay features an interesting profile. For an initial period, the relaxation is quick and spins experience a collapse to a smaller but finite value. Following this, the relaxation becomes very slow and the spin vector spirals around the axis of FM order of the LO solution (Fig. \ref{fig:nlo_FM_xz}). This situation is similar to the phenomenon of prethermalization in the quench dynamics of many-body quantum systems \cite{Berges_PRL04,marino2022dynamical,Babadi_PRX15,marino2012relaxation,marino2014nonequilibrium,lorenzo2018remnants,rodriguez2022far,marcuzzi2016prethermalization,marcuzzi2013prethermalization,bertini2015prethermalization,schutz2014prethermalization}, although here the system can be open. The prethermal behavior is also seen in the time evolution of photon number $n(t)$ shown in Fig. \ref{fig:nlo_FM_N_adia}. During the prethermal plateau of spins, $n(t)$ has a nearly stationary value with a slow growth (visible in the overall slope of the light curve in Fig.~\ref{fig:nlo_n_vs_S}) towards the true equilibrium state. Photon losses only qualitatively affect spin dynamics, by weakly accelerating the process of relaxation.

As we emphasized before, labeling the system as a FM for $g>g_c$ is only valid at the MF level. It is clear from the previous section that the system is not truly a FM as the magnetization decays with time. In the upcoming sections, we will provide evidence that the system is, in fact, a spin glass (SG) in this regime. However, it is not possible to uniquely determine a SG by only looking at single spin observables. The only indirect evidence for SG that we have so far is the slow relaxation of magnetization. We again remark that our results are expected to be correct for $\eta\gtrsim 1$ where the model in Eq.~(\ref{H}) is expected to host a SG~\cite{Strack_PRL11,Gopal_PRL11}.

\begin{figure*}[!t]
    \centering
    \hspace{-10pt}\subfloat[\label{fig:sx_2pi}]{\includegraphics[width=.334\textwidth]{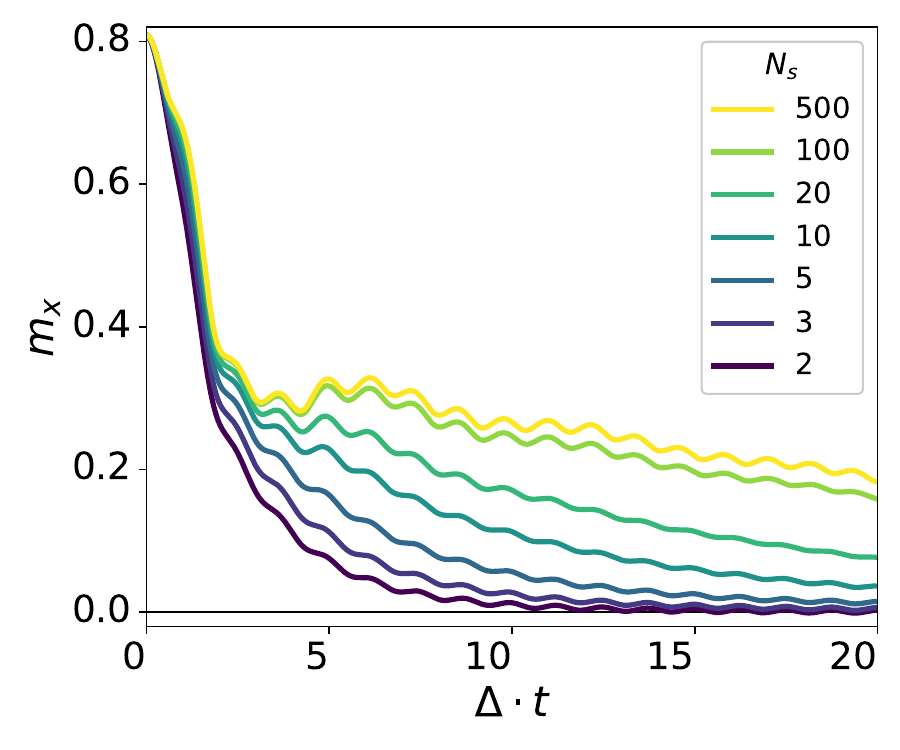}}
    \subfloat[\label{fig:sx_dtwa}]{\includegraphics[width=.33\textwidth]{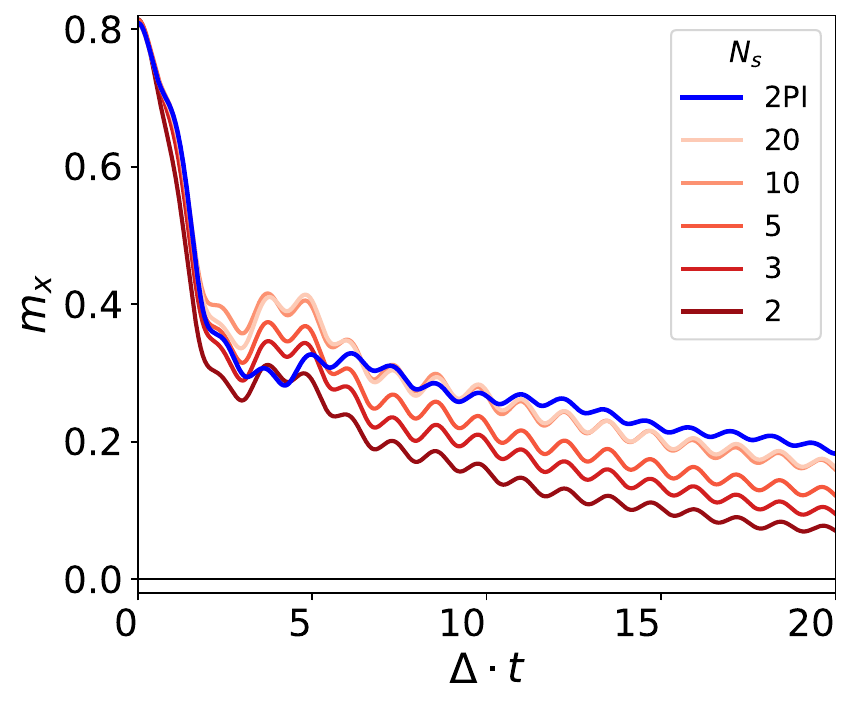}}
    \subfloat[\label{fig:nlo_xz_vs_S}]{\includegraphics[width=.33\textwidth]{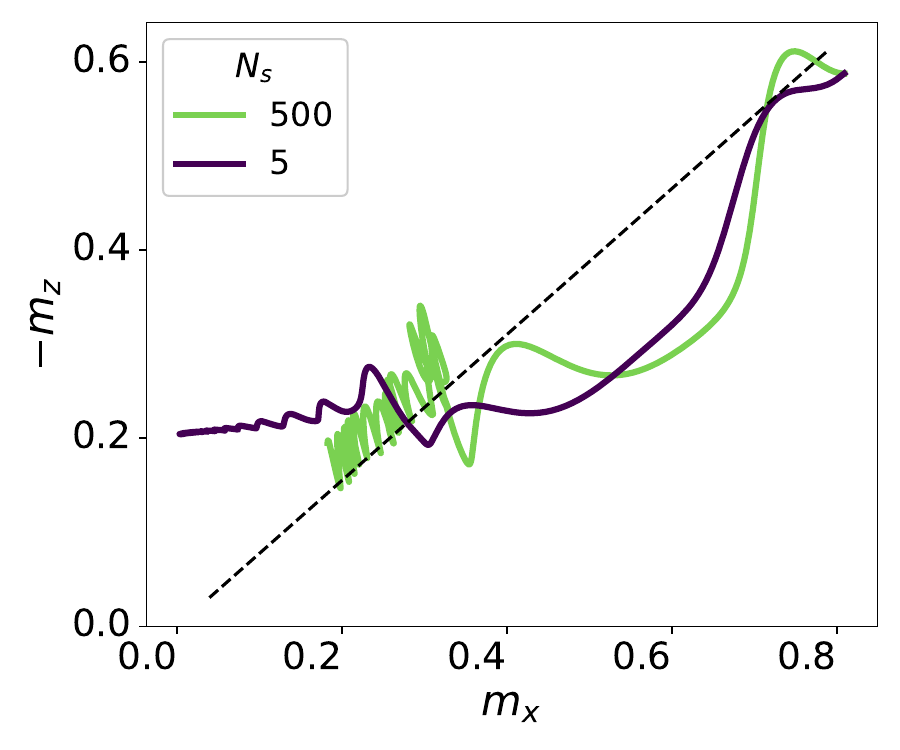}}
    \caption{\blue{Spin dynamics for different numbers of spin sizes $S=N_s/2$, for $g/g_c=1.13$ and $\omega_c/\Delta=5$ without losses. (a) 2PI results show faster relaxation for smaller spins sizes due to enhanced fluctuations. The prethermal magnetization plateau disappears when $N_s$ is decreased due to stronger fluctuations and for $N_s =2,\, 3$ the system is PM, as demonstrated in Fig.~\ref{fig:Q_vs_S} (b) Dynamics obtained from DTWA. Relaxation is less sensitive to spin size. The blue curve is 2PI result for $N_S=500$, showing good agreement with DTWA for large $N_s$. (c) Comparison of spin vectors projected to the $xz$ plane for small and large cluster sizes, in adiabatic $\omega_c/\Delta=5$ and for $\kappa/\Delta=0$. The FM spiral of large spin ensembles crosses over to a quicker, smooth relaxation for smaller values of $N_s$.}}
\end{figure*}

\subsubsection{Dependence of the spin dynamics on cluster size and the semi-classical limit}\label{sec:spin_size}
The number of spins per cluster $N_s$ is not only a control parameter for the expansion of the effective action, but it also determines the strength of fluctuations in the system. This assumption stems from the conventional understanding that larger spins often exhibit a more classical behavior compared to small spins \cite{marino2022dynamical,hosseinshort}. As was said before, the size of the spin for each cluster is determined by the number of spins inside each cluster $N_s$ via $S=N_s/2$. Our solution of the problem also supports that fluctuations are stronger for smaller values of spins per cluster. This can be seen, for example, in Eq. (\ref{Sigma_NLO}) where the fermion self-energy scales as $N_s^{-1}$.

\blue{The dynamics of spins at NLO are shown in fig. \ref{fig:sx_2pi} for different cluster sizes. The main effect of lowering $N_s$ is a faster relaxation of spins. As $N_s$ is decreased, quantum fluctuations become stronger and the transient magnetization lasts shorter. In the other limit $S \gg 1$, dynamics become classical~\cite{hosseinshort}. This can be verified by comparing 2PI results with semi-classical approximations such as DTWA. As shown in Fig.~\ref{fig:sx_dtwa}, in the limit of large spins, DTWA approaches a limiting value, and agrees well with 2PI even quantitatively. Since DTWA becomes exact for large spins~\cite{huber2021phase}, this agreement shows the validity of our diagrammatic expansion in the large spin limit~\cite{Berges2004introduction}. DTWA results display limited sensitivity to spin size, and only predict a qualitative modification of the magnetization plateau. This is due to the fact that as a semi-classical method, DTWA misses quantum tunneling effects which are important for small values of $N_s$. Both 2PI and DTWA predict a prethermal magnetization plateau at large $N_s$. Further comparison of 2PI and DTWA is provided in Appendix.~\ref{app:2pi_dtwa}. For sufficiently small $N_s$, the prethermal state is completely bypassed and the system experiences a quick relaxation towards  equilibrium (Fig. \ref{fig:nlo_xz_vs_S}). Similarly, photon number relaxes faster for smaller $N_s$, while for larger cluster sizes it has weakly damped oscillations (Fig. \ref{fig:nlo_n_vs_S}).}

\subsection{Dynamics of spin glass}\label{sec:dynamics_of_SG}
The results of NLO approximation given in previous sections   indicate that the strong coupling limit of the model in Eq. (\ref{H}) is not a FM and corrections due to frustration and fluctuations drastically change the MF phase diagram. From the $N_s=1$ limit of the problem \cite{Gopal_PRL11,Strack_PRL11,Buchhold_PRA13}, we know that the system hosts a SG at sufficiently low temperatures and strong couplings. The main questions are the following:

(i) What are the physical signatures of SG in and out of equilibrium?

(ii) Can our formalism capture the far from equilibrium dynamics of SG?

In the upcoming sections, we will discuss two direct measures of SG phase. First, we consider the disorder average of the square of local magnetization of each spin ensemble which gives a proxy of the presence of frozen spin configurations. The second one is related to the ability of the system to retain its memory of the far past and is connected to the phenomenon of aging, encoded in 2-point correlation functions of the system for large time separations. We will demonstrate that 2PI can access both of these measures and is able to track their evolution in real time.

\subsubsection{Characteristics of spin glass}

According to Landau's theory of phase transitions (PT), different phases of matter can be classified according to their symmetries \cite{LandauLifshitz_stat1,sachdev_book,kardar_2007}. A phase transition corresponds to a change in the symmetry group of a physical system. For example, the PM to FM transition of an Ising FM corresponds to the spontaneous breaking of the $Z_2$ symmetry of the Hamiltonian. The occurrence of the PT is signaled by the continuous growth of a parameter which vanishes on one side of the transition. Such parameter is termed as the order parameter and the way it grows with the control parameter of the transition provides us with important information about the nature of the PT \cite{kardar_2007,sachdev_book,altland2010condensed}. While the PM to SG phase transition fits into the symmetry breaking paradigm, the proper definition of an order parameter for SG has been a subject of debate for decades, with various candidates \cite{SG_RMP,mezard1987spin,Edwards_75}. In the following, we will consider two of these proposed order parameters to study the formation of SG in our system.

\begin{figure}[!t]
    \centering
    \subfloat[\label{fig:q_vs_k_5}]{\includegraphics[width=.48\textwidth,trim={0.3cm 0 0 0},clip]{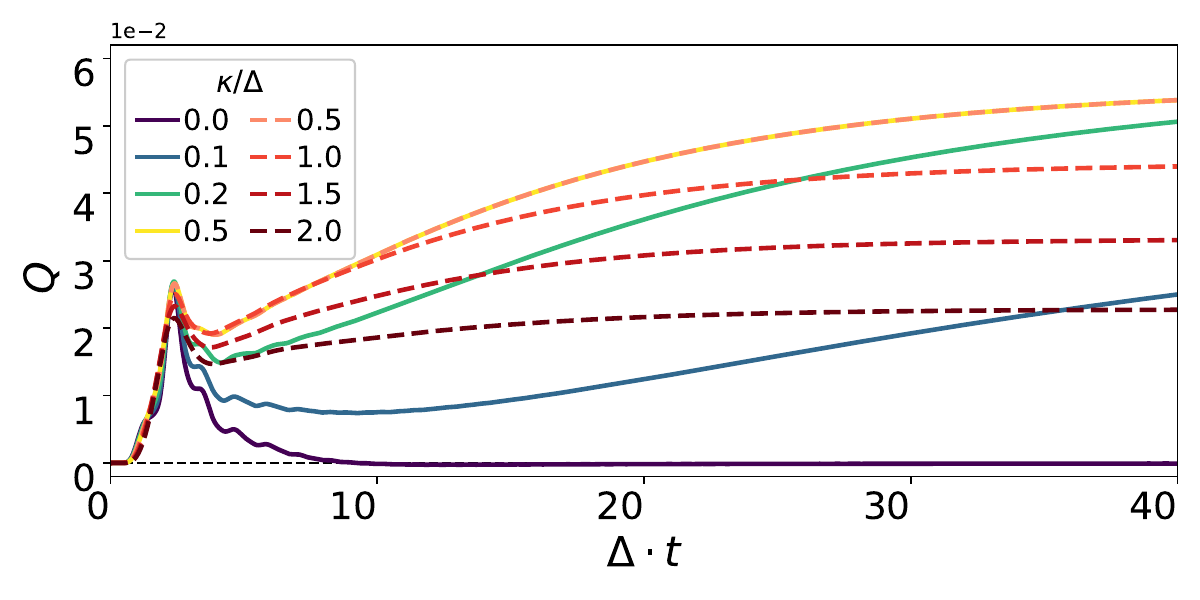}}\\
    \subfloat[\label{fig:q_vs_k_7}]{\includegraphics[width=.48\textwidth,trim={0.3cm 0 0 0},clip]{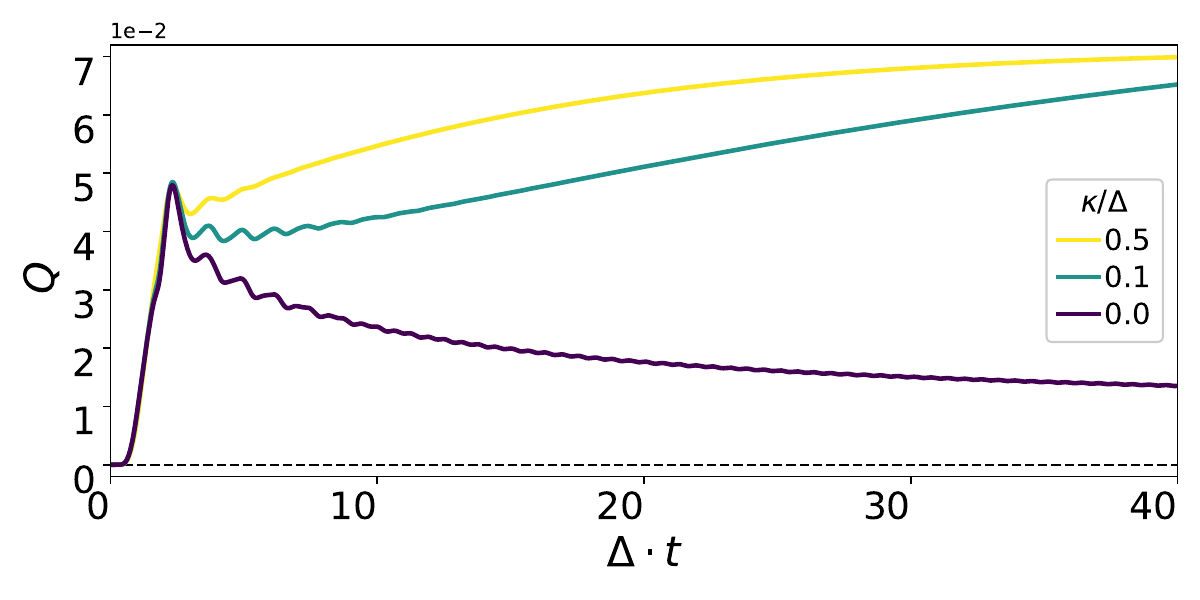}}
    \caption{\blue{Dynamics of the overlap parameter $Q$ for different values of system-bath coupling $\kappa$. (a) For $\theta_0=0.5\pi$, the energy density after the quench is large and the final temperature is too high to realize a SG, unless the system is allowed to cool down via photon loss. However, for $\kappa\gtrsim \Delta$ (dashed lines), photon loss weakens SG order. (b) For $\theta_0=0.7\pi$, energy density is smaller and the system enters a weak SG state without coupling to the bath, but it saturates to a smaller value of $Q$, and hence, glass order is weaker in this case. The other parameters are $g/g_c=1.13$, $\omega_c/\Delta=5.0$, $N_s=5$ and $\eta=1$.}}
\end{figure}

\blue{\subsubsection{Statistical correlations between similar samples}\label{sec:q_def}
One way to detect SG order is to compare several systems, also known as replicas~\cite{SG_RMP,mezard1987spin} , which share the same pattern of the couplings $g_{\alpha,i}$ in Eq.~\eqref{H}. In the SG phase, we expect to find a finite and stationary correlation between the configuration of spins in different replicas. In the simplest case, consider two replicas, which we label as A and B, whose state is given by the density matrix
\begin{equation}
    \rho(t)=\rho_A\big(\qty{g_{\alpha,i}},t\big) \otimes \rho_B\big(\qty{g_{\alpha,i}},t\big).
\end{equation}
$\rho(t)$ remains a separable state since replicas do not physically interact with each other. We initialize both systems in the same state and let them evolve with time. Since both replicas share the same disorder profile and initial states, they will have the same state given by the same density matrix $\rho_A(t)=\rho_B(t)$. However, to compare the profile of magnetization in the two systems and to obtain correlations between them, we have to measure spins in the local basis of $S^x_i$ operators in both replicas. The outcomes of the measurements are not necessarily the same between the two systems. Nevertheless, we can expect a finite overlap between the measured spin configurations of the replicas in the SG phase for each disorder pattern and also after averaging the outcome over the disorder. The simplest overlap is given by
\begin{equation}
    Q(t)=\frac{1}{N_s^2}\overline{\expval{S^x_{i,A}(t)S^x_{i,B}(t)}_c},
\end{equation}
where we have included a normalization factor. As long as the systems are not in FM phase, a finite value for $Q$ at long times implies SG order. An extra textbook condition for the viability of $Q$ is that the $\mathbb{Z}_2$ symmetry should be broken explicitly either by a small term in the Hamiltonian or by the initial state~\cite{chaikin1995principles,kardar_2007,altland2010condensed}. We will take the latter route below, by starting from spin states with finite $\expval{S^x}$. The decay of magnetization shown in Sec.~\ref{sec:dyn_FM} guarantees that the system is not FM at long times and hence, a finite $Q$ means the system is SG. We explained how to extract $Q$ using the 2PI formalism in Sec. \ref{sec:q_eval}. In the following we will show the dynamics of $Q$ as the light-matter coupling $g$ and the photon loss $\kappa$ are varied.\\}

\begin{figure}[!t]
    \centering
    \includegraphics[width=.48\textwidth,trim={0.2cm 0 0 0},clip]{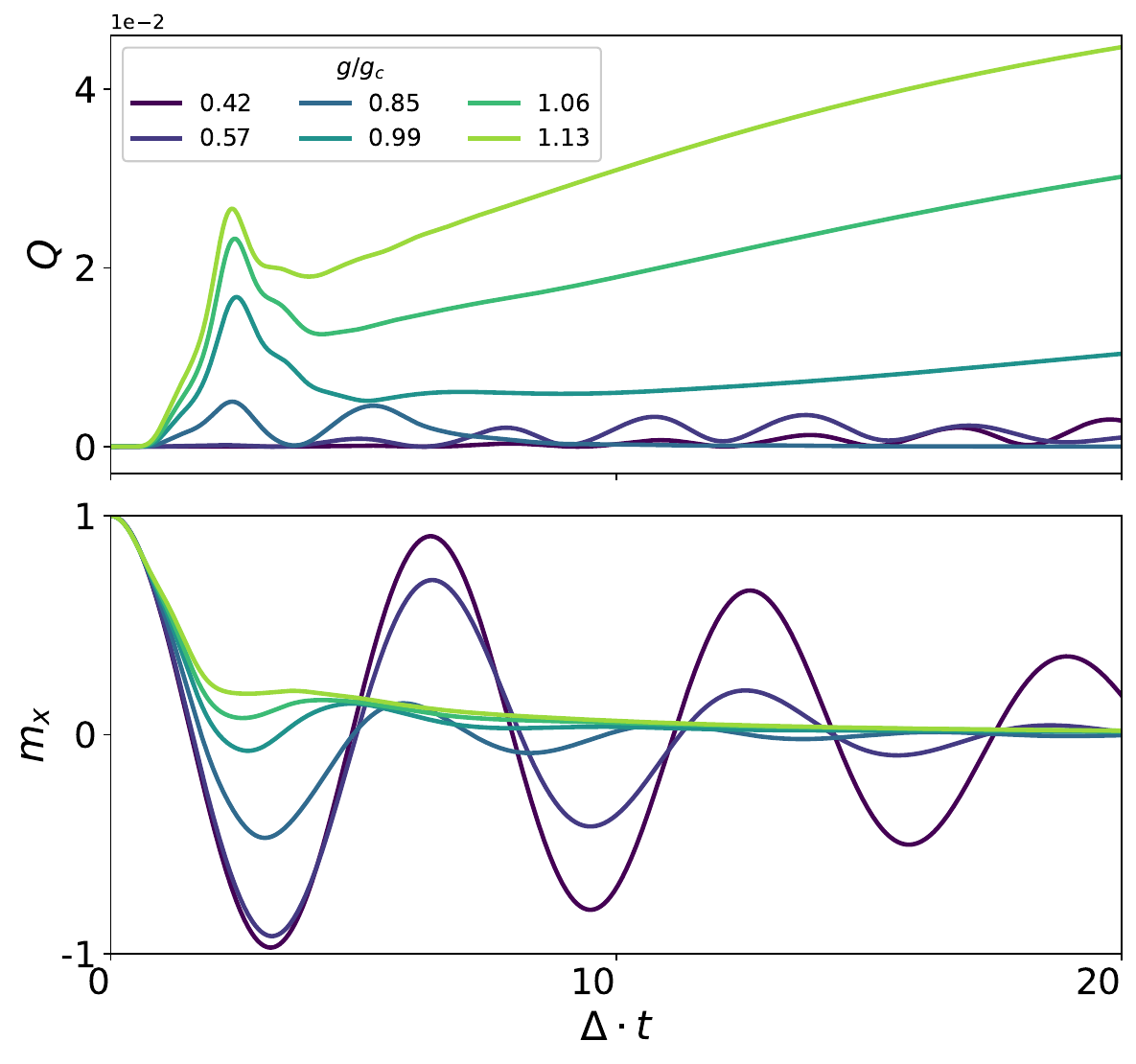}
    \caption{ Dynamics of SG order parameter (top panel) and total magnetization (bottom panel) for different values of coupling strength. $g_c$ is the critical coupling in the MF theory. This plot shows that the observed value of $g_c$ is smaller than the one obtained at LO approximation. The other parameters are $\kappa/\Delta=0.5$, $\omega_c/\Delta=5.0$, $N_s=5$ and $\eta=1$.}
    \label{fig:q_vs_g}
\end{figure}

\begin{figure}[!t]
    \centering
    \subfloat[\label{fig:C_PM_SG_k}]{\includegraphics[width=.48\textwidth]{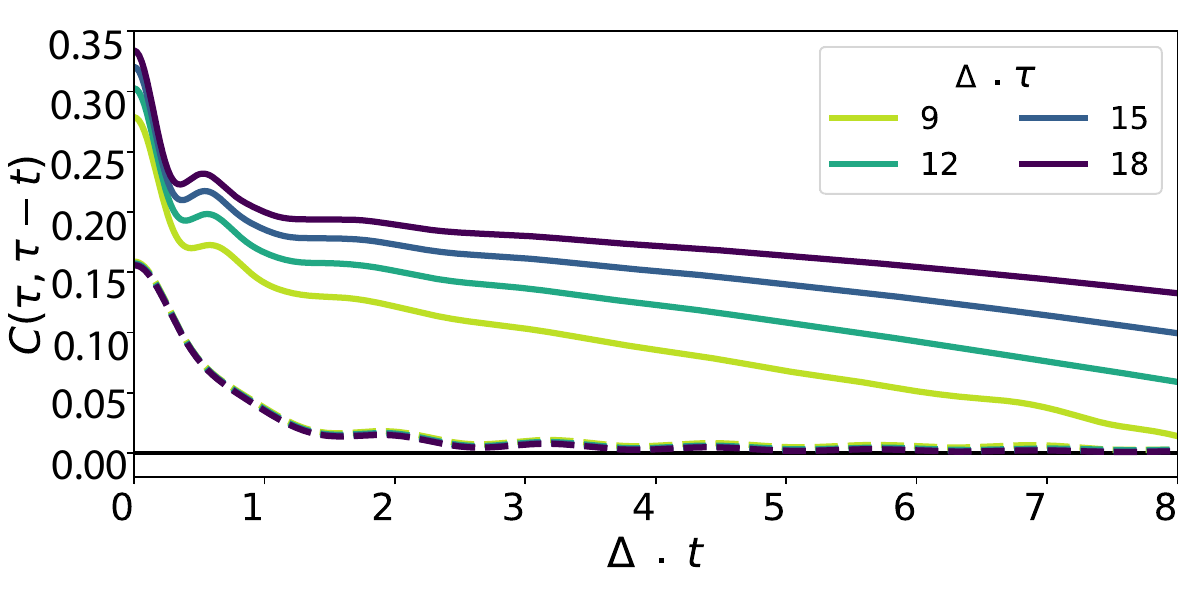}}\\
    \subfloat[\label{fig:C_SG_SG_k}]{\includegraphics[width=.48\textwidth]{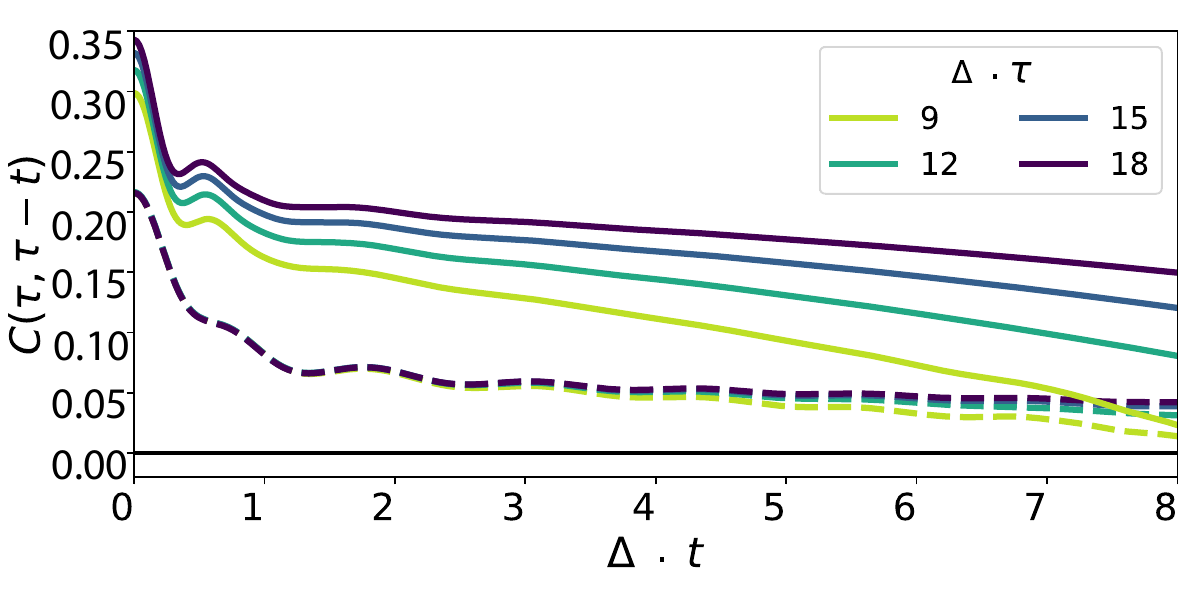}}
    \caption{(a) Symmetric correlation function $C$ at different waiting times for quenches from an initial state with high energy density $\theta_0=\pi/2$. For $\kappa=0$ (dashed lines), the system heats up to PM state and $C$ quickly decays to zero for $t\to \infty$ (lines are almost on top of each other). For $\kappa/\Delta=0.5$ (solid lines) the system cools down to SG where $C$ remains finite for $t\to \infty$. (b) $C$ for an initial state with low energy density $\theta_0=0.7\pi$. Regardless of $\kappa$ the system ends up in the glass phase, however, SG order is stronger for $\kappa/\Delta=0.5$ (solid lines) where the system is cooled down, compared to $\kappa=0$ (dashed lines). For both figures $\omega_c/\Delta=5.0$, $g/g_c=1.27$, $\eta=1$ and $N_s=5$. }
\end{figure}

We initialize the system in a state with $m^x_i=1$ and quench $g$ from zero to a sufficiently large value. We monitor the dynamics of $Q$ for different values of photon loss from $\kappa=0$ (closed system) to $\kappa=0.5\Delta$. As can be seen in Fig. \ref{fig:q_vs_k_5}, after some fluctuations caused by the quench,  $Q$ approaches zero at long times when the system is not coupled to the bath. An explanation for this behavior is that the initial energy  $E_i$ of the system with respect to the post-quench Hamiltonian is large enough to put the system in a high temperature equilibrium PM state if the system is closed. For finite photon loss, Q starts to grow after $t\gtrsim \kappa^{-1}$, the timescale of cooling by photon loss. For $\kappa\lesssim \Delta$, the sole effect of photon loss is cooling, which enhances the growth of $Q$. For $\kappa\gtrsim \Delta$, dissipation becomes detrimental for SG, as shown by dashed lines in Fig.~\ref{fig:q_vs_k_5}. This behavior can be explained by resorting to the stochastic interpretation of Lindblad dynamics~\cite{breuer2002theory,Daley_trajectroies2014}. For weak losses $\kappa\lesssim \Delta$, cavity loss induces dephasing of spins in the x-direction given by the jump operator $S^x$. Dephasing can be modeled exactly by considering a non-Hermitian Hamiltonian in conjunction with repeated weak projections into the eigenstates of $S^x$ to conserve the norm of the state~\cite{breuer2002theory,Daley_trajectroies2014}. These repeated projections compete with the transverse field $\Delta$ and stabilize the ordering of spins in the x-direction. For $\kappa \gtrsim \Delta$, cavity loss induces atomic decay and stimulation with different amplitudes~\cite{jager2022lindblad,Marsh2023} which in turn, suppress spin ordering in the x-direction.

The explanation given above for the role of photon loss in the dynamics is supported by the behavior of quenches with smaller $E_i$. We note that $E_i$ can be calculated easily for the initial states considered in this work:
\begin{equation}\label{E_i}
    E_i = \frac12 NN_s \Delta \cos \theta_0.
\end{equation}
since we have $g=0$ in the pre-quench hamiltonian. 
By initializing spins closer to $\theta_0=\pi$, we can reduce the energy density and attain a lower temperature final state for isolated quenches. For instance, for $\theta_0=0.7\pi$ the system enters the SG phase even without losses (Fig. \ref{fig:q_vs_k_7}), although $Q$ is smaller for $\kappa=0$ compared to $\kappa >0$ due to the absence of cooling.

At last, we consider the emergence of SG order as the coupling $g$ is increased. Taking $N_s=5$ and starting from $\theta_0=0.5\pi$ while connecting the system to an external bath by switching on the photon loss, we look at $Q$ for quenches to different values of $g$ as shown in the top panel of Fig. \ref{fig:q_vs_g}. For weak couplings, $Q(t)$ oscillates and decays to zero. When $g$ becomes large enough, $Q$ grows to a finite value at long times. We see in the bottom panel of Fig. \ref{fig:q_vs_g} that spin dynamics changes from underdamped to overdamped at the same coupling where $Q$ becomes finite. In Fig. \ref{fig:q_vs_g} the coupling $g$ is scaled with $g_c$ which was the critical coupling of the MF limit in Eq. (\ref{gc}). A finite value of $Q$ for $g/g_c <1$ indicates that NLO contributions shift the boundary of PM and SG phases towards SG.

\begin{figure*}[!t]
    \centering
    \subfloat[\label{fig:fdr_PM}]{\includegraphics[width=.35\textwidth]{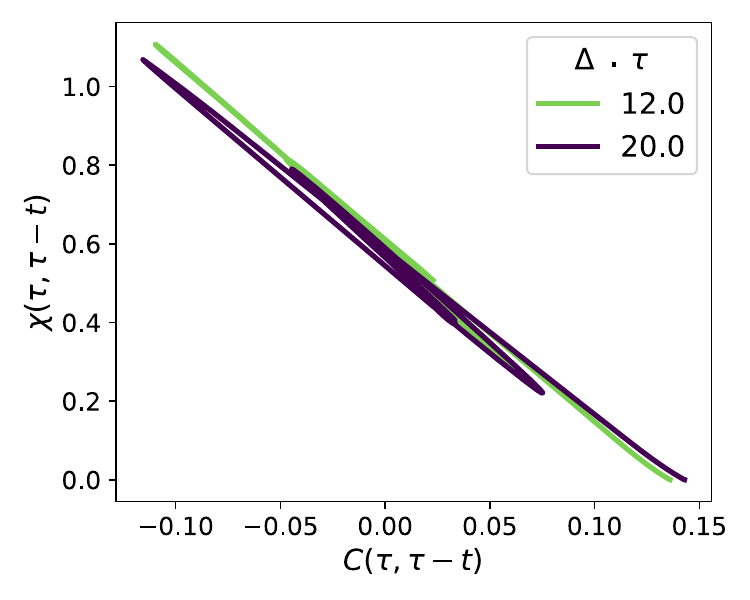}}
    \subfloat[\label{fig:fdr_SG}]{\includegraphics[width=.35\textwidth]{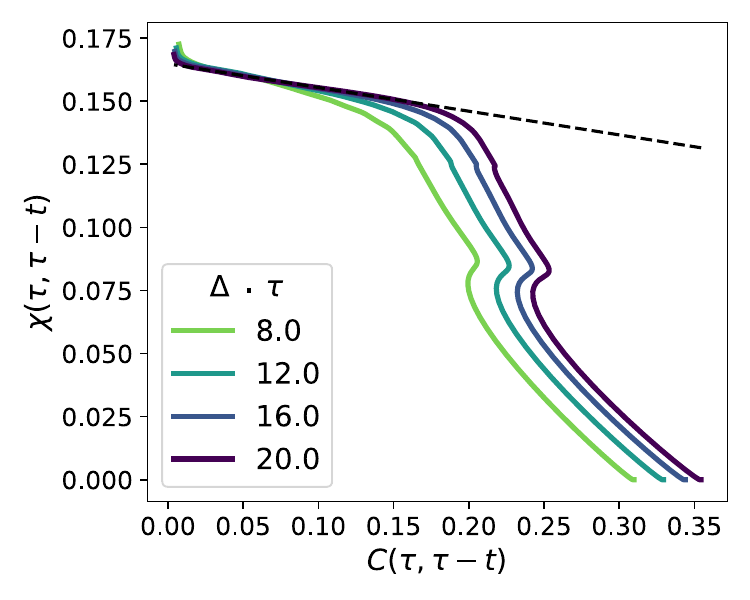}}
    \caption{Integrated response $\chi$ versus $C$ for different waiting times $\tau$. (a) In PM phase, the slope is constant and is proportional to the inverse of temperature. (b) In SG state, dynamics have multiple time scales and the slope changes with $t$. The parameters are $\Delta/\omega_c=0.2$, $\kappa/\Delta=0.5$, $g/g_c=1.27$ and $N_s=5$.}
\end{figure*}

\subsubsection{Temporal correlations and aging phenomena} \label{sec:edwards-anderson}

As discussed before, the diagnosis of SG phase requires the evidence of a frozen spin configuration which breaks the symmetry of the Hamiltonian and at the same time, has no global magnetization. In the previous section we discussed (the square of) the instantaneous magnetization for each site and showed that it is finite in the glass phase while magnetization relaxes to zero. The latter feature distinguished SG from FM. Another way to confirm a frozen spin state in the system is to look at the temporal correlation between the magnetization of each site at large time separations. For example, we take a cluster and its total spin operator $S^x$ at two different times $t$ and $t'$. If spins are frozen, they do not fluctuate strongly in time and the overlap of magnetization for the same site at two different times is large. In fact, it remains finite even if $\abs{t-t'}\to \infty$. For classical degrees of freedom one has
\begin{equation}\label{q_EA_cl}
    q_\mathrm{EA} \equiv \lim_{\abs{t-t'}\to\infty} \overline{\expval{S^x(t)S^x(t')}}>0,
\end{equation}
which is the Edwards-Anderson (EA) order parameter \cite{Edwards_75,SG_RMP}. It is clear that the above condition can also imply FM order in the system. In order to exclude FM, the disorder averaged magnetization $\overline{\expval{S^x}}$ should also vanish. For the case of quantum operators in our problem, the correlator in Eq. (\ref{q_EA_cl}) can be generalized to
\begin{equation}\label{C}
     C(\tau,\tau-t)\equiv \frac{1}{N_s^2}\overline{\expval{\acomm{S_i^x(\tau)}{S_i^x(\tau-t)}}}_c.
\end{equation}
 In the following, we consider the behavior of $C$ for the same quenches discussed in the previous section. It will be shown that $C$ changes its behavior at the same points where $Q$ in Eq. (\ref{q_def}) becomes finite. Therefore, both $C$ and $Q$ are consistent measures of SG order in the system.

We again look at the effect of photon loss on quenches from hot (with $\theta_0=0.5\pi$) and cold (with $\theta_0=0.7\pi$) initial states of the system, according to Eq. (\ref{E_i}). We consider $C(\tau,\tau-t)$ for different values of the ``waiting time" $\tau$ after the quench as a function of $t$. In Fig. \ref{fig:C_PM_SG_k}, we have shown $C$ for $\theta_0=\pi/2$ and for zero and finite $\kappa$. We see that $C$ behaves differently depending on whether the system is isolated or not. When the system is closed (dashed lines in Fig. \ref{fig:C_PM_SG_k}), it ends up in a high temperature PM state with correlations that relax quickly to equilibrium and show weak dependence on the waiting time after the quench $\tau$. The correlations also decay quickly with $t$, meaning that local magnetization loses the memory of its past quickly in the PM phase. When the system is allowed to cool down by emitting photons to the outside of the cavity, a SG phase emerges (solid lines in Fig. \ref{fig:C_PM_SG_k}). $C$ becomes strongly dependent on the waiting time $\tau$ and there is a crossover in the behavior of $C$ for $t\ll \tau$ and $t \approx \tau$. The latter behavior is termed aging \cite{bouchaud_aging92,Calabrese_2005} and is another feature of quenches into SG phases. In aging, the decay of $C(\tau,\tau-t)$ with $t$ becomes slower for ``older" systems with larger $\tau$. In this picture, there is a plateau in $C$ for $t\ll \tau$ (Fig. \ref{fig:C_PM_SG_k}) whose height gives the EA order parameter. The plateau is more visible in figures shown in upcoming sections.

Aging is closely related to the lack of thermalization in spin glasses \cite{bouchaud_aging92,Cugliandolo_SKDyn94,Marinari_FDT_RSB98,cugliandolo1999real}. The system is unable to reach equilibrium because of its highly rugged energy landscape in which adjacent energy configurations, connected by local spin flips, are separated by large energy barriers. As a result, ergodicity is broken and the system will not thermalize \cite{Cugliandolo_SKDyn94}. The relation between aging and thermalization can be made more explicit by using the fluctuation dissipation theorem (FDT) and fluctuation dissipation ratio (FDR) defined below. At thermal equilibrium, the symmetric correlation function $C(t,t')$ and the response function $R(t,t')$ of an observable $\mathcal{O}$ are connected to each other via FDT \cite{kamenev,Rammer_2007,altland2010condensed}:
\begin{equation}
    R(t-t')= -i \int \frac{d\omega}{2\pi}e^{-i\omega(t-t')}\tanh(\frac{\omega}{2T})\tilde{C}(\omega),
\end{equation}
where $\tilde{C}(\omega)=\int dt \exp(i\omega t) C(t)$ is the Fourier transform of $C$ and $T$ is the temperature. At long times or in the classical limit $\omega \ll T$, we can approximate $\tanh(\omega/2T)\approx \omega/2T$ to get for $t>t'$
\begin{equation}
    R(t,t')\approx -\frac{1}{2T}\partial_{t'}C(t,t').
\end{equation}
By integration we get
\begin{equation}\label{FDT_t}
    C(t,t')-C(t,t)=2T \chi(t,t'),
\end{equation}
where $\chi$ is the integrated response defined as
\begin{equation}\label{chi_def}
     \chi(t,t')\equiv \int_{t'}^t R(t,t'')\,dt''.
\end{equation}
Eq. (\ref{FDT_t}) can be generalized to out of equilibrium regimes as~\cite{Cugliandolo_SK93,Cugliandolo_SKDyn94,Cugliandolo_FDR97,foini2011fluctuation}
\begin{equation}
    C(\tau,\tau-t) - C(\tau,\tau) = 2 T_\mathrm{eff}(\tau,t) \chi(\tau,\tau-t),
\end{equation}
where $T_\mathrm{eff}$ is the time-dependent effective temperature. $T_\mathrm{eff}$ can be read from the slope of the plot of $\chi$ versus $C$. If a system is coupled to an external bath with temperature $T$ the fluctuation dissipation ratio $X(\tau,t)$ is defined as
\begin{equation}
    X(\tau,t)\equiv \frac{T}{T_\mathrm{eff}(\tau,t)}.
\end{equation}
Therefore, $X$ measures the deviation of the system from true equilibrium at temperature $T$ \cite{Cugliandolo_SKDyn94,cugliandolo1999real,Calabrese_2005}. For generic systems that thermalize efficiently, $T_\mathrm{eff}\to T$ and $X \to 1$. For glassy systems, on the other hand, $X$ can show a multistage behavior, where the plot of $\chi$ versus $C$ changes its slope. Furthermore, FDR may never reach the limit $X\to 1$, and the system will not thermalize at all.

In Fig. \ref{fig:fdr_PM} we have shown $\chi$ in terms of $C$ for the PM phase of our model. We see that the plots have an almost constant slope (small deviations are mostly due to the error of numerical integration in Eq. (\ref{chi_def})) and therefore, the PM phase has a unique temperature. For quenches inside the SG phase (Fig. \ref{fig:fdr_SG}), $\chi$ acquires a distinct profile. The long-time effective temperature (related to the slope of the dashed line in Fig. \ref{fig:fdr_SG}) and the short-time effective temperature are different \cite{Marinari_FDT_RSB98,cugliandolo1999real,Georges_HeisenbergGlass2000}. This is a direct consequence of aging in the system. \blue{$q_\mathrm{EA}$ can be read from the value of $C$ at which the change of slope happens~\cite{biroli2002out,Marinari_FDT_RSB98}. For larger $\tau$, the vertical section of the curve lasts longer which also corresponds to a wider plateau of $C(\tau,\tau-t)$.} The effective temperature of SG in Fig. \ref{fig:fdr_SG}) is larger than that of the PM in Fig. \ref{fig:fdr_PM} by about a factor of $5$, and hence $X\approx 0.2$. We remark that there is a close connection between the violation of FDT and replica symmetry breaking (RSB) in spin glasses \cite{Marinari_FDT_RSB98}, which suggest that the model under consideration hosts RSB, in agreement with the recent cavity QED experiment of Ref. \cite{kroeze2023replica}.

While photon loss was necessary to realize SG for the high energy initial state with $\theta_0=\pi/2$, if we reduce the initial energy density by starting from a lower energy state with $\theta_0=0.7\pi$, the system becomes SG regardless of coupling it to the bath (Fig. \ref{fig:C_SG_SG_k}). However, the height of the plateau for $\kappa=0$ is smaller, as the quench inside the SG phase is shallower in this case. This is in complete agreement with the behavior of $Q$ given in Figs. \ref{fig:q_vs_k_5} and \ref{fig:q_vs_k_7}.

\blue{\subsubsection{Effect of spin size on spin glass formation}\label{sec:spin_size_SG}
In Section.~\ref{sec:spin_size}, it was shown that smaller spins are more susceptible to quantum fluctuations by considering the evolution of total magnetization of the system. Here, we briefly address the imprint of quantum fluctuations on SG dynamics as viewed through the lens of the overlap parameter $Q$. Our analysis will be more qualitative here, compared to the comprehensive approach of our other work~\cite{hosseinshort} where the quantum to classical crossover of SG was studied by looking at aging dynamics.}

\begin{figure}[!t]
    \centering
    \includegraphics[width=.48\textwidth,trim={0.3cm 0 0 0},clip]{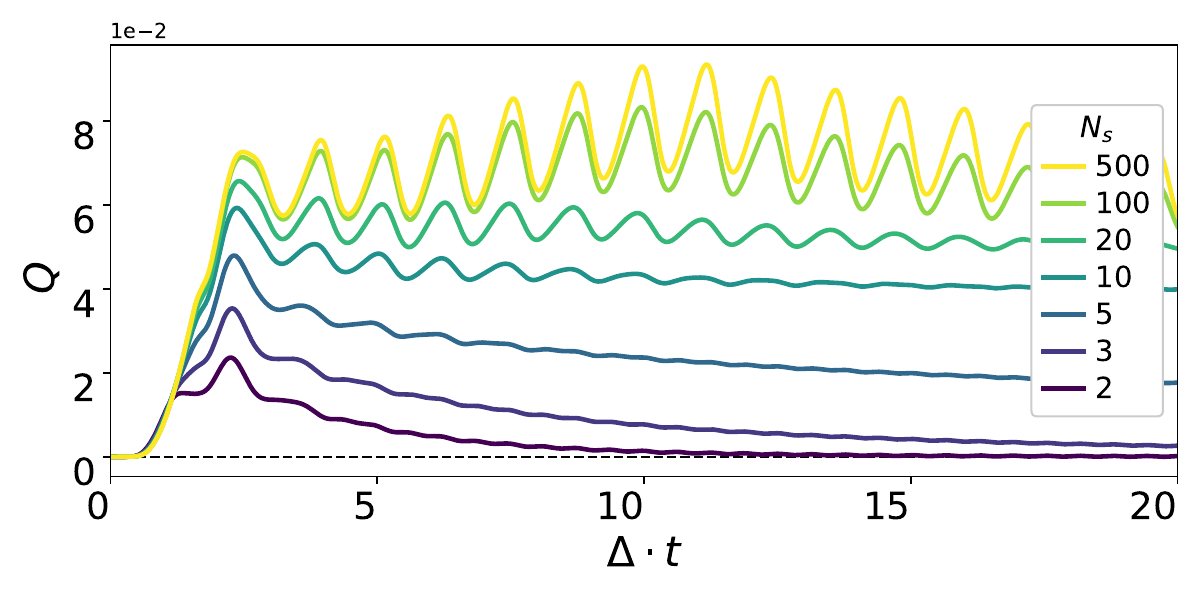}
    \caption{\blue{Spin overlap $Q$ after interaction quenches to $g=1.13g_c$ starting from $\theta_0=0.7\pi$ without photon loss and for different spin sizes. Overlap decreases for smaller $N_s$ and due to heating, vanishes for $N_s=2,\,3$ for which the critical temperature is smaller. Other parameters are $\Delta/\omega_c=0.2$.  }}
    \label{fig:Q_vs_S}
\end{figure}

\blue{We take an initial state with $\theta_0=0.7\pi$ and $\kappa=0$. For systems with different spin sizes, we monitor the overlap $Q$ which is shown in Fig.~\ref{fig:Q_vs_S}. For $N_s\le 3$, the overlap vanishes at $t\to \infty$, indicating a PM state phase due to excessive heat generated by the quench, while $Q$ remains finite for larger $N_s$. This is consistent with mean-field theory calculations of Ref.~\cite{Bray_1980} which predict a lower PM to SG transition temperature for smaller spins. Fig.~\ref{fig:Q_vs_S} also shows that $Q$ is more sensitive to spin size for small $S$. In the opposite limit of large $N_s$, corresponding to a classical SG, $Q$ is bounded from above by its upper limit at $N_S\to \infty$. The mean-field like oscillations of $Q$ at $N_S \gg 1$ further support the classical behavior of this limit. Hence, the profile of $Q$ is qualitatively consistent with the findings of Ref.~\cite{hosseinshort} on the crossover between quantum and classical SG.}

\subsubsection{Spin glass away from the adiabatic limit} \label{sec:resonance}

At last, we consider the effect of changing photon frequency $\omega_c$ on the glass phase. 2PI formalism allows us to treat the problem in all ranges of frequencies and not just in the regime of fast photons where photons can be adiabatically eliminated~\cite{damanet2019atom,jager2022lindblad,PhysRevResearch.3.L032016}. We follow the same tradition as previous sections and scale $g$ with $g_c$ in Eq. (\ref{gc}). However, as will be explained later, this scaling does not alter the general picture given below.

\begin{figure}[!t]
    \centering
    \includegraphics[width=.38\textwidth]{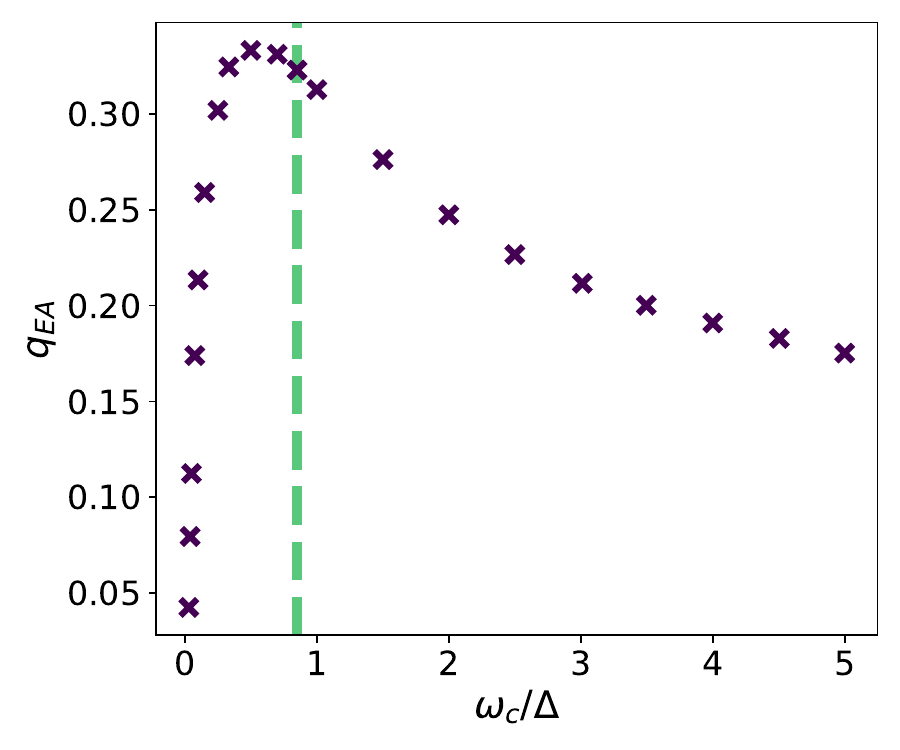}
    \caption{Dependence of SG order parameter at a fixed waiting time after the quench $\Delta \cdot \tau=12.0$ on photon frequency $\omega_c$. SG is strongest close to a resonance frequency near $\Delta$. Dashed line indicates the analytical estimate of the resonance frequency $\omega^\star$. The other parameters are $g/g_c=1.27$, $\kappa/\Delta=0.5$, $N_s=20$ and $\eta=1$.}
    \label{fig:resonance}
\end{figure}

As explained before, initializing the system in a symmetric spin state helps it to reach the steady state earlier. This reduces the required numerical resources to access the SG phase in steady state {(for a discussion of the numerical costs see Appendix \ref{app:DE_sol})}. We look at $C(\tau,\tau-t)$ for sufficiently large values of $\tau$ and a fixed $t$. This gives us an estimate of the EA order parameter $q_\mathrm{EA}$ as a measure of glassiness in the system. $q_\mathrm{EA}$ is shown in Fig. \ref{fig:resonance} for simulations with different photon frequencies. $q_\mathrm{EA}$ displays a peak below $\omega_c = \Delta$, vanishes quickly as $\omega_c \to 0$ and saturates in the adiabatic limit $\omega_c \gg \Delta$. The quick decay of SG order for small photon frequencies is an expected feature. For instance, phonons tend to relieve magnetic or density orderings in solid state platforms, as they have smaller energy gaps and are thermally excited easily. The observed peak in $q_\mathrm{EA}$ can be explained as a resonance effect by looking at the coherent part of the frequency-dependent interaction between spins following integrating out photons, which has the form 
\begin{equation}\label{V_eff}
H_\mathrm{int} = \sum_{i,j}^N \sum_{\lambda,\lambda'}^{N_s} \int \frac{d\omega}{2\pi}\sigma_{i\lambda}^x(-\omega)V^{ij}_\mathrm{eff}(\omega)\sigma_{j\lambda'}^x(\omega),   
\end{equation}
with
\begin{equation}
    V^{ij}_\mathrm{eff}(\omega) \sim \sum_{\alpha}^M g_{\alpha,i}g_{\alpha,j}\omega_c \Re D^R(\omega).
\end{equation}
$D^R$ is the bare retarded Green's function of photons given by
\begin{equation}
    D^R(\omega)=\frac{1}{(\omega+i\kappa)^2-\omega_c^2}.
\end{equation}
{Eq.~(\ref{V_eff}) is obtained by integrating out photons in the original action, resulting in the following spin-spin interaction in the Keldysh action
\begin{multline}\label{S_sigsig}
S_{\sigma \sigma}=-\sum_\alpha^M \sum_{i,j}^N\sum_{\lambda,\lambda'}^{N_s} g_{\alpha,i}g_{\alpha,j} \omega_c \int \frac{d\omega}{2\pi} \\ \begin{pmatrix}
\sigma^x_c(-\omega) & \sigma^x_q(-\omega)
\end{pmatrix}_{i\lambda}   \begin{pmatrix}
0 & D^A(\omega) \\ D^R(\omega) & D^K(\omega)
\end{pmatrix} \begin{pmatrix}
\sigma^x_c(\omega) \\ \sigma^x_q(\omega)
\end{pmatrix}_{j\lambda'}.
\end{multline}
We decompose the kernel matrix into its real and imaginary parts. The former is Hermitian and can be attributed to an effective Hamiltonian given by Eq.~(\ref{V_eff}), while the imaginary part describes dissipation~\cite{Sieberer_2016}. The $(c,q)$ indices in Eq.~(\ref{S_sigsig}) are defined in Appendix~\ref{app:DE} in terms of contour indices.} To evaluate the strength of the interaction we can put $g_{\alpha,i}g_{\beta,j}\sim g^2$ to get
\begin{equation}\label{V_eff_explicit}
    V_\mathrm{eff}(\omega) \sim g^2 \omega_c \frac{\omega^2-\omega_c^2-\kappa^2}{(\omega^2-\omega_c^2-\kappa^2)^2+4\kappa^2 \omega^2}.
\end{equation}

In the adiabatic limit the relevant energy scales in the problem are small compared to photonic energy scales ($\omega \ll \omega_c,\kappa$) and we get
\begin{equation}\label{V_eff_adia}
     V_\mathrm{eff}(\omega\to 0)\sim -g^2\omega_c/(\omega_c^2+\kappa^2) \sim - (\frac{g}{g_c})^2 \Delta.
\end{equation}
If we scale $g$ with $g_c$ such that $g^2 = r g_c^2$ where $r$ is a dimensionless number, the effective interaction becomes insensitive to $\omega_c$. Therefore, changing $\omega_c$ should not affect SG order in the system. This is consistent with Fig. (\ref{fig:resonance}) where $q_\mathrm{EA}$ approaches a constant value at large frequencies. For smaller values of $\omega_c$ and $\kappa$, however, the dependence of effective interaction in Eq. (\ref{V_eff_explicit}) on $\omega$ becomes important. One possible approximation is to use the ``on-shell approximation"  and substitute $\omega\approx\Delta$. Using this approximation and expanding the effective spin-spin interaction to non-zero leading order in $\Delta$, recovers the atom-only description of Ref. \cite{jager2022lindblad} which also captures the dissipative part of the Lindblad dynamics. Nevertheless, this approach fails to describe the dependence of spin order on $\omega_c$, and in particular, the resonance behavior in Fig. \ref{fig:resonance}. However, if we use the on-shell approximation and simply substitute $|\omega|=\Delta$ in $V_\mathrm{eff}(\omega)$ and do not expand it in $\Delta$, we find that $V_\mathrm{eff}$ has two peaks close to $\omega^\star$ given by
\begin{equation}
    \omega^\star\equiv \sqrt{\Delta^2 - \kappa^2},
\end{equation}
with attractive ($V_\mathrm{eff}>0$) and repulsive ($V_\mathrm{eff}<0)$ sides for $\omega_c > \omega^\star$ and $\omega_c < \omega^\star$, respectively. $\omega^\star$ is marked in Fig. \ref{fig:resonance} and is in good agreement with the numerics.

\begin{figure}[!t]
    \centering
    \includegraphics[width=.38\textwidth]{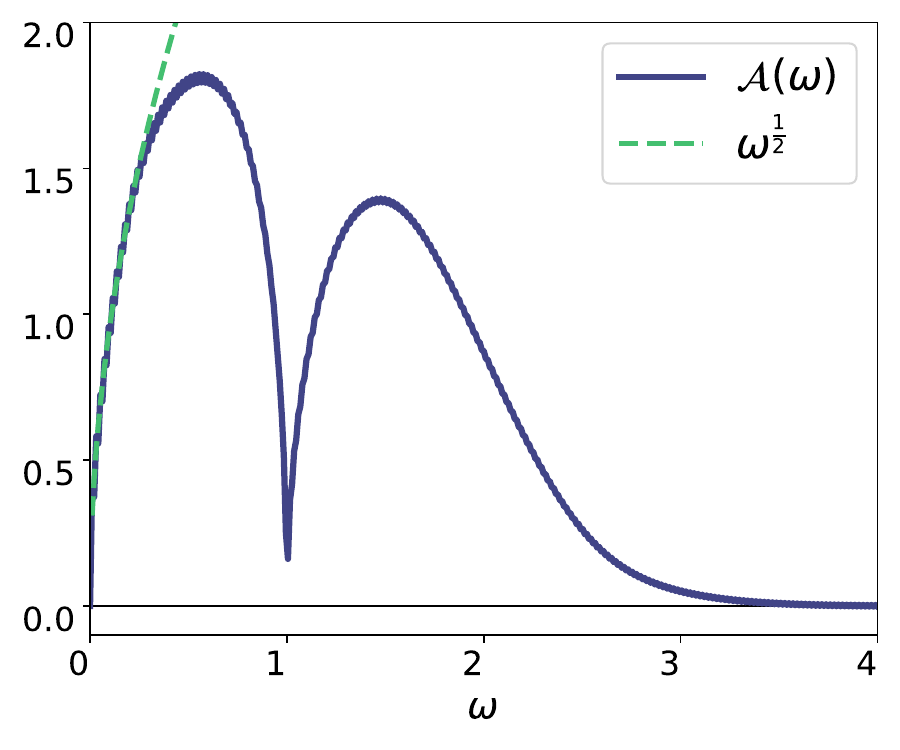}
    \caption{Atom spectral density in the glass phase for a long waiting time after the quench $\Delta \cdot \tau= 60.0$ as function of frequency. Atoms form a continuum of sub-Ohmic modes at small energies. The minimum is due to vacuum Rabi splitting. The other parameters are $g/g_c=1.27$, $\kappa=0$, $\theta_0=\pi$, $N_s=5$ and $\eta=1$.}
    \label{fig:A_omega}
\end{figure}

Nevertheless, there is a feature that cannot be straightforwardly explained by the on-shell approximation used above. $V_\mathrm{eff}(\Delta)$ vanishes at $\omega_c=\omega^\star$, but SG order does not show any signs of the suppression of interactions. A possible explanation for this discrepancy is that due to strong interactions, the spin spectral density defined in terms of the imaginary part of the response function (Eq. (\ref{ss_to_V}))
\begin{equation}
    \mathcal{A}(\omega)=-\frac{(N+M)N_s}{4}\Im V^{22}(\omega),
\end{equation}
is modified compared to the non-interacting limit and the atomic peak at $\omega = \Delta$ is highly broadened. It is believed that quantum SGs are critical with a gapless spectrum of excitations \cite{Georges_HeisenbergGlass2000,Strack_PRL11,Andreanov_QuantumSK2012,Buchhold_PRA13}. For SK  and disordered $SU(N)$ Heisenberg models, Refs. \cite{Andreanov_QuantumSK2012} and \cite{Georges_HeisenbergGlass2000} obtained an Ohmic spectrum with $\sim \omega$. For closed Dicke SG at zero temperature, Ref. \cite{Strack_PRL11} predicted a similar Ohmic profile, while for Dicke SG connected to a Markovian bath Ref. \cite{Buchhold_PRA13} obtained a sub-Ohmic $\sim \omega^\frac12$ dependence for  $\omega \lesssim \kappa$. We have plotted the atomic spectral density for a low temperature quench \textit{without photon losses} within our approach in Fig. \ref{fig:A_omega}. We see that $\mathcal{A}(\omega)$ displays a gapless spectrum at low frequencies, but with a sub-Ohmic dispersion $\sim \omega^\frac12$, even though the system is not coupled to a Markovian bath. One explanation, although unlikely due to the similar symmetry of this model to the SK model, is that the system is in a different universality class with different exponents. Another possibility is that the final state long after the quench is at finite temperature with many emitted photons existing in the cavity, which act as an effective bath for spins. The exponent can be read experimentally using RF spectroscopy to compare with theoretical predictions.

\section{Conclusions}\label{sec:conclusion}

In this work we have provided a diagrammatic derivation of non-perturbative DE  suited to describe spin glass formation far from equilibrium in the cavity QED  platform of   Refs.~\cite{kroeze2022high,vaidya2018tunable,kollar2015adjustable,PhysRevLett.122.193601}.
Very recently, the same group has reported for the first time the observation of SG order~\cite{kroeze2023replica} and associated replica symmetry breaking, which is a notoriously hard experimental task.
We predict that SG order in this system can be obstructed by transient, competing ferromagnetism when the atomic ensembles trapped in the cavity are largely occupied. We show that SG order is enhanced by strong quantum fluctuations (few atoms per ensemble)  and when photons are active and resonant with the atomic transitions. 
The set of DE derived here have the flexibility to explore simultaneously a broad set of parameters' regimes (weak/strong coupling, variable ranges of $N_s$, adiabatic elimination and active photons) without resorting to descriptions valid only in a corner of parameter space.

Our work sets also the stage for studying the crossover from strongly correlated regimes to semi-classical dynamics in other cavity QED platforms  with tunable loading capabilities~\cite{periwal2021programmable}. The dynamics of these systems are at reach of state-of-art 2PI-DE since interactions, although inhomogeneous, are all-to-all and this allows for more controlled diagrammatic expansions and numerical integration.
At the same time, many of the new-generation cavity QED experiments   combine short and all-to-all interactions, introducing a notion of dimensionality and lattice spacing that would make the numerical solution of the 2PI-DE more challenging.\\

We are currently making progress in this direction by exploring various experimental platforms:\\

\textit{Rydberg arrays integrated in optical cavities -- }
The experiment in Ref.~\cite{kong2021melting} marked the first combination of strong, short-range, Rydberg interactions with conventional photon-mediated long-range coupling. Analyzing these systems beyond mean-field theory opens up possibilities for exploring exotic phases of matter. These include spin liquids that are protected from dissipation~\cite{kong2021melting} or limit cycles~\cite{gelhausen2016quantum} that persist in the presence of strong classical and quantum noise. This line of research should also provide novel opportunities to explore topological order and lattice gauge theories in the context of atom-light interfaces~\cite{kong2021melting}.\\

\textit{Cavity QED with programable spin exchange interactions -- }
 Ref.~\cite{periwal2021programmable} reports the realization of programmable non-local interactions in an array of atomic ensembles within an optical cavity. This experiment introduces a $U(1)$ variant~\cite{bentsen2019integrable} to the work discussed here, as it focuses on spin-exchange interactions rather than ferromagnetic ($Z_2$-invariant) ones. The ability to program the distance-dependent interactions with a sophisticated combination of Raman sidebands and magnetic field gradients enables the engineering of dimensionality, topology, and metric as needed. This opens the door to studying quantum optimization problems on tree-like~\cite{bentsen2019treelike} or fully connected geometries in cavity QED, bridging the gap between quantum information and many-body quantum optics.\\

 \textit{Natural and synthetic correlated emission --} In recent years, there has been a surge of interest in the study of dissipative spin problems that describe correlated emission in atomic ensembles. The key idea is that emission into free space involves multiple scattering and interference effects, making the phenomenon more complex than in traditional textbook quantum optics~\cite{asenjo2017exponential,masson2020many,PhysRevA.99.023802,albrecht2019subradiant,asenjo2019optical}. This mechanism is typically encoded in   a Lindblad equation that is non-local in space, or equivalently non-diagonal in jump operators. While solutions for a few atomic excitations are accessible~\cite{albrecht2019subradiant}, the many-body regime of correlated emission remains largely unexplored. This would be of significance   for Rydberg   experiments realizing atomic mirrors via this mechanism~\cite{srakaew2023subwavelength,rui2020subradiant} and for applications in the cavity QED experiment of Ref.~\cite{periwal2021programmable,davis2019photon}, where correlated emission could be artificially engineered~\cite{seetharam2022dynamical,marino2022universality,seetharam2022correlation}. This would enable the synthesis of many-body entangled states using dissipation, by leveraging the full potent of programmable cavity QED in the domain of non-unitary dynamics.\\
 
 On the interdisciplinary front, 2PI-DE for dissipative spin dynamics could also serve to guide dissipative quantum state preparation   in hybrid AMO-spintronics platforms which rely on correlated emission to entangle NV centers~\cite{li2023solid}. \\

 This list of subjects is by no means exhaustive. For instance,   2PI-DE may be one of the few methods with sufficient versatility  to treat the intrinsically strongly correlated dynamics of  {fermionic cavity QED experiments}~\cite{sauerwein2023engineering,uhrich2023cavity,roux2020strongly}. We hope that this set of potential applications will motivate   readers from different communities to embark in the fascinating challenge to study the next generation of many-body quantum optics experiments using dynamical field theory methods. 

 \section{Acknowledgements}
 We are thankful to A. N. Mikheev for carefully proof reading the manuscript, and to D. Gribben for early contributions to this project. 
  HH and JM acknowledge financial  support by the Deutsche Forschungsgemeinschaft (DFG, German Research Foundation): through  Project-ID 429529648, TRR 306 QuCoLiMa (“Quantum Cooperativity of Light and Matter”) and through TRR 288 - 422213477 (project B09).   This project has been supported by  the QuantERA II Programme that has received funding from the European Union’s Horizon 2020 research and innovation programme  under Grant Agreement No 101017733 ('QuSiED') and by the DFG (project number 499037529). DC acknowledges support from the European Union, under European Research Council grant agreement No 101002107 (NEWSPIN); the Government of Spain under the Severo Ochoa Grant CEX2019-000910-S [MCIN/AEI/10.13039/501100011033]); QuantERA II project QuSiED, co-funded by the European Union Horizon 2020 research and innovation programme (No 101017733) and the Government of Spain (European Union NextGenerationEU/PRTR PCI2022-132945 funded by MCIN/AEI/10.13039/501100011033); Generalitat de Catalunya (CERCA program and AGAUR Project No. 2021 SGR 01442); Fundaci\'{o} Cellex, and Fundaci\'{o} Mir-Puig.

\appendix

\blue{
\section{Schwinger bosons and Abrikosov fermions}\label{app:schwinger_bosons}
In this appendix, we compare the Schwinger boson and the Abrikosov fermion representations of spins for treating dynamics and demonstrate that the latter is substantially more accurate for 2PI, if we limit ourselves to Gaussian initial states. We remark that the Majorana fermion representation can be written in terms of complex fermions and therefore, provides similar advantages and additionally, requires less numerical resources due to working with a smaller number of Green's functions.}

\blue{Consider the Schwinger boson representation~\cite{auerbach1998interacting,sachdev_book}
\begin{align}
    S^z &=\frac12 \qty(a^\dagger a - b^\dagger b), \label{schwinger_Sz}\\
    S^+ &= a^\dagger b,
\end{align}
together with the constraint
\begin{equation}\label{schwinger_constraint}
    a^\dagger a + b^\dagger b = 2S.
\end{equation}
We take the spin state $\ket{M_z=S}$ such that $S^z\ket{S}=S\ket{S}$. This corresponds to the following bosonic state
\begin{equation}
    \ket{\psi}=\ket{n_a=2S}\otimes\ket{n_b=0}.
\end{equation}
However, $\ket{n_a=2S}$ is not a Gaussian state, as can be checked easily. We can approximate it by a mixed Gaussian state
\begin{equation}
    \rho_{G_b} \equiv \lim_{\epsilon \to \infty} \frac{e^{-H_{G_b}(\epsilon)}}{\Tr(e^{-H_{G_b}(\epsilon)})},
\end{equation}
where
\begin{equation}
    H_{G_b}(\epsilon) \equiv \ln(1+\frac{1}{2S}) \,a^\dagger a + \epsilon \,b^\dagger b,
\end{equation}
$H_{G_b}(\epsilon)$ has been chosen such that 
\begin{align}
    \Tr(\rho_G\, a^\dagger a) &= 2S,\\
    \Tr(\rho_G\, b^\dagger b) &= 0,
\end{align}
in order to satisfy Eqs.~\eqref{schwinger_Sz} and \eqref{schwinger_constraint} at the level of expectation values. However, the squares of Eqs.~\eqref{schwinger_Sz} and \eqref{schwinger_constraint} are not satisfied
\begin{align}
    \frac14 \expval{(a^\dagger a - b^\dagger b)^2}_G &= \frac{S}{2}(1+4S) \neq S^2,\\
    \expval{(a^\dagger a + b^\dagger b)^2}_G  &= 2S(1 + 4S) \neq 4S^2.
\end{align}
We note that the errors are larger than the quantities themselves. If we could use coherent states for bosons, the errors would be $\mathcal{O}(S)$ and therefore, sub-leading at least for large $S$. But this is not permissible, as coherent states break the local $U(1)$ gauge symmetry of the Schwinger bosons given by $(a_i,b_i) \to (a_i e^{i\psi_i},b_i e^{i\psi_i})$.}

 \blue{Alternatively, we can express a spin of size $S$ using Abrikosov fermions~\cite{fradkin_2013}
 \begin{align}
     S^z &= \frac12 \sum_{n=1}^{2S} \qty(f_n^\dagger f_n - c^\dagger_n c_n),\\
     S^+ &= \sum_{n=1}^{2S} f^\dagger_n c_n,
 \end{align}
 where $f_n$ and $c_n$ are fermion annihilation operators that satisfy the usual anti-commutation relations.  Physical states should satisfy the following constraint
 \begin{equation}\label{absrikosov_constraint}
     f^\dagger_n f_n + c^\dagger_n c_n = 1,
 \end{equation}
 which implies
 \begin{equation}\label{abrikosov_constraint2}
     \sum_n \qty(f^\dagger_n f_n + c^\dagger_n c_n) = 2S.
 \end{equation}
 For the spin state $\ket{M_z=S}$ we can use the following Gaussian fermionic state
 \begin{equation}\label{rho_gauss_f}
    \rho_{G_f} \equiv \lim_{\epsilon \to \infty} \frac{e^{-H_{G_f}(\epsilon)}}{\Tr(e^{-H_{G_f}(\epsilon)})},
 \end{equation}
where
\begin{equation}
    H_{G_f}(\epsilon) \equiv \epsilon \sum_n \qty( c^\dagger_n c_n - f^\dagger_n f_n).
\end{equation}
The advantage of fermionic spinons is that $\rho_{G_f}$ is in fact, a pure state for $ \epsilon \to \infty$. In this case, it is easy to see that $\rho_{G_f}$ satisfies Eqs.~\ref{absrikosov_constraint} and \ref{abrikosov_constraint2} exactly and not only at the level of expectation values. Note that while we assumed a fully polarized spin state in the $+\hat{z}$ direction above, the argument is general and holds for any other spin coherent states after a proper unitary transformation. Therefore, fermionic Gaussian states can exactly represent any spin coherent state without introducing errors.}

\section{Hubbard-Stratonovich transformation}\label{app:HS}
We use the following identity
\begin{equation}
1=\int \mathcal{D}\qty[\chi^1,\chi^2]e^{iS_\chi},
\end{equation}
where $S_\chi$ is defined in Eq.~(\ref{S_chi}). This identity follows from the fact that the integral is Gaussian. Then, we shift the integration variable according to
\begin{equation}
\vec{\chi}_{\alpha i} \to \vec{\chi}_{\alpha i} + \hat{W}_0 \otimes \vec{A}_{\alpha i}.
\end{equation}
$\otimes$ is matrix multiplication over all indices (time and species). $\hat{W}_0$ was given by Eq.~(\ref{W0}) and $\vec{A}_{\alpha i}$ is defined as
\begin{equation}
\vec{A}_{\alpha i}(t_c)\equiv \begin{pmatrix}
-\frac{2i}{\sqrt{N_s}}\sum_\lambda \psi^y_{i\lambda}(t_c) \psi^z_{i\lambda}(t_c) \\ \\ \sqrt{2\omega_c} g_{\alpha i} \phi_\alpha(t_c)
\end{pmatrix}.
\end{equation}
Applying the shift yields
\begin{equation}
1= \int \mathcal{D}\qty[\chi^1,\chi^2]e^{iS_\chi+iS_{\chi \psi}+iS_{g\chi \phi}-iS_\mathrm{int}},
\end{equation}
where $S_{\chi \psi}$, $S_{g \chi \phi}$ and $S_\mathrm{int}$ are respectively given by Eqs.~(\ref{S_chipsi}), (\ref{S_chiphi}) and (\ref{S_int}). Since $S_\mathrm{int}$ does not depend on $\chi$ we can move it to the LHS to get
\begin{equation}
e^{i S_\mathrm{int}}=\int \mathcal{D}\qty[\chi^1,\chi^2]e^{iS_\chi+iS_{\chi \psi}+iS_{g\chi \phi}}.
\end{equation}
Therefore, we can substitute the RHS of the above expression in the original action.

\blue{To prove the connection between the spin correlation function and the Green's function of HS field in Eq.~\eqref{ss_to_V}, we add source terms to the Keldysh action
\begin{equation}
    S \to S + \sum_i \oint J_i S^x_i\,dt.
\end{equation}
The spin correlation function can be found from
\begin{equation}
    \expval{S^x_i(t)S^x_i(t')} = - \frac{\delta^2 Z\qty[J]}{\delta J_i(t) \delta J_i(t')},
\end{equation}
where $Z$ is the generating functional
\begin{equation}
    Z\qty[J] \equiv \int \mathcal{D}\qty[\psi,\phi,\pi,\chi] e^{iS - i \sum_j \int J_j S^x_j\,dt}.
\end{equation}
The source term can be absorbed into $S_{\chi \psi}$ as
\begin{equation}
    S_{\chi \psi} \to -\frac{2i}{\sqrt{N_s}} \sum_{\alpha,i,\lambda}\oint dt_c\, \qty(\chi^1_{\alpha i} - \frac{\sqrt{N_s}}{2M} J_i)\psi^y_{i \lambda}\psi^z_{i \lambda}.
\end{equation}
We shift $\chi^1$ according to
\begin{equation}
    \chi^1_{\alpha i}\to \chi^1_{\alpha i} + \frac{\sqrt{N_s}}{2M} J_i,
\end{equation}
which generates a coupling between $J$ and $\chi^2$
\begin{equation}
    S_{\chi J} = \frac{(N+M)}{2M}Ns\sum_\alpha^M \sum_i^N \oint dt \, \chi^2_{\alpha i}(t) \, J_i(t).
\end{equation}
Now we take functional derivatives with respect to the source and get
\begin{equation}
    \expval{S^x_i(t)S^x_i(t')} = \frac{N+M}{4 M^2}N_s \sum_{\alpha,\beta}^M \expval{\chi^2_{\alpha i}(t) \chi^2_{\beta i}(t')}.
\end{equation}
Using permutation symmetry we have $\expval{\chi^2_{\alpha i}(t) \chi^2_{\beta i}(t')}=iU^{22}(t,t')=iV^{22}(t,t')$ which yields
\begin{equation}
     \expval{S^x_i(t)S^x_i(t')} = i\frac{(N+M)N_s}{4} V^{22}(t,t').
\end{equation}
}

\section{Dyson equations}\label{app:DE}

For any practical calculation, we need to write equations in terms of normal time variables. DE are more transparent if we write fields in the classical/quantum basis defined as \cite{kamenev}
\begin{align}
    \phi_c(t)&\equiv \frac{1}{\sqrt{2}}\qty(\phi_+(t)+\phi_-(t)),\\
    \phi_q(t)&\equiv \frac{1}{\sqrt{2}}\qty(\phi_+(t)-\phi_-(t)),
\end{align}
where $t$ is a normal time variable. The retarded, advanced and Keldysh (RAK) Green's functions are defined as \cite{kamenev,Rammer_2007}
\begin{align}
    {G}^R(t,t')&\equiv -i \expval{\phi_c(t)\phi_q(t')}, \\
    {G}^A(t,t')&\equiv -i \expval{\phi_q(t)\phi_c(t')}, \\
    {G}^K(t,t')&\equiv -i \expval{\phi_c(t)\phi_c(t')}.
\end{align}

Accordingly, we define the 6-component fermion field 
\begin{equation}
    \Psi_{i\lambda}^T(t) \equiv \qty(
        \psi^x_c(t), \psi^y_c(t), \psi^z_c(t), \psi^x_q(t), \psi^y_q(t), \psi^z_q(t)).
\end{equation}
Where we have omitted the $i\lambda$ index on RHS for simplicity. Then, the action for free fermions $S_\sigma$ in Eq. (\ref{S_f}) reads
\begin{equation}\label{S_f_RAK}
    S_\sigma = \frac12 \sum_i^N\sum_\lambda^{N_s}\int_0^t \Psi^T_{i\lambda} \hat{G}_0^{-1} \Psi_{i\lambda} \, dt,
\end{equation}
with
\begin{align}
    \hat{G}_0^{-1}&\equiv \begin{bmatrix}
        0 & (\hat{G}_0^A)^{-1} \\ (\hat{G}_0^R)^{-1} & 0
    \end{bmatrix}, \\
    (\hat{G}_0^R)^{-1} = \qty[(\hat{G}_0^A)^{-1}]^\dagger &\equiv \begin{bmatrix}
        i\partial_t & i \Delta & 0 \\ -i \Delta & i\partial_t & 0 \\ 0 & 0 & i\partial_t
    \end{bmatrix}.
\end{align}
For self-energies we define
\begin{align}
    \hat{\Sigma}^R&\equiv\frac12 \qty(\hat{\Sigma}^{++}-\hat{\Sigma}^{+-}-\hat{\Sigma}^{--}+\hat{\Sigma}^{-+}), \label{sigmaR}\\
    \hat{\Sigma}^A&\equiv\frac12 \qty(\hat{\Sigma}^{++}+\hat{\Sigma}^{+-}-\hat{\Sigma}^{--}-\hat{\Sigma}^{-+}),\\
    \hat{\Sigma}^K&\equiv\frac12 \qty(\hat{\Sigma}^{++}+\hat{\Sigma}^{+-}+\hat{\Sigma}^{--}+\hat{\Sigma}^{-+}).\label{sigmaK}
\end{align}

DE have a simple structure in terms of RAK functions
\begin{equation}
    \begin{bmatrix}
        0 & (\hat{G}_0^A)^{-1} - \hat{\Sigma}^A \\ (\hat{G}_0^R)^{-1} -  \hat{\Sigma}^R &  -\hat{\Sigma}^K
    \end{bmatrix} \begin{bmatrix}
        \hat{G}^K & \hat{G}^R \\ \hat{G^A} & 0
    \end{bmatrix} = \mathbb{1},
\end{equation}
\begin{equation}
    (\hat{G}_0^R)^{-1} \hat{G}^R = \mathbb{1} +  \hat{\Sigma}^R \hat{G}^R,
\end{equation}
\begin{equation}
    (\hat{G}_0^R)^{-1} \hat{G}^K =   \hat{\Sigma}^R \hat{G}^K + \hat{\Sigma}^K \hat{G}^A. 
\end{equation}
 The fermion self-energy matrix at NLO reads 
 \begin{equation}
     \hat{\Sigma}_\mathrm{NLO}^{ss'}=\begin{bmatrix}
         0 &  0 &  0 \\  0 &  \tilde{\Sigma}^{ss'}_{yy} &  \tilde{\Sigma}^{ss'}_{yz} \\ 0 &  \tilde{\Sigma}^{ss'}_{zy} &  \tilde{\Sigma}^{ss'}_{zz}
     \end{bmatrix}, \quad (s,s'=\pm).
 \end{equation}
\begin{widetext}
Following Eq.(\ref{Sigma_NLO}), the expanded forms of the components of $\Sigma$ are given by
   \begin{align}
     \tilde{\Sigma}^{ss'}_{yy}(t,t')&=+\frac{4i}{N_s} \qty(M V^{11}_{ss'}(t,t') + M(M-1) U^{11}_{ss'}(t,t')) G_{zz}^{ss'}(t,t'),\\
     \tilde{\Sigma}^{ss'}_{zz}(t,t')&=+\frac{4i}{N_s} \qty(M V^{11}_{ss'}(t,t') + M(M-1) U^{11}_{ss'}(t,t')) G_{yy}^{ss'}(t,t'),\\
     \tilde{\Sigma}^{ss'}_{yz}(t,t')&=-\frac{4i}{N_s} \qty(M V^{11}_{ss'}(t,t') + M(M-1) U^{11}_{ss'}(t,t')) G_{zy}^{ss'}(t,t'),\\
     \tilde{\Sigma}^{ss'}_{zy}(t,t')&=-\frac{4i}{N_s} \qty(M V^{11}_{ss'}(t,t') + M(M-1) U^{11}_{ss'}(t,t')) G_{yz}^{ss'}(t,t').
 \end{align}
 We do not write RAK self energies in terms of RAK Green's functions as it is numerically less costly to find them from Eqs. (\ref{sigmaR})- (\ref{sigmaK}). Defining $B_x(t)$ as
 \begin{equation}
     B_x(t)\equiv \frac{2M}{\sqrt{N_s}}\tilde{\chi}^1(t),
 \end{equation}
 which captures the fermion self energy at LO (Eq. \ref{Sigma_LO}). Then, the expanded DE for retarded fermion Green's functions are given by
 \begin{align}
     &i\partial_t G^R_{xx}(t,t')+ i \Delta G^R_{yx}(t,t') = \delta(t-t'), \label{DE_F_first}\\
     &i\partial_t G^R_{xy}(t,t')+ i \Delta G^R_{yy}(t,t') = 0,\\
     &i\partial_t G^R_{xz}(t,t')+ i \Delta G^R_{yz}(t,t') = 0,\\
     &i\partial_t G^R_{yx}(t,t')- i \Delta G^R_{xx}(t,t') - i B_x(t) G^R_{zx}(t,t') = \int_0^t \qty(\tilde{\Sigma}^R_{yy}(t,t'')G^R_{yx}(t'',t') + \tilde{\Sigma}^R_{yz}(t,t'')G^R_{zx}(t'',t'))\, dt'',\\
     &i\partial_t G^R_{yy}(t,t')- i \Delta G^R_{xy}(t,t') - i B_x(t) G^R_{zy}(t,t') = \delta(t-t') +  \int_0^t \qty(\tilde{\Sigma}^R_{yy}(t,t'')G^R_{yy}(t'',t') + \tilde{\Sigma}^R_{yz}(t,t'')G^R_{zy}(t'',t'))\, dt'',\\
     &i\partial_t G^R_{yz}(t,t')- i \Delta G^R_{xz}(t,t') - i B_x(t) G^R_{zz}(t,t') = \int_0^t \qty(\tilde{\Sigma}^R_{yy}(t,t'')G^R_{yz}(t'',t') + \tilde{\Sigma}^R_{yz}(t,t'')G^R_{zz}(t'',t'))\, dt'',\\
     &i\partial_t G^R_{zx}(t,t')+ i B_x(t) G^R_{yx}(t,t') = \int_0^t \qty(\tilde{\Sigma}^R_{zy}(t,t'')G^R_{yx}(t'',t') + \tilde{\Sigma}^R_{zz}(t,t'')G^R_{zx}(t'',t'))\, dt'',\\
     &i\partial_t G^R_{zy}(t,t')+ i B_x(t) G^R_{yy}(t,t') = \int_0^t \qty(\tilde{\Sigma}^R_{zy}(t,t'')G^R_{yy}(t'',t') + \tilde{\Sigma}^R_{zz}(t,t'')G^R_{zy}(t'',t'))\, dt'',\\
     &i\partial_t G^R_{zz}(t,t')+ i B_x(t) G^R_{yz}(t,t') =\delta(t-t') + \int_0^t \qty(\tilde{\Sigma}^R_{zy}(t,t'')G^R_{yz}(t'',t') + \tilde{\Sigma}^R_{zz}(t,t'')G^R_{zz}(t'',t'))\, dt''.
 \end{align}
 For Keldysh components we have
 \begin{align}
    i\partial_t G^K_{xx}(t,t')+ i \Delta G^K_{yx}(t,t') &= 0, \label{GK_xx_eom}\\
     i\partial_t G^K_{xy}(t,t')+ i \Delta G^K_{yy}(t,t') &= 0, \label{GK_xy_eom}\\
     i\partial_t G^K_{xz}(t,t')+ i \Delta G^K_{yz}(t,t') &= 0,
\end{align}
\begin{multline}\label{GK_yx_eom}
   i\partial_t G^K_{yx}(t,t')- i \Delta G^K_{xx}(t,t') - i B_x(t) G^K_{zx}(t,t') = \int_0^t \Big(\tilde{\Sigma}^R_{yy}(t,t'')G^K_{yx}(t'',t') + \tilde{\Sigma}^R_{yz}(t,t'')G^K_{zx}(t'',t') \\ + \tilde{\Sigma}^K_{yy}(t,t'')G^A_{yx}(t'',t') + \tilde{\Sigma}^K_{yz}(t,t'')G^A_{zx}(t'',t')\Big)\, dt'',
\end{multline}
\begin{multline}
   i\partial_t G^K_{yy}(t,t')- i \Delta G^K_{xy}(t,t') - i B_x(t) G^K_{zy}(t,t') = \int_0^t \Big(\tilde{\Sigma}^R_{yy}(t,t'')G^K_{yy}(t'',t') + \tilde{\Sigma}^R_{yz}(t,t'')G^K_{zy}(t'',t') \\ + \tilde{\Sigma}^K_{yy}(t,t'')G^A_{yy}(t'',t') + \tilde{\Sigma}^K_{yz}(t,t'')G^A_{zy}(t'',t')\Big)\, dt'', 
\end{multline}
\begin{multline}
   i\partial_t G^K_{yz}(t,t')- i \Delta G^K_{xz}(t,t') - i B_x(t) G^K_{zz}(t,t') = \int_0^t \Big(\tilde{\Sigma}^R_{yy}(t,t'')G^K_{yz}(t'',t') + \tilde{\Sigma}^R_{yz}(t,t'')G^K_{zz}(t'',t') \\ + \tilde{\Sigma}^K_{yy}(t,t'')G^A_{yz}(t'',t') + \tilde{\Sigma}^K_{yz}(t,t'')G^A_{zz}(t'',t')\Big)\, dt'', 
\end{multline}
\begin{multline}
   i\partial_t G^K_{zx}(t,t') + i B_x(t) G^K_{yx}(t,t') = \int_0^t \Big(\tilde{\Sigma}^R_{zy}(t,t'')G^K_{yx}(t'',t') + \tilde{\Sigma}^R_{zz}(t,t'')G^K_{zx}(t'',t') \\ + \tilde{\Sigma}^K_{zy}(t,t'')G^A_{yx}(t'',t') + \tilde{\Sigma}^K_{zz}(t,t'')G^A_{zx}(t'',t')\Big)\, dt'', 
\end{multline}
\begin{multline}
   i\partial_t G^K_{zy}(t,t') + i B_x(t) G^K_{yy}(t,t') = \int_0^t \Big(\tilde{\Sigma}^R_{zy}(t,t'')G^K_{yy}(t'',t') + \tilde{\Sigma}^R_{zz}(t,t'')G^K_{zy}(t'',t') \\ + \tilde{\Sigma}^K_{zy}(t,t'')G^A_{yy}(t'',t') + \tilde{\Sigma}^K_{zz}(t,t'')G^A_{zy}(t'',t')\Big)\, dt'', 
\end{multline}
\begin{multline}
   i\partial_t G^K_{zz}(t,t') + i B_x(t) G^K_{yz}(t,t') = \int_0^t \Big(\tilde{\Sigma}^R_{zy}(t,t'')G^K_{yz}(t'',t') + \tilde{\Sigma}^R_{zz}(t,t'')G^K_{zz}(t'',t') \\ + \tilde{\Sigma}^K_{zy}(t,t'')G^A_{yz}(t'',t') + \tilde{\Sigma}^K_{zz}(t,t'')G^A_{zz}(t'',t')\Big)\, dt''. \label{GK_zz_eom}
\end{multline}

 Advanced functions can be obtained from $\hat{G}^A=\qty[\hat{G}^R]^\dagger$ and do not require separate computation.
 
Photons fields in $(c,q)$ basis are cast into a multi-component field defined by
\begin{equation}
    \Phi^T_\alpha(t) \equiv \qty(\phi_{\alpha,c},\pi_{\alpha,c},\phi_{\alpha,q},\pi_{\alpha,q}),
\end{equation}
with the free action (Eq. (\ref{S_phi})
\begin{equation}\label{S_phi_RAK}
    S_\mathrm{ph}= \frac12 \sum_\alpha^M\int_0^t \Phi_\alpha^T \hat{D}_0^{-1} \Phi_\alpha \, dt,
\end{equation}
where 
\begin{align}
    \hat{D}_0^{-1} & \equiv \begin{bmatrix}
        0 & (\hat{D}_0^{-1})^A \\ (\hat{D}_0^{-1})^R & (\hat{D}_0^{-1})^K
    \end{bmatrix},\\
    (\hat{D}_0^{-1})^R = \qty[(\hat{D}_0^{-1})^A]^\dagger &\equiv \begin{bmatrix}
        -\omega_c^2 & -\kappa -\partial_t \\ \kappa + \partial_t & -1
    \end{bmatrix}, \\
    (\hat{D}_0^{-1})^K &\equiv \begin{bmatrix}
        2i\kappa \omega_c & 0 \\ 0 & 2i\kappa/\omega_c
    \end{bmatrix}.
\end{align}

Note that $(\hat{D}_0^{-1})^K$ is not the inverse of the bare Keldysh function, for details see Ref. \cite{kamenev}. Similar to fermions, DE has the general structure
\begin{equation}
    \begin{bmatrix}
        0 & (\hat{D}_0^A)^{-1} - \hat{\Pi}^A \\ (\hat{D}_0^R)^{-1} -  \hat{\Pi}^R &   (\hat{D}_0^K)^{-1} -\hat{\Pi}^K
    \end{bmatrix} \begin{bmatrix}
        \hat{D}^K & \hat{D}^R \\ \hat{D^A} & 0
    \end{bmatrix} = \mathbb{1},
\end{equation}
\begin{align}
    &(\hat{D}_0^R)^{-1} \hat{D}^R = \mathbb{1} +  \hat{\Pi}^R \hat{D}^R,\\
    &(\hat{D}_0^R)^{-1} \hat{D}^K + (\hat{D}_0^K)^{-1} \hat{D}^A =   \hat{\Pi}^R \hat{D}^K + \hat{\Pi}^K \hat{D}^A. 
\end{align}
The photon self energy matrix $\hat{\Pi}$ only has one non-zero entry in $(\phi,\pi)$ basis
\begin{equation}
    \hat{\Pi}^{R/A/K}=\begin{bmatrix}
        \Pi^{R/A/K} && 0 \\ 0 && 0
    \end{bmatrix},
\end{equation}
where $\Pi$ was given in Eq. (\ref{Pi_NLO}). The expanded DE for retarded photon functions are given by

\begin{align}
    &-(\partial_t+\kappa)D^R_{\pi \phi}(t,t')-\omega_c^2 D^R_{\phi \phi}(t,t') = \delta(t-t') + \int_0^t \Pi^R(t,t'')D^R_{\phi \phi}(t'',t')\, dt'', \label{DE_P_first}\\
    &-(\partial_t+\kappa)D^R_{\pi \pi}(t,t')-\omega_c^2 D^R_{\phi \pi}(t,t') =  \int_0^t \Pi^R(t,t'')D^R_{\phi \pi}(t'',t')\, dt'',\\
    &+(\partial_t+\kappa)D^R_{\phi \phi}(t,t')- D^R_{\pi \phi}(t,t') = 0,\\
    &+(\partial_t+\kappa)D^R_{\phi \pi}(t,t')- D^R_{\pi \pi}(t,t') = \delta(t-t'),
\end{align}
and for Keldysh functions:
\begin{align}
    &-(\partial_t+\kappa)D^K_{\pi \phi}(t,t')-\omega_c^2 D^K_{\phi \phi}(t,t') + 2i\kappa \omega_c D^A_{\phi \phi}(t,t') =   \int_0^t \qty(\Pi^R(t,t'')D^K_{\phi \phi}(t'',t') + \Pi^K(t,t'')D^A_{\phi \phi}(t'',t'))\, dt'',\\
    &-(\partial_t+\kappa)D^K_{\pi \pi}(t,t')-\omega_c^2 D^K_{\phi \pi}(t,t')  + 2i\kappa \omega_c D^A_{\phi \pi}(t,t') = \int_0^t \qty(\Pi^R(t,t'')D^K_{\phi \pi}(t'',t') + \Pi^K(t,t'')D^A_{\phi \pi}(t'',t'))\, dt'',\\
    &+(\partial_t+\kappa)D^K_{\phi \phi}(t,t')- D^K_{\pi \phi}(t,t') + \frac{2i\kappa }{\omega_c} D^A_{\pi \phi}(t,t') = 0,\\
    &+(\partial_t+\kappa)D^K_{\phi \pi}(t,t')- D^K_{\pi \pi}(t,t')+ \frac{2i\kappa }{\omega_c} D^A_{\pi \pi}(t,t') = 0. \label{DE_P_last}
\end{align}

As mentioned in the main text, the off diagonal ($\alpha \neq \beta$) elements of Ising field Green's functions have to be retained and DE for Ising field Green's functions couple diagonal ($V$) and off diagonal ($U$) elements of $W$. The compact form of DE for $W$ is 
\begin{equation}
    \begin{bmatrix}
        0 & (\hat{W}_0^A)^{-1} - \hat{\Omega}^A \\ (\hat{W}_0^R)^{-1} - \hat{\Omega}^R &  - \hat{\Omega}^K
    \end{bmatrix}\begin{bmatrix}
        \hat{W}^K & \hat{W}^R \\ \hat{W}^A & 0
    \end{bmatrix}=\mathbb{1}.
\end{equation}
The self energy matrices are diagonal in the two dimensional space of the Ising field components
\begin{equation}
    \hat{\Omega}^{R/A/K}= \begin{bmatrix}
        \Omega_{11}^{R/A/K} & 0 \\ 0 & \Omega_{22}^{R/A/K}
    \end{bmatrix},
\end{equation}
where $\Omega_{11}$ and $\Omega_{22}$ were given in Eqs. (\ref{Omega_11_NLO}) and (\ref{Omega_22_NLO}). It is convenient to decompose $W$ into its free and renormalized parts
\begin{equation}
    W\equiv W_0 + \tilde{W},
\end{equation}
yielding 
\begin{equation}\label{DE_WtildeR}
    \tilde{W}^R = W_0^R \Omega^R W_0^R + W_0^R \Omega^R \tilde{W}^R.
\end{equation}
\begin{equation}\label{DE_WtildeK}
    \tilde{W}^K = W_0^R \Omega^K W_0^A + W_0^R \Omega^R \tilde{W}^K + W_0^R \Omega^K \tilde{W}^A.
\end{equation}
The expanded form of Eq. (\ref{DE_WtildeR}) in the retarded sector is given by (note that $U_0=V^{11}_0=V^{22}_0=0$)
\begin{align}
    &\tilde{V}^R_{11}(t,t')= \frac{1}{N+M} \Omega_{22}^R(t,t') + \frac{1}{\sqrt{N+M}}\int_0^t \Omega^R_{22}(t,t'')\tilde{V}^{21}_R(t'',t')\,dt'', \label{DE_W_first}\\
    &\tilde{U}^R_{11}(t,t')= \frac{1}{\sqrt{N+M}} \int_0^t \Omega_{22}^R(t,t'')\tilde{U}^R_{21}(t'',t')\, dt'', \\
    &\tilde{V}^R_{12}(t,t')=\tilde{U}^R_{12}(t,t')= \frac{1}{\sqrt{N+M}} \int_0^t \Omega_{22}^R(t,t'')\tilde{V}^R_{22}(t'',t')\, dt'', \\
    &\tilde{V}^R_{21}(t,t')=\tilde{U}^R_{21}(t,t')=\frac{1}{\sqrt{N+M}} \int_0^t \Omega_{11}^R(t,t'')\qty(\tilde{V}^R_{11}(t'',t')+(M-1)\tilde{U}^R_{11}(t'',t'))\, dt'', \\
    &\tilde{V}^R_{22}(t,t')=\tilde{U}^R_{22}(t,t')=\frac{1}{N+M} \Omega_{11}^R(t,t') + \frac{1}{\sqrt{N+M}} \int_0^t \Omega_{11}^R(t,t'')\qty(\tilde{V}^R_{12}(t'',t')+(M-1)\tilde{U}^R_{12}(t'',t'))\, dt'',
\end{align}
and for Eq. (\ref{DE_WtildeK}) we have
\begin{align}
    &\tilde{V}^K_{11}(t,t')= \frac{1}{N+M} \Omega_{22}^K(t,t') + \frac{1}{\sqrt{N+M}}\int_0^t \qty(\Omega^R_{22}(t,t'')\tilde{V}_{21}^R(t'',t')+\Omega^K_{22}(t,t'')\tilde{V}_{21}^A(t'',t'))\,dt'',\\
    &\tilde{U}^K_{11}(t,t')= \frac{1}{\sqrt{N+M}} \int_0^t \qty(\Omega_{22}^R(t,t'')\tilde{U}^K_{21}(t'',t')+\Omega_{22}^K(t,t'')\tilde{U}^A_{21}(t'',t'))\, dt'',\\
    &\tilde{V}^K_{12}(t,t')=\tilde{U}^K_{12}(t,t')= \frac{1}{\sqrt{N+M}} \int_0^t \qty(\Omega_{22}^R(t,t'')\tilde{V}^K_{22}(t'',t')+\Omega_{22}^K(t,t'')\tilde{V}^A_{22}(t'',t'))\, dt'',
\end{align}
\begin{multline}
    \tilde{V}^K_{21}(t,t')=\tilde{U}^K_{21}(t,t')=\frac{1}{\sqrt{N+M}} \int_0^t \Omega_{11}^R(t,t'')\qty(\tilde{V}^K_{11}(t'',t')+(M-1)\tilde{U}^K_{11}(t'',t'))\,dt'' \\  +\int_0^t \Omega_{11}^K(t,t'')\qty(\tilde{V}^A_{11}(t'',t')+(M-1)\tilde{U}^A_{11}(t'',t'))\, dt'',
\end{multline}
\begin{multline}\label{DE_W_last}
    \tilde{V}^K_{22}(t,t')=\tilde{U}^K_{22}(t,t')= \frac{1}{N+M}\Omega_{11}^K(t,t')+\frac{1}{\sqrt{N+M}} \int_0^t \Omega_{11}^R(t,t'')\qty(\tilde{V}^K_{12}(t'',t')+(M-1)\tilde{U}^K_{12}(t'',t'))\,dt'' \\  +\int_0^t \Omega_{11}^K(t,t'')\qty(\tilde{V}^A_{12}(t'',t')+(M-1)\tilde{U}^A_{12}(t'',t'))\, dt''.
\end{multline}
\end{widetext}
DE have a causal structure \cite{Eberlein_PRB17,schuckert2018nonequilibrium,HH_PRB23} with solutions that propagate in time such that the values of Green's functions at each time step only depend on their values at previous times, allowing us to efficiently solve them using conventional numerical methods for differential equations. 

\section{Evaluation of the overlap parameter}\label{app:q}
After  the procedure discussed in Section \ref{sec:q_eval}, fields of different replicas will be coupled to each other. Seemingly, this increases the complexity of the problem as one usually has to consider larger Green's function matrices. However, the inter-replica couplings do not affect the dynamics of replica diagonal correlators. Therefore, the dynamics generated by DE given above are still valid. To evaluate replica off-diagonal Green's functions, we have to derive extra dynamical equations that take replica diagonal Green's functions as inputs. Further simplifications occur by noting that replica off diagonal fermion Green's functions vanish due to the local $Z_2$ gauge symmetry of the fermionic representation for spins $\psi \to - \psi$. The only non-zero replica off diagonal self energies are 
\begin{equation}
    \underaccent{\sim}{\Pi}_\mathrm{LO}(t_c,t'_c) = -2iNg^2 \omega_c \tilde{\chi}^2(t_c)\tilde{\chi}^2(t'_c),
\end{equation}
\begin{equation}
    \underaccent{\sim}{\Pi}_\mathrm{NLO}(t_c,t'_c) = 2Ng^2 \omega_c \underaccent{\sim}{V}^{22}(t_c,t'_c),
\end{equation}
\begin{equation}
    \underaccent{\sim}{\Omega}^{22}(t_c,t'_c) = 2g^2 \omega_c \underaccent{\sim}{D}^{\phi\phi}(t_c,t'_c).
\end{equation}
We also note that off diagonal Green's functions only have Keldysh (symmetric) components. This is natural, as replicas are only mathematical entities without real physical interactions with each other. DE governing $\underaccent{\sim}{D}$ are given by
\begin{equation}
    -(\partial_t + \kappa) \underaccent{\sim}{D}^K_{\phi\phi}(t,t')+ \underaccent{\sim}{D}^K_{\pi \phi}(t,t')=0,
\end{equation}
\begin{equation}
    -(\partial_t + \kappa) \underaccent{\sim}{D}^K_{\phi\pi}(t,t')+ \underaccent{\sim}{D}^K_{\pi \pi}(t,t')=0,
\end{equation}
\begin{multline}
    -(\partial_t + \kappa) \underaccent{\sim}{D}^K_{\pi\phi}(t,t')- \omega_c^2 \underaccent{\sim}{D}^K_{\phi \phi}(t,t')= \\\int_0^t \qty(\Pi^R(t,t'')\underaccent{\sim}{D}^K_{\phi\phi}(t'',t')+\underaccent{\sim}{\Pi}^K(t,t'')D^A_{\phi\phi}(t'',t'))\,dt''.
\end{multline}
\begin{multline}
    -(\partial_t + \kappa) \underaccent{\sim}{D}^K_{\pi\pi}(t,t')- \omega_c^2 \underaccent{\sim}{D}^K_{\phi \pi}(t,t')= \\\int_0^t \qty(\Pi^R(t,t'')\underaccent{\sim}{D}^K_{\phi\pi}(t'',t')+\underaccent{\sim}{\Pi}^K(t,t'')D^A_{\phi\pi}(t'',t'))\,dt''.
\end{multline}
For Ising field correlators we have
\begin{multline}
    \underaccent{\sim}{U}_{11}^K(t,t')= \frac{1}{\sqrt{M+N}}  \int_0^t \Big(\Omega^R_{22}(t,t'')\underaccent{\sim}{V}_{21}^K(t'',t') \\  + \underaccent{\sim}{\Omega}^K_{22}(t,t'')V_{21}^A(t'',t')\Big)\, dt'',
\end{multline}
\begin{equation}
    \underaccent{\sim}{V}_{11}^K(t,t')=\frac{1}{M+N} \underaccent{\sim}{\Omega}^K_{22}(t,t') + \underaccent{\sim}{U}_{11}^K(t,t')
\end{equation}

\begin{multline}
    \underaccent{\sim}{V}^K_{12}(t,t')=\underaccent{\sim}{U}^K_{12}(t,t')= \frac{1}{\sqrt{N+M}} \\ \times \int_0^t \Big( \Omega^R_{22}(t,t'')\underaccent{\sim}{V}^K_{22}(t'',t')   + \underaccent{\sim}{\Omega}^K_{22}(t,t'')V^A_{22}(t'',t') \Big) \, dt'',
\end{multline}
\begin{multline}
    \underaccent{\sim}{V}^K_{21}(t,t')=\underaccent{\sim}{U}^K_{21}(t,t')=\frac{1}{\sqrt{N+M}}\int_0^t \Omega^R_{11}(t,t'')\\ \times \Big(\underaccent{\sim}{V}^K_{11}(t'',t')+ (M-1) \underaccent{\sim}{U}^K_{11}(t'',t')\Big)\, dt'',
\end{multline}
\begin{equation}
    \underaccent{\sim}{V}^K_{22}(t,t')=\frac{M}{\sqrt{N+M}}\int_0^t \Omega^R_{11}(t,t'')\underaccent{\sim}{V}^K_{12}(t'',t')\, dt''.
\end{equation}
These equations can be numerically solved using the same approach explained in Appendix~\ref{app:DE_sol}.

\section{Numerical solution of Dyson equations}\label{app:DE_sol}
Solving Dyson equations (DE) is only possible using numerical integration. To solve DE numerically, we discretize both time variables of the correlation functions according to
\begin{equation}
G(t,t') \to G(i,j)
\end{equation}
 and solve the problem on a two dimensional time grid with time spacing $\Delta t$. Assuming that correlation functions are known for $i,j \le l$, which is an $l \times l$ square in the time grid, we evaluate the time derivates for fermion and photon Green's functions using Dyson equations in Eqs.~(\ref{DE_F_first})-(\ref{GK_zz_eom}) and Eqs.~(\ref{DE_P_first})-(\ref{DE_P_last}) . The memory integrals only depend on the values of correlation functions at $i,j \le l$ and can be evaluated using various numerical integration methods to be chosen depending on the desired accuracy. The equations for HS correlation functions in Eqs.~(\ref{DE_W_first})-(\ref{DE_W_last}) involve  no time derivatives and HS correlation functions are directly found from their equations at each step. Using the values for the time derivatives allows us to predict the values of Green's functions for $i,j \le l+1$ using Euler's method. However, the predicted value is usually not accurate enough and leads to numerical instabilities for long time evolutions. To remedy this issue, one can use more accurate approaches such as Runge-Kutta or predictor-corrector methods. For this work, we used a two-step predictor-corrector method. The reader is referred to an appendix in Ref.~\cite{schuckert2018nonequilibrium} for a detailed discussion of the two-step predictor-corrector method for DE.
 
 The main numerical costs are due to memory integrals on the RHS of DE. At each step of the integration, the numerical cost of evaluating memory integrals scales with $l^2$, where $l$ is the (discretized) evolution time since the initial state. The quadratic scaling with $l$ stems from the fact that at each step, memory integrals should be evaluated for $l$ points on the edge of the square given by $i,j \le l$ and for each point, the cost of numerical integration scales with $l$. We can reduce the numerical costs by using the analytical properties of Green's functions by only evaluating Green's functions on one side of the diagonal ($i \ge j$) explicitly and finding their values for $i < j$ using a combination of complex conjugation or transposition. Nevertheless, the numerical costs increase with the evolution time. To reduce the cost further, one may truncate memory integrals and keep the memory of the system up to a certain point in the past. However, this truncation is not permissible for our glassy system since the integrands have significant contributions at long times due to strong memory effects in the SG phase, and omitting memory effects prevents us to capture glassy dynamics. In this case, the overall cost of evolving the system for $l$ steps using serial computation scales as $l^3$. Using parallel computation, we could achieve a slower increase of the runtime with $l$ given by
 \begin{equation}
 t_\mathrm{runtime}\sim l^\beta,
 \end{equation}
 where $2<\beta<3$. \blue{For this work, each simulation took about 3 minutes for time evolutions up to $\Delta\cdot t =20$. Longer simulations like those in Figs.~\ref{fig:q_vs_k_5} and \ref{fig:q_vs_k_7} took about 30 minutes to complete. To obtain the spectral density at $\Delta \cdot \tau = 60$ in Fig.~\ref{fig:A_omega}, the simulation took about two hours. The computation time and resources would be less if the inter-replica overlap parameter did not need to be evaluated. All of the simulations have been done on a personal computer.}

\subsection{Equations of motion at LO}\label{app:lo_eom}
At LO fermion self energies will only have local terms captured by $B_x$ and the RHS of Eqs.~(\ref{GK_xx_eom})-(\ref{GK_zz_eom}) vanish. We want to find the time derivatives of spin expectation values $m^\alpha = \expval{\sigma^\alpha}$. Using Eq.(\ref{Majorana_rep}) we have
\begin{multline}
    \dv{}{t}m^\alpha(t)=-i\ \epsilon_{\alpha\beta\gamma} \dv{}{t} \expval{\psi^\beta(t) \psi^\gamma(t)}\\ = \frac12 \epsilon_{\alpha \beta \gamma} \dv{}{t} G^K_{\beta\gamma}(t,t).
\end{multline}
We use the chain rule to write
\begin{equation}
    \dv{}{t} G^K_{\beta\gamma}(t,t)=\qty(\partial_t + \partial_{t'})  G^K_{\beta\gamma}(t,t')|_{t'=t}.
\end{equation}
In DE we have   time derivatives of Green's functions only with respect to $t$. However, $\partial_{t'}G$ can be obtained easily by noting that
\begin{equation}
    G^{K}_{\alpha\beta}(t,t')=-G^K_{\beta\alpha}(t',t),
\end{equation}
\begin{equation}
    \to \partial_{t'}G^{K}_{\alpha\beta}(t,t')|_{t=t'}=-\partial_{t}G^K_{\beta\alpha}(t',t)|_{t'=t}.
\end{equation}
For example, to find $\mathrm{d}\expval{\sigma^z}/\mathrm{d}t$ we use Eq.~ (\ref{GK_xy_eom})
\begin{equation}
    \partial_t G^K_{xy}(t,t')|_{t'=t}=-\Delta G^K_{yy}(t,t),
\end{equation}
and Eq.~(\ref{GK_yx_eom})
\begin{multline}
    \partial_{t'}G^K_{xy}(t,t')|_{t'=t}=-\partial_{t}G^K_{yx}(t',t)|_{t'=t}\\=-\Delta G^K_{xx}(t,t)-B_x(t)G^K_{zx}(t,t),
\end{multline}
to get
\begin{equation}
    \dv{}{t}G^K_{xy}(t,t)=-B_x(t)G^K_{zx}(t,t),
\end{equation}
where we have used $G^K_{\alpha\alpha}(t,t)=0$. Using $G^K_{zx}(t,t)=m_y(t)$ we get
\begin{equation}
    \dv{}{t}m_z(t)=-B_x(t) m_y(t).
\end{equation}
Using $B_x$ from Eq.(\ref{Bx_def}) we get
\begin{equation}
    \dv{}{t} m_z(t) =  J\omega_c^2 \int_0^t k^y(t) D^{\phi \phi}_R(t,t') m_x(t') \,dt',
\end{equation}
which is the LO equation of motion for $m_z$ given in Eq.~(\ref{LO_Sz_EOM}). Eqs.~(\ref{LO_Sx_EOM}) and (\ref{LO_Sy_EOM}) can be obtained similarly using DE at LO for other components of $G^K$.

\begin{figure}[!t]
    \centering
    \includegraphics[width=.48\textwidth,trim={0 1.8cm 0 0},clip]{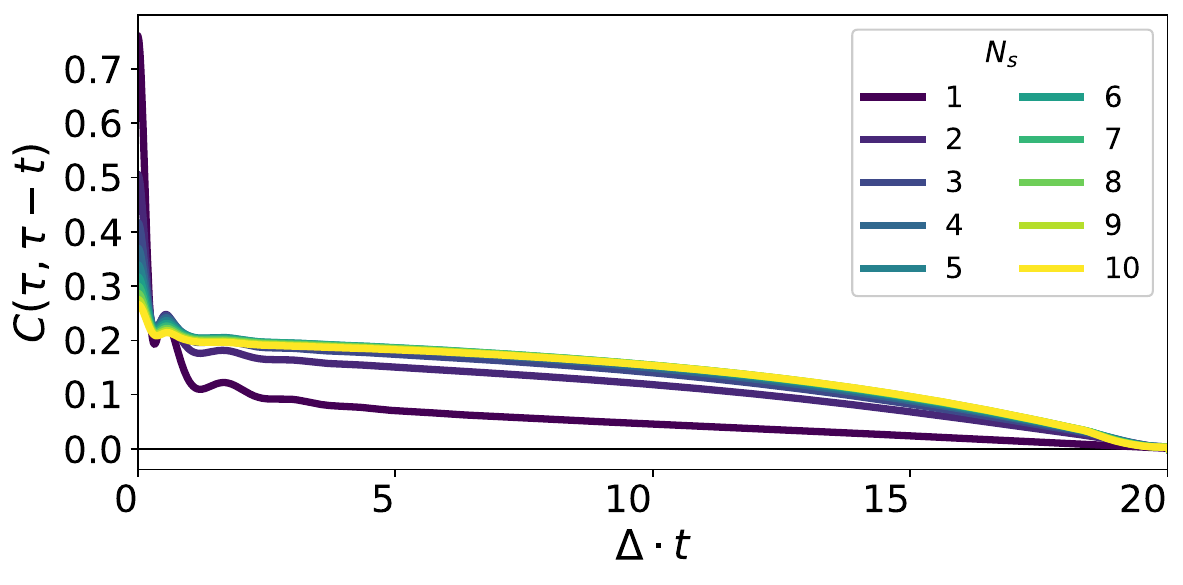}
    \includegraphics[width=.48\textwidth]{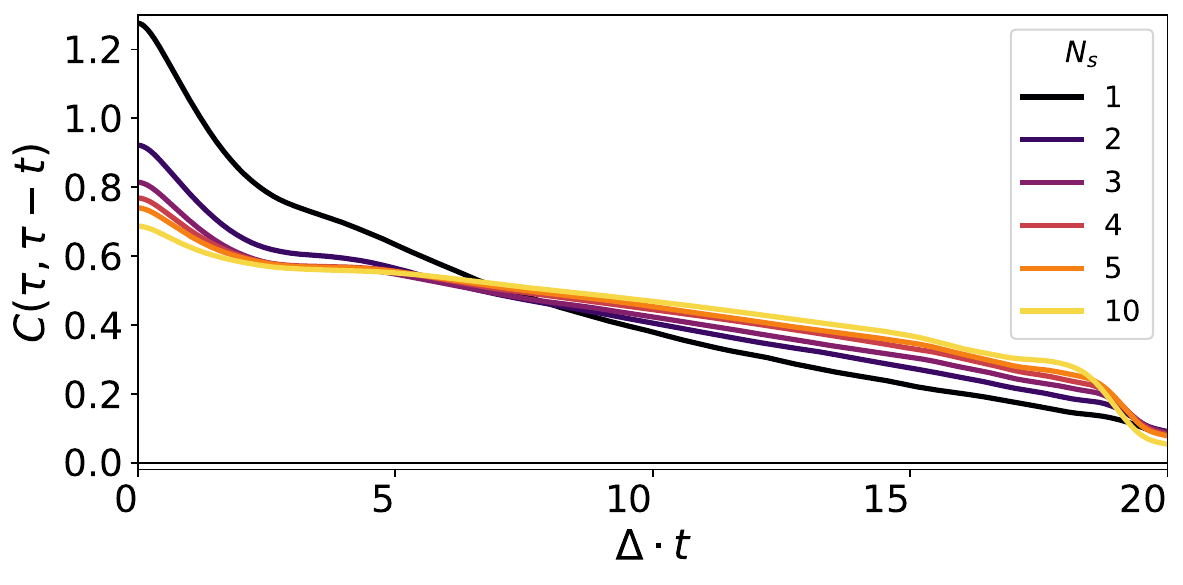}
    \blue{\caption{Comparison of aging dynamics obtained from 2PI (top panel) and DTWA (bottom panel) for different cluster sizes. Parameters are $g/g_c=1.27$, $\Delta/\omega_c=0.2$, $\kappa/\Delta=0.5$.}}
    \label{fig:C_2pi_vs_dtwa}
\end{figure}

\blue{\section{Comparison between 2PI and DTWA}\label{app:2pi_dtwa}
In this appendix, we further compare the results of 2PI and DTWA for aging dynamics. We consider finite cavity loss $\kappa\neq0$ which is implemented by solving a stochastic Langevin equation for cavity modes. Aging dynamics, given by the behavior of the correlation function in Eq.~\eqref{C}, is shown in Fig.~\ref{fig:C_2pi_vs_dtwa}. The 2PI results, which have been given in another work by us~\cite{hosseinshort}, are shown here over a longer timescale for a better comparison. 2PI predicts that $C$ approaches a limiting value as $N_s$ is increased. This limit corresponds to a classical SG with larger correlations at long times. For small $N_s$ correlation are strongly suppressed by quantum fluctuations  although the SG phase still survives. DTWA shows a qualitatively similar behavior for large $N_s$, where $C$ approaches the same limit for $N_s \gtrsim 10$, with a relatively flat profile for a wide range of $t$ . However, the agreement for large $N_s$ is qualitative for aging dynamics, in contrast to the quantitative agreement of the two methods for magnetization dynamics presented in Sec.~\ref{sec:spin_size}. Similar to the magnetization dynamics, DTWA predicts less sensitivity to spin size down to $N_s=1$, although it still predicts weaker SG. This has to be compared with 2PI, where $C$ drastically changes for small spins, and particularly for $N_s=1$. Furthermore, 2PI predicts a different correlation profile at between short ($t\lesssim \Delta^{-1}$) and long ($t\gtrsim \Delta^{-1}$) time separations, with a crossover of behavior at $t\sim \Delta^{-1}$. This sharp crossover was attributed to the small size of the local Hilbert space for small spins in Ref.~\cite{hosseinshort}, where quantum effects are more important. DTWA shows a crossover as well, but the timescale does not match $\Delta^{-1}$ which is the natural timescale for quantum fluctuations due to the transverse field $\Delta$ to affect local dynamics. Furthermore, the crossover is weaker and more smooth.}

\section{2PI for the single mode Dicke model without disorder}\label{app:Dicke}
For the sake of completeness, we give a concise derivation of far from equilibrium dynamics for the singe-mode Dicke model~\cite{kirton2019introduction} below, using a systematic expansion of the 2PI effective action in powers of the system size $N$.

\begin{figure}[!t]
    \centering
    \includegraphics{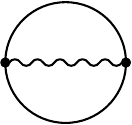}
    \caption{The NLO contribution to 2PI action for the Dicke model.}
    \label{fig:Gamma_dicke}
\end{figure}

For the Dicke model defined by
\begin{multline}\label{H_dicke}
    H=\frac{\Delta}{2} \sum_{i}^N \sigma^z_{i}+ \omega_c a^\dagger a   + \frac{g}{\sqrt{N}}\sum_{i}  \qty(a+a^\dagger) \sigma^x_{i},
\end{multline}
the Keldysh actions for spins and photons are the same as Eqs. (\ref{S_f}) and (\ref{S_phi}) together with their RAK representations in Eqs. (\ref{S_f_RAK}) and (\ref{S_phi_RAK}). The main difference is the spin-photon coupling part of the action
\begin{equation}\label{S_dice_int}
    S_\mathrm{int}=2ig\sqrt{\frac{2\omega_c}{N}}\sum_{i}^N\oint dt_c\, \phi\psi^y_{i}\psi^z_{i}.
\end{equation}
In this case, the expectation value of the photon field $\varphi \equiv \expval{\phi}$ can be finite. It is straightforward to show that at NLO, the 2PI action reads:
\begin{multline}\label{Gamma_dicke}
    \Gamma\qty[\varphi,\pi,D,G]=S_\mathrm{cl}[\varphi,\pi] + \frac{i}{2}\Tr \ln D^{-1} + \frac{i}{2} \Tr(D_0^{-1}D) \\ - \frac{i}{2} \Tr \ln G^{-1} - \frac{i}{2} \Tr(\tilde{G}_0^{-1}G) + \Gamma_2\qty[D,G].
\end{multline}
$S_\mathrm{cl}$ is obtained by substituting ${\varphi}$ and ${\pi}$ in $S_\mathrm{ph}$ in Eq. (\ref{S_phi}). $\tilde{G}_0$ contains the contribution of ${\varphi}$ to spin dynamics
\begin{align}
    \tilde{G}_0^{-1}&\equiv \begin{bmatrix}
        0 & (\tilde{G}_0^A)^{-1} \\ (\tilde{G}_0^R)^{-1} & 0
    \end{bmatrix}, \\
    (\tilde{G}_0^R)^{-1} &\equiv \begin{bmatrix}
        i\partial_t & i \Delta & 0 \\ -i \Delta & i\partial_t & 2ig\sqrt{2\omega_c/N}{\varphi} \\ 0 & -2ig\sqrt{2\omega_c/N}{\varphi} & i\partial_t
    \end{bmatrix}, \\
     (\tilde{G}_0^A)^{-1} &= \qty[(\tilde{G}_0^R)^{-1}]^\dagger.
\end{align}
The last term $\Gamma_2$ is given  at NLO ($\mathcal{O}( N^0)$) by the diagram in Fig.~\ref{fig:Gamma_dicke} and reads as
\begin{multline}
    \Gamma_2^\mathrm{NLO} = \frac{4\omega_c g^2}{N}\sum_j  \oint \oint  dt_c \, dt'_c \, D^{\phi \phi}(t_c,t'_c) \\ \times \big(G^{yy}_{jj}(t_c,t'_c) G^{zz}_{jj}(t_c,t'_c) - G^{yz}_{jj}(t_c,t'_c) G^{zy}_{jj}(t_c,t'_c) \big).
\end{multline}

DE for fermions are similar to those for the disordered model, provided that we substitute
\begin{equation}
    B_x(t)=-2g\sqrt{\frac{2\omega_c}{N}}{\varphi}(t),
\end{equation}
and use the following fermion self energies in the basis of contour branches
\begin{align}
    \Sigma^{ss'}_{yz}(t,t')&=-\frac{8i\omega_c g^2}{N} \,  D^{ss'}_{\phi \phi}(t,t')G^{ss'}_{zy}(t,t'), \\
    \Sigma^{ss'}_{zy}(t,t')&=-\frac{8i\omega_c g^2}{N}\,  D^{ss'}_{\phi \phi}(t,t')G^{ss'}_{yz}(t,t'),\\
    \Sigma^{ss'}_{yy}(t,t')&=+\frac{8i\omega_c g^2}{N}\,  D^{ss'}_{\phi \phi}(t,t')G^{ss'}_{zz}(t,t'),\\
    \Sigma^{ss'}_{zz}(t,t')&=+\frac{8i\omega_c g^2}{N}\,  D^{ss'}_{\phi \phi}(t,t')G^{ss'}_{yy}(t,t').
\end{align}

\begin{figure}[!t]
    \centering
    \includegraphics[width=0.48\textwidth]{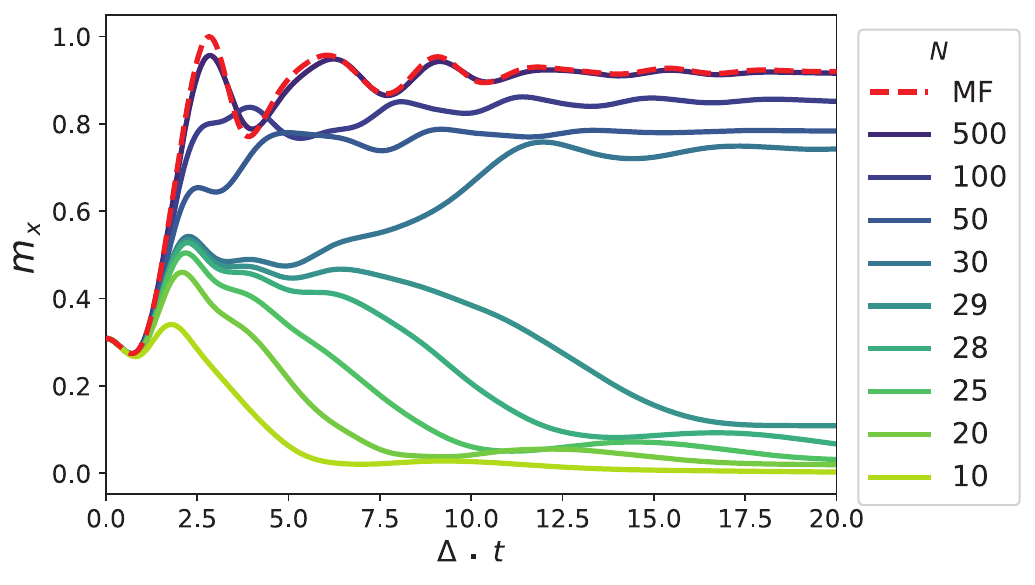}
    \caption{Spin dynamics for a quench into the SR phase of the Dicke model with photon loss, starting from $\theta_0=0.9\pi$. For $N<N^\star$ where $N^\star=30$, the symmetry is restored by fluctuations. $N^\star$ depends on the degree of symmetry breaking in the initial state. The other parameters are $g/g_c=1.6$, $\Delta/\omega_c=0.4$ and $\kappa/\omega_c=0.2$. }
    \label{fig:sx_Dicke}
\end{figure}

The equations of motion (EOM) for classical fields are given by
\begin{equation}
    \qty(\partial_t + \kappa){\varphi} = {\pi},
\end{equation}
\begin{equation}
    \qty(\partial_t + \kappa) {\pi} + \omega_0^2 {\varphi} +g\sqrt{2N\omega_0}  G^{K}_{zy}(t,t)=0.
\end{equation}
DE for photons are identical to those for the disordered system with the photon self energy
\begin{multline}
    \Pi^{ss'}(t,t')=-8i \omega_c g^2 \, \bigg(G^{ss'}_{yz}(t,t')G^{ss'}_{zy}(t,t')\\-G^{ss'}_{yy}(t,t')G^{ss'}_{zz}(t,t')\bigg),
\end{multline}
given  in the basis of contour branches.

Dynamics at LO are equivalent to the mean field treatment of the Dicke model with the critical coupling
\begin{equation}
    g_c=\frac12 \sqrt{\frac{\Delta(\omega_c^2+\kappa^2)}{\omega_c}}.
\end{equation}
At NLO the inclusion of fluctuations modifies the dynamics for smaller system sizes. For instance, after quenching the coupling into the SR regime $g>g_c$, the system may not end up in a symmetry broken state with $\expval{a}\neq 0$, even if the initial state is not symmetric (Fig. \ref{fig:sx_Dicke}). For smaller values of $N$, one should choose a larger $m_x()$, otherwise the symmetry will be restored. The symmetry restoration is due to the fact that for smaller values of $N$ fluctuation are stronger, resulting in the enhancement of the tunneling between the different minima in the energy of the system. This does not mean that the system is not in a SR state for small $N$. The photon population still saturates at a finite value $\sim N$.

\bibliography{refs}
\end{document}